\documentclass[prd,reprint,superscriptaddress,longbibliography,eqsecnum,nofootinbib,amsmath,amssymb,aps,floatfix]{revtex4-2}

\usepackage[utf8]{inputenc}
\usepackage[T1]{fontenc}
\usepackage[dvipsnames]{xcolor}
\usepackage{graphicx}
\usepackage{bm}
\usepackage[colorlinks=true,allcolors=blue]{hyperref}
\usepackage{mathrsfs}
\usepackage{leftidx}

\newcommand{\mem}{\mathrm{mem}}
\newcommand{\PN}{\mathrm{PN}}
\newcommand{\surr}{\mathrm{surr}}
\newcommand{\fit}{\mathrm{fit}}
\newcommand{\comp}{\mathrm{comp}}
\newcommand{\EMRI}{\mathrm{EMRI}}
\newcommand{\GW}{\mathrm{GW}}
\newcommand{\ISCO}{\mathrm{ISCO}}

\newcommand{\UVA}{Department of Physics, University of Virginia, P.O.~Box 400714, Charlottesville, Virginia 22904-7414, USA}

\begin{document}

\title{Waveform models for the gravitational-wave memory effect:\\
Extreme mass-ratio limit and final memory offset}

\author{Arwa Elhashash}
\email{aze5tn@virginia.edu}
\affiliation{\UVA}%

\author{David A.~Nichols}%
\email{david.nichols@virginia.edu}
\affiliation{\UVA}

\date{\today}

\begin{abstract}
The gravitational-wave (GW) memory effect is a strong-field relativistic phenomenon that is associated with a persistent change in the GW strain after the passage of a GW. The nonlinear effect arises from interactions of GWs themselves in the wave zone and is an observable effect connected to the infrared properties of general relativity. The detection of the GW memory effect is possible with LIGO and Virgo in a population of binary-black-hole (BBH) mergers or from individual events with next-generation ground- and space-based GW detectors or pulsar timing arrays. Matched-filtering-based searches for the GW memory require accurate, and preferably rapid-to-evaluate waveform models of the memory effect's GW signal. One important element of such a waveform model is a model for the final memory offset---namely, the net change in strain between early and late times. In this paper, we construct a model for the final memory offset from the merger of nonspinning BBH systems in quasicircular orbits. A novel ingredient of this model is that we first compute the memory signal for extreme mass-ratio inspirals using a high post-Newtonian-order analytic calculation, and we use this analytical result to fix the coefficient in the fit which is linear in the mass-ratio. The resulting memory-offset fit could be used for detecting the GW memory for binaries that merge on a timescale that is short relative to the inverse of the low-frequency cutoff of a GW detector. Additionally, this fit will be useful for analytic waveform models of the GW memory signals in the time and frequency domains.
\end{abstract}

\maketitle

\tableofcontents

\section{Introduction} \label{sec:intro}

Gravitational waves (GWs) from the mergers of nearly 100 binary black holes (BBHs) have been detected during the first three observing runs of LIGO, Virgo, and KAGRA~\cite{LIGOScientific:2018mvr,LIGOScientific:2020ibl,KAGRA:2021vkt}.
Based on the rate of detection candidates during the fourth observing run~\cite{web:GraceDB}, the number of confirmed detections is likely to double.
These GW observations have had important implications for both astrophysics and fundamental physics.
The increasing number of detections has led to a more precise characterization of the population of BBH mergers, including the distributions of the masses and spins of the individual BHs and the overall merger rate~\cite{LIGOScientific:2018jsj,LIGOScientific:2020kqk,KAGRA:2021duu}.
BBH mergers also have allowed the predictions of general relativity (GR) to be tested in a strong-gravity and high-luminosity regime of the theory that is challenging to explore with other methods and systems (see, e.g.,~\cite{LIGOScientific:2019fpa,LIGOScientific:2020tif,LIGOScientific:2021sio}).
The high luminosities associated with BBH mergers produce sufficiently strong gravitational waves that nonlinear interactions of the waves in the asymptotic wave zone around the source become significant; this produces features in the gravitational waveforms that are absent in the linearized approximation to GR.
The chief nonlinear feature that will be discussed in this paper is the nonlinear GW memory effect~\cite{Christodoulou:1991cr,Blanchet:1992br}, which is a distinctive and observable example of this type of nonlinear gravitational interaction.

The nonlinear GW memory effect, like the linear effect~\cite{Zeldovich:1974gvh}, leads to a lasting offset in the GW strain after a burst of GWs pass by a detector far from an isolated source.
This offset is produced by unbound stress energy carried by massive~\cite{Zeldovich:1974gvh} or massless~\cite{Epstein:1978dv,Turner:1978jj} particles, which includes the effective stress-energy of gravitational waves~\cite{Thorne:1992sdb}.
The memory is also part of an ``infrared triangle'' that relates the memory effect to the Bondi-Metzner-Sachs supertranslation symmetries~\cite{Bondi:1962px,Sachs:1962zza} of asymptotically flat spacetimes~\cite{Bondi:1962px,Sachs:1962wk} and to Weinberg's soft theorem~\cite{Weinberg:1965nx} (see, e.g.,~\cite{Strominger:2013jfa,Strominger:2014pwa,Strominger:2017zoo}).
This perspective of GW memory as being closely tied to the properties of classical and quantum gravitational scattering serves as an additional motivating factor for observational studies of the GW memory effect.
Analogous memory effects and infrared triangles exist in Yang-Mills theories (see, e.g.,~\cite{Strominger:2017zoo}), but the gravitational-wave memory---despite the relatively weak coupling of gravity compared to other fundamental interactions in nature---may have the best chance of being detected (as we discuss in more detail in the next paragraph).
Given that memory effects are generic predictions of gauge and gravitational theories which have not yet been measured, this is a compelling reason to search for these effects and verify experimentally that they are consistent with their theoretical predictions.

There are several studies of the detection prospects of the GW memory effect prior to the detection of GWs from BBH mergers (e.g.,~\cite{Braginsky:1985vlg,Braginsky:1987kwo,Kennefick:1994nw,Favata:2009ii,Pollney:2010hs}); however, an important change in viewpoint on memory detection occurred after the detection of the first BBH merger, GW150914~\cite{LIGOScientific:2016aoc}.
Despite the GW150914 event being a relatively high signal-to-noise-ratio (SNR) event (the network SNR was roughly 24~\cite{LIGOScientific:2016aoc}), the GW signature of the memory effect for such an event is small compared to the dominant harmonic (from which the SNR was computed) that was used in the detection and parameter estimation.
After the first GW150914 event, it was proposed in~\cite{Lasky:2016knh} that one could instead search for evidence of the GW memory effect in a population of BBH mergers in which each individual memory signal was below the threshold of detection.
Forecasts that take into account the now-better-constrained BBH-population properties and merger rates (namely,~\cite{LIGOScientific:2018jsj,LIGOScientific:2020kqk,KAGRA:2021duu}) have been performed that show the memory could be detected during the fifth observing run~\cite{Hubner:2019sly,Boersma:2020gxx,Grant:2022bla} when the LIGO detector is in its A+ configuration~\cite{KAGRA:2013rdx}.\footnote{The next generation of ground-based GW detectors, Einstein Telescope and Cosmic Explorer, will likely be able to detect the memory effect from single events~\cite{Grant:2022bla} and will be able to determine if the amplitude of the memory signal is consistent with the value predicted by general relativity to a few percent accuracy~\cite{Goncharov:2023woe}.}
Dedicated pipelines that search of the GW memory that make use of template-based searches have been applied to the events from all three gravitational-wave transient catalogues~\cite{Hubner:2019sly,Hubner:2021amk,Cheung:2024zow}, and other searches using minimally modeled methods have also been implemented~\cite{Ebersold:2020zah}.

Measuring the GW memory requires a well-defined notion of the memory signal (see, e.g.,~\cite{Grant:2021hga,Siddhant:2024nft}), and for template-based searches, a signal model that is fast to evaluate.
For comparable-mass BBH mergers most of the memory signal accumulates around the time of the merger, which typically implies that input from numerical-relativity (NR) simulations will be needed.
Gravitational waveforms computed using extrapolation methods fail to extract the memory from such simulations (see, e.g.,~\cite{Boyle:2019kee}), whereas those that use Cauchy-characteristic extraction~\cite{Bishop:1996gt,Bishop:1997ik} are capable of resolving the memory signal, and different NR codes have computed the signal from BBH mergers~\cite{Pollney:2010hs,Mitman:2020pbt} (see also the review~\cite{Mitman:2024uss}).
Because generating numerical-relativity waveforms is too computationally intensive to use to generate a sufficient number of GW templates for GW searches and parameter estimation, gravitational waveform models that take NR data as input and interpolate over the parameter space of binary mass ratios and spins are required.

Most efforts to compute the memory signal rely on the fact that it can be computed from oscillatory waveform modes (for which there are time- or frequency-domain waveform models) by using the effective stress-energy tensor of GWs~\cite{Favata:2009ii,Talbot:2018sgr} or the continuity equations for charges associated with asymptotic supertranslation symmetries (the supermomenta~\cite{Geroch:1977big,Wald:1999wa}), as described in~\cite{Nichols:2017rqr,Ashtekar:2019viz,Mitman:2020bjf}. 
These additional post-processing steps add a computational overhead to searches for the memory effect, and require more delicate signal processing to perform frequency-domain GW data analysis~\cite{Chen:2024ieh}.
There has been recent progress in making surrogate~\cite{Yoo:2023spi} and phenomenological frequency-domain waveform models~\cite{Valencia:2024zhi} that include the $l=2$, $m=0$ spin-weighted spherical-harmonic mode of the GWs, which is the dominant mode in which the memory signal appears for nonprecessing BBH systems.
However, the $l=2$, $m=0$ mode contains both memory and quasinormal-mode ringing; thus, they do not represent a ``memory-only'' signal, that would be required to perform a (Bayesian) model comparison between a signal with or without memory, which is the currently implemented method that is used to assess the significance of the GW memory effect in GW data.

This paper is a first step towards developing a stand-alone waveform model for the GW memory effect from nonspinning BBH mergers, which contains information only about the memory effect and not other linear or nonlinear GW phenomena.
Our focus will be on two aspects of the memory signal: the limit of extreme mass-ratio inspirals (EMRIs) and the final memory offset that is accumulated after the ringdown stage of a BBH merger.
The two will not be independent; the EMRI calculation will feed into the calculation of the final memory offset.
In addition to being an input for the full memory signal model, we can see a few other applications of both the EMRI calculation and the final memory offset. 

EMRIs, in which a stellar-mass compact-object inspirals into a supermassive black hole, are one well-studied class of sources that the space-based detector LISA~\cite{Amaro-Seoane:2017ADS} likely will observe~\cite{LISA:2022yao}.
They are of interest both in terms of their astrophysics~\cite{LISA:2022yao} and because the gravitational waves can map out the spacetime geometry precisely, in a result that is referred to as ``Ryan's theorem''~\cite{Ryan:1995wh}.
The calculations in this paper for EMRIs on quasicircular orbits will show that the memory accumulates mostly during the late inspiral, but over a timescale that for a ``typical'' EMRI (e.g., a 10$M_\odot$ BH inspiralling into a $10^6 M_\odot$ BH), will be slow compared to the longest period that the LISA detector can accurately measure.
Given that EMRIs likely will form with some residual eccentricity~\cite{LISA:2022yao}, and the memory signal from highly eccentric EMRI systems has a more complicated structure~\cite{Burko:2020gse}, the nonspinning, quasi-circular assumptions that are used in this paper should be revisited for future studies of the memory signals from EMRIs. 

Pulsar timing arrays are also on the cusp of a detection of the stochastic background of GWs at low frequencies (see, e.g.,~\cite{NANOGrav:2023gor,Reardon:2023gzh,EPTA:2023fyk,Xu:2023wog}).
NANOGrav and the Parkes Pulsar Timing Array perform searches for GW bursts with memory, where the bursts accumulate on a timescale that is short compared with the shortest period that they can measure~\cite{NANOGrav:2023vfo,Wang:2014zls}.
These searches do not specify a source for the memory signal, but they constrain the amplitude of the burst with memory.
While forecasts are more optimistic about detecting memory with LISA than pulsar timing arrays~\cite{Islo:2019qht} (see also~\cite{Gasparotto:2023fcg,Goncharov:2023woe,Inchauspe:2024ibs}), the final memory offset fit that we construct in this paper could be used to interpret the amplitude of a future pulsar-timing-array detection of the memory effect under the hypothesis that the burst with memory was produced by the merger of a supermassive BBH system.\footnote{Note that our fit for the $l=2$, $m=0$ spherical harmonic mode is a function of the total mass, (symmetric) mass ratio and luminosity distance to the source.
To use it in the context of interpreting a pulsar-timing-array detection, one would need to take into account not only the dependence of the amplitude on these three parameters in the fit, but also the dependence of the binary's inclination and the array's response to the memory signal as a function of sky position and polarization.}

\subsection{Summary and organization of this paper}

We now give a brief overview of the organization and main results of this paper.
In Sec.~\ref{sec:memory-multipolar}, we review our notation and our prescription for computing the memory signal in the $l=2$, $m=0$ spherical-harmonic mode from oscillatory ($m\neq 0$) modes.
Section~\ref{sec:waveform-multipolar} then describes the waveform modes used to compute the memory.
Because for comparable mass ratios, most of the memory accumulates during the late inspiral, merger and ringdown, we review in Sec.~\ref{sec:memory_comp} the numerical-relativity hybrid surrogate model NRHybSur3dq8~\cite{Varma:2018mmi} that we use to compute the memory signal (for nonspinning BBHs with mass ratios $1\leq q\leq 8$) in this regime.
Although the oscillatory modes of this surrogate model have been hybridized with Effective-One-Body waveforms, to speed up the calculation of the memory signal, we directly hybridize the memory computed from the NRHybSur3dq8 surrogate model with a 3PN memory waveform~\cite{Favata:2008yd} that accounts for the memory accumulated before the first time at which we evaluate the surrogate model.
This PN waveform is also discussed in Sec.~\ref{sec:memory_comp}.
Section~\ref{subsec:EMRIs} contains a discussion of the 22PN-accurate resummed, factorized waveforms for EMRIs on quasicircular orbits that were computed by Fujita in~\cite{Fujita:2012cm}, and an argument for why the inspiral waveforms there are sufficient to compute the memory to leading order in the mass-ratio expansion.
We also discuss how we use energy balance to compute the memory signal as a function of a PN parameter $v$ and of time $t$ along the worldline of the small compact object in the EMRI.

Section~\ref{sec:signals} contains some of the main results of the paper.
The first part, Sec.~\ref{sec:memory_comp_hybridization}, shows the result of hybridizing the surrogate NRHybSur3dq8 with a 3PN waveform that accounts for the memory accumulated from past infinity up to the starting time of the surrogate.
We also provide a polynomial fit (in the symmetric mass ratio) for the offset that must be added to the surrogate at its starting time, which also is valid for mass ratios within the range of hybridization ($1\leq q \leq 8$).
There is a similar polynomial fit for the time-of-coalescence parameter $t_c$ in the post-Newtonian waveform that is required for the hybridization, too.
The next part, Sec.~\ref{sec:EMRI_surrogate}, gives a more detailed illustration of the increasing importance of the inspiral contribution to the memory signal as the mass ratio becomes more extreme.
The remaining parts of Sec.~\ref{sec:signals} show the memory signal as a function of time and velocity and the impact of computing different numbers of oscillatory multipole moments and different PN orders on the memory signal and the final memory offset.

In Sec.~\ref{sec:memory-fit}, we construct two polynomial fits in symmetric mass ratio for the final memory offset from nonspinning BBH mergers on quasicircular orbits.
The first fit uses the hybridized surrogate, whereas the second uses the same surrogate data, but also fixes the coefficient linear in the symmetric mass ratio to be the value computed from the 22PN-order EMRI calculation.
The fit using comparable-mass data only overestimates the memory in the EMRI limit by about ten percent, and it performs similarly to the fit with EMRI information in the comparable-mass limit.
We conclude in Sec.~\ref{sec:conclusions}, and there are two Appendices~\ref{app:memory-integrand} and~\ref{app:memory-fit} that contain some supplementary results about the properties of the memory integrand and the final memory offset fit.

\section{Multipolar expansion of the memory signal}
\label{sec:memory-multipolar}

We describe how we compute the multipolar expansion of the memory signal in terms of a multipolar expansion of the oscillatory GW strain in this section.
We leave the discussion of which oscillatory GW modes we use to Sec.~\ref{sec:waveform-multipolar}.

We denote the multipole moments of the gravitational wave strain by $h_{lm}$, which are the multipoles that arise in the expansion of $h \equiv h_+ - i h_\times$ in terms of spin-weighted spherical harmonics:
\begin{equation}
    h \equiv h_+ - i h_\times = \sum_{l=2}^\infty \sum_{m=-l}^l h_{lm} (_{-2}Y_{lm}) .
\end{equation}
The coefficients $h_{lm}$ are functions of retarded time $u$, and the spherical harmonics are functions of the polar and azimuthal angles $(\theta,\phi)$, respectively.
Note that waveforms are often parameterized by a time $t$, where $t$ is the ordinary time at a fixed radius $r$ from the source (for example, at the detector).
Numerical-relativity waveforms often use $h_{lm}$, whereas those in the post-Newtonian approximation alternately use the radiative mass and current moments $U_{lm}$ and $V_{lm}$, which are related to $h_{lm}$ as follows:
\begin{align}
    \label{eq:strain_radiative_multipole_moments_relation}
    h_{lm} = \frac{1}{r\sqrt{2}}(U_{lm}-iV_{lm}) .
\end{align}
The multipoles $U_{lm}$ and $V_{lm}$ satisfy the relationship that $U_{l(-m)} = (-1)^m \bar U_{lm}$ (and similarly for $V_{lm}$), where the overline represents complex conjugation.
For the non-spinning binaries that we will consider in this paper, the transformation of $h_{lm}$ can also be written as $h_{l(-m)} = (-1)^l \bar h_{lm}$ because $U_{lm}$ is nonvanishing when $l+m$ is even and $V_{lm}$ is nonvanishing when $l+m$ is odd.

We compute the displacement memory signal from the balance laws for the flux of supermomentum, as described in~\cite{Nichols:2017rqr,Ashtekar:2019viz,Mitman:2020bjf,Grant:2022bla}.
The prescription used in those references calculates just the nonlinear contribution from the ``oscillatory'' (namely $m\neq 0$) multipole moments of the gravitational wave strain, which do not contain the memory effect.
The multipole moments of the memory strain are obtained from integrating a term quadratic in the time derivative of the strain with a spin-weighted spherical harmonic.
Because the strain itself is expanded in spin-weighted harmonics, the resulting angular integral will involve the integral of three spin-weighted harmonics. 
We use the notation of~\cite{Nichols:2017rqr} (based on that in~\cite{Beyer:2013loa}) for these integrals. 
Specifically, we define
\begin{align} \label{eq:CldefIntegral}
& C_l(s',l',m';s'',l'',m'') \equiv \nonumber\\
& \int d^2\Omega \, (_{s'+s''}\bar{Y}_{lm'+m''})(_{s'}{Y}_{l'm'})(_{s''}{Y}_{l''m''}) ,
\end{align}
where the coefficients are nonvanishing only for $l$ in the set $\Lambda$ defined by 
\begin{align} \label{eq:Lambda-set}
&\Lambda \equiv \nonumber \\
&\{\max(|l'-l''|,|m'+m''|,|s'+s'' |),...,l'+l''-1,l'+l''\} \, .
\end{align}
To compute these coefficients numerically, we will use the fact that they can be written in terms of Clebsch-Gordon coefficients:
\begin{align} \label{eq:CldefClebschGordon}
C_l(s',l',m';s'',l'',m'')= (-1)^{l+l'+l''} \sqrt{\frac{(2l'+1)(2l''+1)}{4 \pi (2l+1)}} \nonumber\\
\times \left< l',s';l'',s''|l,s'+s'' \right> \left< l',m';l'',m''|l,m'+m'' \right> .
\end{align}
The conventions for the Clebsch-Gordon coefficients that we use are those implemented in \textsc{Mathematica}. 

When written in terms of the moments $U_{lm}$ and $V_{lm}$ the multipole moments of the memory strain, $h_{lm}^\mem$, are given by
\begin{align} \label{eq:hlm-mem}
    h_{lm}^{\mem}(u) = {} & \frac 1{4r} \sqrt{\frac{(l-2)!}{(l+2)!}} \sum_{l',l'',m',m''} \! \! C_l(-2,l',m';2,l'',m'')\nonumber\\
    &\times \int_{-\infty}^{u} du' \, \Big[ 2i s^{l,(-)}_{l';l''} \dot{U}_{l'm'}\dot{V}_{l''m''}\nonumber\\
    &+s^{l,(+)}_{l';l''}(\dot{U}_{l'm'}\dot{U}_{l'',m''}+\dot{V}_{l'm'}\dot{V}_{l'',m''})\Big] .
\end{align}
We also defined the coefficients $s^{l,(\pm)}_{l';l''}$ by
\begin{equation}
        s^{l,(\pm)}_{l';l''} = 1 \pm (-1)^{l+l'+l''} .
\end{equation}
The sum over the indices $l'$, $l''$, $m'$ and $m''$ in Eq.~\eqref{eq:hlm-mem} must satisfy the constraints that $l'$, $l''\geq 2$ as well as $|m'| \leq l'$ and $|m''| < l''$. 
However, for a fixed $l$ and $m$ on the left-hand side of Eq.~\eqref{eq:hlm-mem}, the coefficients $C_l(-2,l',m';2,l'',m'')$ in the sum will only be nonzero when $m = m' + m''$ as well as when $l$, $l'$ and $l''$ satisfy the relationships in Eq.~\eqref{eq:Lambda-set}.
This will decrease the number of modes required to compute the memory signal for particular values of $l$ and $m$.

In this paper, we focus on the $l=2$, $m=0$ mode of the memory signal.
This will require that $m'' = -m'$.
Specializing Eq.~\eqref{eq:hlm-mem} to this case gives
\begin{align} \label{eq:Deltah20UV}
    h^{\mem}_{20}(u) = {} & \frac{\sqrt{6}}{48r}\sum_{l',l'',m'} \! \! C_2(-2,l',m';2,l'',-m') \nonumber\\
    &\times \int_{-\infty}^{u} du' \, \Big\{ 2i s^{2,(-)}_{l';l''} \dot{U}_{l'm'}\dot{V}_{l''(-m')} \nonumber\\
    &+s^{2,(+)}_{l';l''} \big[\dot{U}_{l'm'}\dot{U}_{l''(-m')}+\dot{V}_{l'm'}\dot{V}_{l''(-m')} \big] \Big\} .
\end{align}

We will be suppressing the ``mem'' superscript in the rest of this paper for the $(2,0)$ mode, because we only use oscillatory modes with $m\neq 0$ on the right-hand side of Eq.~\eqref{eq:Deltah20UV} (and there will be no ambiguity that the $m=0$ modes are the ''memory modes'').
In addition, specializing to $l=2$ requires that the magnitude of the difference of $l'$ and $l''$ (i.e., $|l'-l''|$), is at most 2 for these modes to contribute to the $(2,0)$ memory mode. 

In the discussion that follows, we find it convenient to commute the sum and integral in Eq.~\eqref{eq:Deltah20UV} to write the memory signal in the form 
\begin{equation} \label{eq:Deltah20_h20dot}
    h_{20}(u) = \int_{-\infty}^{u} \! du' \, \dot{h}_{20} ,
\end{equation}
where the dot means a derivative with respect to $u'$.\footnote{For waveforms parameterized by $t$, we would instead have the analogous expression
\begin{equation} \label{eq:hlm-t}
    h_{20}(t) = \int_{-\infty}^{t} \! dt' \, \dot{h}_{20} ,
\end{equation}
where the dot now denotes a derivative with respect to $t'$.}
The expression for the integrand $\dot h_{20}$ can be inferred from Eq.~\eqref{eq:Deltah20UV}.
We will also find it useful to introduce the notation for the ``final memory offset''
\begin{equation}
    \Delta h_{20} = \lim_{u\rightarrow\infty} h_{20}(u) ,
\end{equation}
which represents the total memory strain accumulated over all times.

\section{Waveform multipole moments and memory signals}
\label{sec:waveform-multipolar}

We now turn to discussing which waveform multipole moments $U_{lm}$ and $V_{lm}$ we input into the right-hand side of Eq.~\eqref{eq:hlm-mem} to compute the memory signal.
We first cover the waveform modes required for the comparable-mass (mass ratios $q=m_1/m_2 < 8)$, and then we turn to the extreme mass-ratio case.

\subsection{Comparable mass ratios}\label{sec:memory_comp}

\begin{table}[t!]
    \centering
    \caption{$m$ values for each $l$ value in the NRHybSur3dq8 surrogate model that are used in the calculations for comparable mass ratios.}
    \begin{tabular}{lcccc}
    \hline
    \hline
       & $l=2$ & $l=3$ & $l=4$ & $l=5$ \\
       \hline
       $m$ values  & $\{\pm 1, \pm 2\}$ & $\{\pm 1, \pm 2\, \pm 3\}$ & $\{\pm 2, \pm 3, \pm 4\}$ & $\{\pm 5\}$ \\
    \hline
    \hline
    \end{tabular}
    \label{tab:lm-modes-sur}
\end{table}

For comparable mass ratios, the memory signal grows most rapidly and accumulates most of its offset close to the merger, when NR (or IMR waveforms fit to NR) are necessary to accurately model the gravitational waves.
Thus, we will need to use an IMR waveform, and we use the NR hybrid surrogate model NRHybSur3dq8~\cite{Varma:2018mmi}, which is calibrated for aligned spin BBHs with mass ratios $1\leq q\leq 8$ (we specialize to nonspinning systems, however).
The oscillatory modes in NRHybSur3dq8 have been hybridized with EOB waveforms to produce waveforms that allow the surrogate to be evaluated for times much longer than the duration of the numerical relativity waveforms from which the surrogate model is built.

The NRHybSur3dq8 model contains only a subset of all the $l\geq 2$ modes in the waveform, which are listed in Table~\ref{tab:lm-modes-sur}.
We compute the contribution to the displacement memory signal in the $(2,0)$ mode from Eq.~\eqref{eq:Deltah20UV}
using the surrogate modes given in Table~\ref{tab:lm-modes-sur}.
The integral for the memory signal has as its lower limit negative infinity, which would require evaluating the surrogate model for an infinitely long time.
This, however, is not feasible, so we instead would like to determine the appropriate initial offset to apply to the surrogate memory at a finite starting time.

We achieve this by hybridizing the memory signal computed from the surrogate with the 3PN memory waveform that had been computed by Favata~\cite{Favata:2008yd}.
The 3PN memory signal in~\cite{Favata:2008yd} is written in terms of the post-Newtonian parameter $x$, which is defined to be
\begin{align} \label{eq:xPN}
    x \equiv (M\Omega)^{2/3} ,
\end{align}
where $M = m_1 + m_2$ is the total mass of the binary with primary mass $m_1$ and secondary mass $m_2$, and $\Omega$ is the orbital frequency of the circular orbit.
At Newtonian order, the PN parameter can be written in terms of time as 
\begin{align} \label{eq:x-of-t}
    x(t) = \frac 14 \left[ \frac{\eta}{5M} (t_c-t) \right]^{-1/4} .
\end{align}
The time here, is usually the coordinate time in the near zone of the source, but one can write it in terms of retarded time $u$, too.
We introduced the parameter $t_c$, which is the time of coalescence, and the symmetric mass ratio, $\eta$.
There are several equivalent expressions for it in terms of the individual masses, $m_1$ and $m_2$ or the mass ratio $q = m_1/m_2$:
\begin{equation}
    \eta = \frac{q}{(q+1)^2} = \frac{m_1 m_2}{M^2} = \frac{\mu}M .
\end{equation}
The last equality used the definition of the reduced mass $\mu = m_1 m_2/M$.
We reproduce Favata's expression for the 3PN memory in terms of $x$:
\begin{widetext}
\begin{align}
    \label{eq:3PN_memory}
    h_{20}^{\PN}(x) = {} & \frac{4}{7} \sqrt{\frac{5\pi}{6}} \eta x \bigg\{1+ x \bigg(-\frac{4075}{4032}+\eta \frac{67}{48}\bigg)
    + x^2 \bigg(-\frac{151877213}{67060224} - \eta \frac{123815}{44352} + \eta^2 \frac{205}{352}\bigg) + \pi x^{5/2} \bigg(-\frac{253}{336} + \eta  \frac{253}{84}\bigg) \nonumber\\
    &  + x^3 \bigg[-\frac{4397711103307}{532580106240} + \eta\bigg(\frac{700464542023}{13948526592} - \frac{205}{96} \pi^2\bigg) + \eta^2\frac{69527951}{166053888}   + \eta^3\frac{1321981}{5930496} \bigg]\bigg\} .
\end{align}
\end{widetext}
We introduced the ``PN'' superscript on the $(2,0)$ mode of the strain to indicate it is only valid in the regime of validity of the PN expansion.

As a brief comment, the $(2,0)$ mode at 3PN order is computed from oscillatory waveform modes with $l \leq 6$, so it contains information about additional $l$ modes that are not included in the surrogate modes in Table~\ref{tab:lm-modes-sur}.
However, it is still possible to achieve a robust hybridization between the surrogate and the 3PN memory as we show in more detail in Sec.~\ref{sec:memory_comp_hybridization}.

\subsection{Extreme mass ratios} \label{subsec:EMRIs}

For extreme mass-ratio systems, we will work to linear order in the symmetric mass ratio $\eta \ll 1$.
Such systems tend to be strongly relativistic, so that perturbation theory about a Schwarzschild background (for non-spinning binaries) is a more suitable approach for modeling these systems (see, e.g.,~\cite{Barack:2018yvs}).
Thus, neither the 3PN memory waveform nor the NRHybSur3dq8 surrogate will be well suited for computing the memory signal from EMRIs.
There is a more recent EMRI surrogate, EMRISur1dq1e4~\cite{Rifat:2019ltp}, which is calibrated up to a mass ratio of $q=10^4$, and spans a duration of time of order $10^4 M$.
This could be used for EMRI systems with more comparable mass ratios.
The rest of this subsection covers how we compute a memory signal that is accurate to linear order in $\eta$.

\subsubsection{Contributions of different stages of an extreme mass-ratio binary coalescence to the memory signal}

The coalescence of an extreme mass-ratio binary takes place in several stages.
The stage with the longest duration is referred to as the ``adiabatic inspiral''; in this stage, the small compact object can be modeled as adiabatically evolving between a sequence of bound orbits because of gravitational radiation reaction.
As the small compact object approaches the innermost stable circular orbit (ISCO) radius, the adiabatic approximation becomes inaccurate, and the binary's evolution is better described by a ``transition from inspiral to plunge''~\cite{Ori:2000zn} (such a transition also occurs for comparable-mass binaries; see, e.g.,~\cite{Buonanno:2000ef}).
As described in~\cite{Compere:2021iwh,Compere:2021zfj,Kuchler:2024esj}, the characteristic timescale to cross the ISCO in this transition stage goes as $M/\eta^{1/5}$. 
Naturally, the ``plunge'' stage follows this transition stage, which occurs after the small compact object crosses the ISCO and before it enters the event horizon of the more massive black hole.
At the end of the plunge, the small compact object passes through the event horizon of the primary black hole, and there will be a merger and ringdown signal in the gravitational waveform associated with this process.

We now argue that for extreme mass-ratio systems, the majority of the GW memory offset will accumulate during the adiabatic inspiral.
For any mass ratio (and all stages of the binary's evolution), the radiative moments $U_{lm}$ and $V_{lm}$ are proportional to the reduced mass $\mu = M \eta$.
Their time derivatives are proportional to $\eta \Omega \sim \eta x^{3/2}/M$ times the radiative moments.
The integrand that arises in the memory integral in Eq.~\eqref{eq:hlm-mem} is proportional to $\eta^2$.
However, the memory accumulates over the radiation-reaction timescale, which goes like $M/\eta$, which is why the memory strain scales as $\eta M/r = \mu/r$ (up to other dimensionless factors).
This can be determined more quantitatively by assuming energy balance, writing the integral with respect to the PN parameter $x$, and converting the integration measure from $dt$ [as in Eq.~\eqref{eq:hlm-t}] to $dx$.
This procedure introduces a factor of $dt/dx$, which scales as $1/\eta$, thereby canceling one factor of $\eta$ in the integrand for the memory.
The integral is evaluated over an $\eta$-independent range of $x$, which leads to the scaling linear in $\eta$.

The transition from inspiral to plunge, followed by the plunge, takes place over the timescale $M/\eta^{1/5}$ described in~\cite{Compere:2021iwh,Compere:2021zfj,Kuchler:2024esj}.
Given that the memory integrand scales with $\eta^2$, a scaling argument (like that described for the adiabatic inspiral above) suggests that the contribution to the GW memory offset should scale as $\eta^{9/5}$ from these stages.
It would be beneficial to perform a more quantitative calculation of the memory signal during these stages in future work, however.
In our subsequent calculations, we will ignore the contributions to the memory from the transition from inspiral to plunge and the plunge itself, because we expect that they contribute to the memory signal at higher orders in the mass ratio than the leading linear effect during the adiabatic inspiral.
We will also truncate the adiabatic inspiral at the ISCO radius rather than at the ISCO radius plus a correction term of order $\eta^{2/5}$ times the ISCO radius (which is the length scale over which the transition takes place), for simplicity.

Finally, during the merger and ringdown stages of the waveform (where we are referring to the merger as the end of the plunge phase), the integrand for the memory signal, with $dt$ as the measure [as in Eq.~\eqref{eq:hlm-t}], should scale as $\eta^2$, but the memory accumulates on a timescale of order the dynamical time of the massive black hole, which is determined by $M$ and is independent of $\eta$.
Thus, the memory signal generated during the merger and ringdown will be of order $\eta^2 M/r = \eta \mu/r$.
In the comparable mass-ratio limit, the symmetric mass ratio is of order $\eta \sim 1/4$, and the more relativistic speeds during the merger and higher radiative losses compensate for the shorter timescale over which the memory accumulates (and in fact, the merger and ringdown stages produce the largest part of the memory signal).
For extreme mass-ratios however, the $\eta^2$ scaling of the memory during the merger and ringdown also makes its contribution negligible (when working to linear order in $\eta$), despite the fact that the merger is nominally the most relativistic stage of a binary merger.
We will provide a visual illustration of the relative importance of the adiabatic inspiral over the other stages of the coalescence in Sec.~\ref{sec:EMRI_surrogate}.
Because the adiabatic inspiral waveforms will be sufficient to compute the memory signal at the accuracy in $\eta$ at which we are working, we next describe the waveform modes that we use during the adiabatic inspiral.

\subsubsection{Factorized post-Newtonian waveform} \label{subsubsec:factorized-hlm}

We use the 22PN analytical waveforms computed by Fujita~\cite{Fujita:2012cm} for Schwarzschild black holes based on the expansion of analytical solutions to the Teukolsky equation~\cite{Teukolsky:1973ha} in the low-frequency limit~\cite{Mano:1996vt}. 
These waveforms are typically written in terms of $v \equiv \sqrt x = (M\Omega)^{1/3}$, as defined in Eq.~\eqref{eq:xPN}. 
The waveforms are written in a resummed, factorized form (see~\cite{Damour:2008gu,Fujita:2010xj}) which we now review.
The discussion below will focus on modes with $m>0$; modes with $m<0$ can be obtained from the fact that for nonprecessing binaries $h_{l(-m)} = (-1)^l \bar h_{lm}$.
The multipole moments of the strain are written as
\begin{equation} \label{eq:hlm-factorized}
    h_{lm} = h_{lm}^{(\mathrm{N},\epsilon_p)} \hat S_\mathrm{eff}^{(\epsilon_p)} T_{lm} e^{i\delta_{lm}} (\rho_{lm})^l .
\end{equation}
The label $\epsilon_p$ is the parity of the mode for nonprecessing binaries:
\begin{equation} \label{eq:epsilon_p}
    \epsilon_p =
    \begin{cases}
         0 & \mbox{for} \ l+m \ \mbox{even}\\
         1 & \mbox{for} \ l+m \ \mbox{odd} 
    \end{cases}
    .
\end{equation}
The first term $h_{lm}^{(\mathrm{N},\epsilon_p)}$ in Eq.~\eqref{eq:hlm-factorized} is the leading-order ``Newtonian'' part of the waveform.
It is linear in $\eta$ and given by 
\begin{equation}
    h_{lm}^{(\mathrm{N},\epsilon_p)} = \frac{\mu}r n^{(\epsilon_p)}_{lm} (-v)^{l+\epsilon_p} Y_{l-\epsilon_p,-m}(\pi/2, \phi) ,
\end{equation}
where the coefficient $n^{(\epsilon_p)}_{lm}$ is defined in the parity even and odd cases, respectively by
\begin{subequations}
\begin{align}
    n^{(0)}_{lm} = {} & \frac{8\pi(im)^l}{(2l+1)!!} \sqrt{\frac{(l+2)(l+1)}{l(l-1)}} , \\
    n^{(1)}_{lm} = & - \frac{16\pi i(im)^l}{(2l+1)!!} \sqrt{\frac{(l+2)(2l+1)(l^2-m^2)}{(2l-1)(l+1)l(l-1)}} .
\end{align}
\end{subequations}

The effective source $S_\mathrm{eff}^{(\epsilon_p)}$ is given by the relativistic reduced energy ($\tilde E > 0$) of stable circular geodesics in the Schwarzschild spacetime for the even-parity case and $v$ times the reduced angular momentum per total mass of the geodesics for the odd-parity case:
\begin{equation} \label{eq:Seff}
    S_\mathrm{eff}^{(\epsilon_p)} = 
    \begin{cases}
        \tilde E = \dfrac{1-2v^2}{\sqrt{1-3v^2}} & \mbox{for} \ \epsilon_p = 0 \\
        \dfrac{v \tilde L}M = \dfrac 1{\sqrt{1-3v^2}} & \mbox{for} \ \epsilon_p = 1 
    \end{cases}
    .
\end{equation}
This effective source can be expanded straightforwardly in a PN series in $v$.
The factor $T_{lm}$ is a ``resummed tail factor'' that is defined to be
\begin{equation}
    T_{lm} = \frac{\Gamma(l+1 - i 2m v^3)}{\Gamma(l+1)} e^{m\pi v^3} e^{i 2 m v^3 \ln(4m v^3/\sqrt{e}) } .
\end{equation}

The natural log of $T_{lm}$, for small $v$, can be expanded in terms of a Taylor series with coefficients involving the polygamma function; it is then straightforward to use the Taylor series for the exponential of this series to obtain the post-Newtonian expansion of $T_{lm}$.
The terms $(\rho_{lm})^l$ and $e^{i\delta_{lm}}$ represent the remaining amplitude and phase of the waveform that cannot be expressed in terms of the Newtonian, effective source, or resummed tail factors.
Post-Newtonian series for $\rho_{lm}$ and $\delta_{lm}$ can be downloaded in a format adapted for \textsc{Mathematica} on the webpage~\cite{web:BHPclub}.
The expressions are quite lengthy series in powers of $v$ (including terms of the form $v^j (\ln v)^k$, for whole numbers $j$ and $k$ with $j > k$); thus, we do not give their explicit expressions here.

The result of this calculation is that we can obtain the radiative modes $U_{lm}$ and $V_{lm}$ or $h_{lm}$ as PN series in the parameter $v$, which can then be used to compute the EMRI memory signal, as described next.

\subsubsection{Evaluating the memory signal} \label{subsubsec:eval-mem}

Once the modes $h_{lm}$ are computed, we can use Eqs.~\eqref{eq:strain_radiative_multipole_moments_relation} and~\eqref{eq:Deltah20UV} to compute the memory signal $h_{20}$ for the inspiral of an EMRI.
Because $\dot v$ is an order $\eta$ term, whereas $\dot \phi = \Omega = v^3/M$ is $\eta$ independent, then the time derivatives of $h_{lm}$ satisfy 
\begin{equation}
    M \dot h_{lm}(v) = im v^3 h_{lm}(v) + O(\eta^2) \, .
\end{equation} 
Thus, the time derivatives are straightforward to compute analytically.
For the expression for the memory signal $h_{20}$ to be accurate to 22 PN order, we need to include oscillatory memory terms up to $l'=25$ (and similarly for $l''$) in Eq.~\eqref{eq:Deltah20UV}; as a result, there are hundreds of terms in the sum that contribute to the memory signal.

The product of two modes in the integrand in Eq.~\eqref{eq:Deltah20UV} always involves one mode with positive $m$ multiplying one with $-m$.
Because the modes $h_{lm}$ satisfy the complex-conjugate relationship $h_{l(-m)} = (-1)^l \bar h_{lm}$, then the terms in the integrand of the form $|\dot U_{lm}|^2$ (or similarly for $V_{lm}$) depend only on the amplitudes of all the terms in the factorized waveform in Eq.~\eqref{eq:hlm-factorized}.
However, for the terms that involve products of $U_{lm}$ (or $V_{lm}$) modes with different $l$, then the phases will contribute, but only the parts of the phases that are dependent on both $l$ and $m$ (for example, the $\phi$ dependence in the Newtonian part of the waveform will not contribute).
This phase dependence then comes from the supplementary phase $\delta_{lm}$ and the phase in the resummed tail $T_{lm} = |T_{lm}|e^{i\tau_{lm}}$.
The resulting memory waveform can then be expressed in terms of the amplitude and the cosine of the difference in the phase $\delta_{lm}+\tau_{lm}$ for modes with different values of $l$.
These cosine terms must be expanded in a PN series, as well.

Finally, to compute the integral in Eq.~\eqref{eq:Deltah20UV}, we postulate the energy balance holds adiabatically.
More specifically, we assume the secondary evolves from a circular geodesic parameterized by a relativistic specific energy $\tilde E > 0$ to another circular orbit with a smaller such energy in response to radiative losses from gravitational-wave emission.
The specific energy $\tilde E$ is that of a test particle moving on a circular orbit around a Schwarzschild black hole, which is the same expression as that given in the $\epsilon_p = 0$ case of Eq.~\eqref{eq:Seff}, namely
\begin{equation} \label{eq:energy}
    \tilde E=\frac{1-2v^2}{\sqrt{1-3v^2}} .
\end{equation}
We will consider losses that include the GW luminosity radiated to infinity and into the horizon:
\begin{equation}\label{eq:dEdu_total}
    \frac{dE_\GW}{dt} = \left(\frac{dE}{dt}\right)_{\!\infty} + \left(\frac{dE}{dt}\right)_{\!H} .
\end{equation}
The time variable $t$ on the left-hand is the time that parameterizes the geodesic followed by the secondary in the EMRI system.
While it is most natural to parameterize the power radiated to infinity by retarded time $u=t-r_*$ (where $r_*$ is the tortoise coordinate in the Schwarzschild spacetime) and the power radiated into the horizon by advanced time $t+r_*$, they can equivalently be parameterized by $d/dt$.\footnote{This follows because a differential change in retarded time is related to the coordinate time at the location of the geodesic by $du = dt - dr_* = dt(1-\dot r_*) \approx dt$, where $\dot r_*$, the change in radial position of the particle is an order $\eta$ change on circular orbits, and we have been consistently ignoring such higher order in $\eta$ terms throughout this paper.}
For our main computations, we will use 22PN-accurate expressions for the GW luminosity at infinity and 22.5PN-accurate expression for the power radiated into the horizon, which are given in~\cite{Fujita:2012cm}; the lengthy expressions for these GW luminosities also can be obtained from~\cite{web:BHPclub}.
We will perform convergence tests in which we consider lower PN orders (which also involves fewer multipole moments of the luminosities---see Sec.~\ref{subsubsec:PN-orders} for further detail).

We next rewrite the integral with respect to $u$ in Eq.~\eqref{eq:Deltah20_h20dot} instead as an integral with respect to the velocity $v = (M\Omega)^{1/3}$.
To do so, we use the chain rule, the fact that $E = \mu \tilde E$ (for the reduced mass $\mu$), the fact that $dt=du$ (as discussed above), and the notion of energy balance,
\begin{equation}
    \frac{dE}{dt} = - \frac{dE_\GW}{dt} .
\end{equation}
This then allows us to write
\begin{align}\label{eq:Deltah20_intg_eq}
    h_{20}(v) = & \int_{0}^{v} dv' \frac{dh_{20}}{dv'} ,
\end{align}
where we defined
\begin{equation}\label{eq:dhdv}
    \frac{dh_{20}}{dv} = \frac{dE}{dv} \left(\frac{dE}{dt}\right)^{-1} \dot{h}_{20} = - \frac{dE}{dv} \left(\frac{dE_\GW}{dt}\right)^{-1} \dot{h}_{20} ,
\end{equation}
We also assumed that the parameter $v$ goes to zero as $u$ goes to minus infinity.

The derivative of the energy $d\tilde E/dv$ is 
\begin{equation}\label{eq:dEdv}
    \frac{d\tilde E}{dv} = \frac{v(6 v^2-1)}{(1 - 3 v^2)^{3/2}} ,
\end{equation}
and can be expanded to the necessary PN order.
When combined with the PN expansions for the inverse of $dE_\GW/dt$ in Eq.~\eqref{eq:dEdu_total} and the PN expansion of $\dot h_{20}$, which can be obtained from Eqs.~\eqref{eq:Deltah20UV} and~\eqref{eq:Deltah20_h20dot}, we can obtain the 22PN accurate expansion of $dh_{20}/dv$ or $h_{20}(v)$.

We will also find it convenient to have a PN expression for $h_{20}(t)$, which we obtain by integrating $dt$ from some reference time $t_i$ in the past:
\begin{equation} \label{eq:t-of-v}
    t-t_i = \int_{t_i}^{t} dt' = -\int_{v_i}^v dv' \left(\frac{dE}{dv'}\right) \left(\frac{dE_\GW}{dt}\right)^{-1}
\end{equation}
to give $t$ as a function of $v$.
We then can construct an interpolation function for $v(t)$, which we substitute into our expression for $h_{20}(v)$ to obtain $h_{20}(t)$.

\subsubsection{Required multipole and PN orders for computing the memory} \label{sec:hmem_EMRI_22PNl25}

We described in Sec.~\ref{subsubsec:eval-mem} how the memory signal can be computed from the factorized waveform modes $h_{lm}$ (see Sec.~\ref{subsubsec:factorized-hlm} for a discussion of how the factorized modes are computed) and the GW luminosity $dE/dt$.
For the luminosity, the part radiated to infinity (into the horizon) has been computed in~\cite{Fujita:2012cm} (respectively, \cite{Fujita:2014eta}) up to 22PN (22.5PN) order beyond the leading quadrupole formula for the power at infinity,
\begin{equation} \label{eq:dEdu_infinity_Newtonian}
    \left(\frac{dE}{dt}\right)_{\! N} = \frac{32}{5} \eta^2 v^{10} .
\end{equation}
To compute the power at infinity, Ref.~\cite{Fujita:2012cm} uses the expression
\begin{equation}
\left(\frac{dE}{dt}\right)_{\!\infty} =\frac{r^2}{16 \pi}\sum_{l,m} \left|\dot{h}_{lm} \right|^2 ,
\end{equation}
where the sum runs over integers $l\geq 2$ and $m\in[-l,l]$.
Having an accuracy of 22PN order requires including $\dot h_{lm}$ modes with $l$ running from 2 to 24, because the scaling of $\dot{h}_{lm}$ with $v$ is $\dot{h}_{lm} \sim v^{l+\epsilon_p+3}$.
For the $l=24$ modes, only the leading ``Newtonian'' part of the mode contributes for $\epsilon_p = 0$; however, for the $l=2$, $m=2$ mode, the full 22PN accuracy of the mode is required.\footnote{Interpolating between these two values, one can find that for a mode $h_{lm}$ with $l\geq 2$ and $m\in[-l,l]$ that $(24-l-\epsilon_p)$ PN orders beyond the leading ``Newtonian'' part of the mode would be required to obtain the contribution of $\dot h_{lm}$ to the radiated power at 22PN orders beyond the quadrupole formula.}
To compute the radiated power at infinity, we use the per mode data that is available on the website~\cite{web:BHPclub}.
That data is given in terms of modes $\eta_{lm}^{(\infty)}$ (which are \emph{not} related to our symmetric mass ratio, $\eta = \mu/M$).
The $\eta_{lm}^{(\infty)}$ related to the GW luminosity at infinity by
\begin{equation} \label{eq:dEdu_infinity}
    \left(\frac{dE}{dt}\right)_{\!\infty} = \left(\frac{dE}{dt}\right)_{\! N} \sum_{l=2}^{24}\sum_{m=1}^{m=l} \eta_{lm}^{(\infty)} .
\end{equation}
Note that the normalization of the expression is such that only positive values $m$ are summed over.

The radiated power into the future event horizon involves a different number of multipole modes, because it is determined by the magnitude squared of different radiative degrees of freedom: specifically, the shear of the generators of the horizon, which is related to the time integral of the Weyl scalar $\Psi_0$~\cite{Hawking:1972hy}.
The PN scaling of each $(l, m)$ mode of the flux goes as $v^{10+4l+2\epsilon_p}$ on the horizon, rather than $v^{6+2l+2\epsilon_p}$ at infinity; this implies that the maximum required value of $l$ will be $l=11$.
In the paper~\cite{Fujita:2014eta} and the data available on the website~\cite{web:BHPclub}, the per mode contributions to the GW luminosity down the horizon, $\eta^{(H)}_{lm}$, are normalized such that
\begin{equation} \label{eq:dEdu_horizon}
    \left(\frac{dE}{dt}\right)_{\! H} = v^5 \left(\frac{dE}{dt}\right)_{\! N}  \sum_{l=2}^{11}\sum_{m=1}^{m=l} \eta^{(H)}_{lm} ,
\end{equation}
where again the sum runs over positive values of $m$ only.

For computing the memory signal $h_{20}$, the required multipoles $\dot h_{lm}$ are those at infinity, but they are not precisely the same as those used for calculating the GW luminosity at infinity, in the following sense.
Because the expression in Eq.~\eqref{eq:Deltah20UV} involves products of modes with $|l'-l''|\leq 2$, then this requires computing many of the oscillatory $h_{lm}$ modes at 0.5 or 1PN order higher than is required for the radiated power.
In addition, it also requires some of the $l=25$ modes (specifically those with $\epsilon_p =0$).
The 1PN-order-higher maximum here is specific to the $l=2$ memory signal; for $l>2$, this would require evaluating some modes at $(l/2)$PN orders higher than is required for the equivalent PN order in the radiated power.

\section{Memory signals for comparable and extreme mass ratios} \label{sec:signals}

In this section, we summarize the main features of the memory signals from nonspinning binary black-hole mergers in comparable, intermediate and extreme mass-ratio binaries.

\subsection{Memory for comparable mass ratios} \label{sec:memory_comp_hybridization}

For comparable mass-ratio binaries, we argued in Sec.~\ref{sec:memory_comp} that to obtain a memory signal of arbitrary length in time, it is more efficient computationally to hybridize the NRHybSur3dq8 signal to a 3PN memory waveform than to evaluate the surrogate for a long stretch and compute the memory waveform from just the surrogate model.
We now describe how we perform this hybridization.

Given the different coordinate conditions in PN and NR calculations, and the fact that there is not a known mapping between the two conditions, some postulates must be made to hybridize the two waveforms.
We assume that the individual masses $m_1$ and $m_2$ have the same values in the PN and NR contexts and that the time variables $t$ have the same meaning in both cases.   
With these assumptions, we can perform the hybridization by adding two free parameters, one in the NR surrogate and one in the PN memory waveform, and specifying the initial and final times over which the hybridization takes place.
Specifically, for the NR surrogate, we add a positive, undetermined constant, $h_0$ to the memory signal, which represents the amount of memory that has accumulated in the surrogate at the earliest time at which it is evaluated ($t/M=-10^4$ here).
For the PN waveform, we treat the time of coalescence $t_c$ as an undetermined parameter in Eq.~\eqref{eq:x-of-t}; when this is substituted into Eq.~\eqref{eq:3PN_memory}, this gives a time-domain PN signal for the memory effect, $h_{20}^\PN(t)$.
There is also freedom in the choice of the initial and final times over which the hybridization takes place, denoted by $t_1$ and $t_2$ respectively.
We choose $t_1=-5000M$ and $t_2=-4000M$, because the interval $t\in[t_1,t_2]$ was the same range used by the NRHybSur3dq8 surrogate model to hybridize between the NR and the EOB waveforms for the oscillatory $m\neq 0$ waveform modes.

\begin{figure}[t!]
    \centering
    \includegraphics[width=0.48\textwidth]{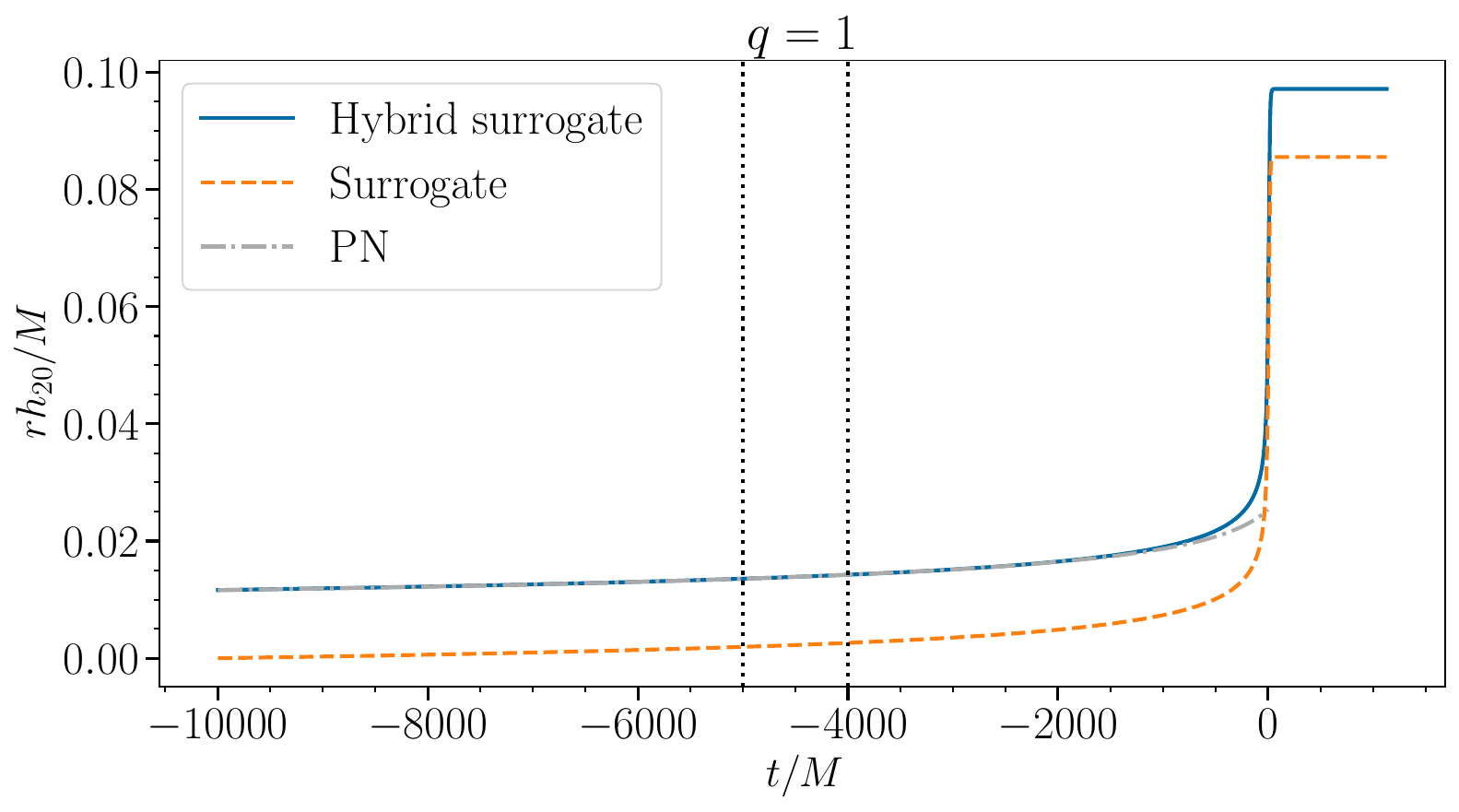}
    \caption{\textbf{Hybridized memory signal versus time}:
    The three curves shown are the PN memory signal (dashed-dotted gray curve), surrogate memory signal computed from the NRHybSur3dq8 surrogate waveform modes (dashed orange curve), and hybridized surrogate memory signal (solid blue curve) for an equal-mass non-spinning BBH merger. 
    The vertical black dotted lines indicate the region over which the PN and surrogate memory signals were hybridized (specifically, from $t_1=-5000M$ to $t_2=-4000M$; see the main text for why this region was selected).
    The surrogate memory has been computed using the modes summarized in Table~\ref{tab:lm-modes-sur}.}
    \label{fig:hybrid_hmem_vs_t}
\end{figure}

We then hybridize the 3PN and NR surrogate waveforms by solving a nonlinear optimization problem for the two free parameters $h_0$ and $t_c$ by minimizing the cost function
\begin{align}
\label{eq:PN_surrogate_cost_function}
    C[h_{\surr},h_{\PN}] = \frac{\displaystyle\int_{t_1}^{t_2} dt |h_{20}^{\surr}(t)-h_{20}^{\PN}(t)|^2}{\displaystyle\int_{t_1}^{t_2} dt |h_{20}^{\surr}(t)|^2} \, .
\end{align}
We hybridize at 50 different values of $q$ between $q=1$ and $q=8$, which we distribute uniformly in $\eta$ (this corresponds to $\eta$ in the range $[0.01, 0.25]$, accurate to the hundredths digit place).
An example of the result of the hybridization procedure is shown in Fig.~\ref{fig:hybrid_hmem_vs_t} for an equal-mass ($q=1$) nonspinning BBH merger.
We truncate the 3PN memory at the peak time of the waveform $t/M = 0$, because it starts to deviate significantly from the NR surrogate memory waveform signal at larger values (which is not surprising, because $t/M=0$ is the time at which the $l=2$, $m=2$ mode of the surrogate waveform reaches its peak value).
The hybridized memory, in principle, could be obtained for an arbitrarily long duration by evaluating the PN waveform from the desired initial time to $t_2$ and the surrogate from $t_1$ to the desired final time.
Over the hybridization interval $t\in[t_1,t_2]$, the PN and surrogate waveforms could be smoothly ``blended'' to obtain the memory signal $h_{20}(t)$.\footnote{The term ``blended'' means that the PN signal should be multiplied by a smooth function that is one at $t\leq t_1$ and goes to zero at $t \geq t_2$, whereas the surrogate should be multiplied by one minus this function.
Adding the PN and surrogate waveforms scaled by these functions then will yield a smooth memory signal, up to small errors in the hybridization procedure that cause differences in the values of the PN and surrogate waveforms over the hybridization interval $[t_1,t_2]$.}

When we require the surrogate memory signal with the final memory offset accumulated in the limit as the past time goes to minus infinity, but only for a short time interval, we do not need to perform the blending described above.
In the hybridization procedure, we evaluated the surrogate waveforms for the same length of time between $t/M=-10^4$ and $t/M=130$ for the different values of $\eta$ that we used.
Therefore, the value of $h_0$ that comes out of the hybridization is the value of the memory signal at the starting time ($t/M=-10^4$) that should be added to the surrogate waveform to compute the remainder of the memory signal consistent with one that extends indefinitely into the past.
For this reason, it is convenient to have an expression for $h_0(q)$ that can be evaluated for any value of $q$ in the range $[1,8]$, rather than the fixed values of $q$ at which the hybridization was performed.
Thus, we find it useful to have a fit for $h_0$ over this range.

In fact, we construct polynomial fits for both $h_0$ and $t_c$ as a function of $\eta$ over this range using a quartic polynomial.\footnote{The choice of quartic order was determined empirically. 
The residuals between the fit and the values of $h_0$ determined through hybridization (at specific values of $\eta$) improved as the polynomial fit order was increase from linear to quartic, but did not decrease as dramatically for higher-order polynomial fitting functions.}
We write the polynomials in the form
\begin{subequations}
\label{eq:tc_h0_fits}
\begin{align}
\label{eq:h0_fit}
    h_0^{\fit}(\eta) = {} & \frac{\mu}{r} \sum_{j=0}^4 a_j \eta^j , \\
    t_c^{\fit}(\eta) = {} & M \sum_{j=0}^4 b_j \eta^j ,
\end{align}
\end{subequations}
where the coefficients $a_j$ and $b_j$ are given in Table~\ref{tab:tc_h0_surr}.
The $t_c$ fit could be used to determine the correct 3PN time-domain waveform to match with the surrogate model, though we do not use it for that purpose in this paper.

\begin{table*}[htb]
    \centering
    \caption{Coefficients for the $h_0$ and $t_c$ polynomial fits in Eq.~\eqref{eq:tc_h0_fits}.
    The fits were constructed by using $50$ BBH systems with mass ratios equally spaced in $\eta$ for a range of mass ratios with the range of validity of the NRHybSur3d18: $1\leq q\leq 8$. 
    The values of $h_0$ and $t_c$ were obtained using the hybridization procedure described in Sec.~\ref{sec:memory_comp_hybridization}.}
    \begin{tabular}{cccccc}
    \hline
    \hline
       Coefficient & $j=0$ & $j=1$ & $j=2$ & $j=3$ & $j=4$  \\
       \hline
       $a_j$   & $5.67\times 10^{-4}$ & $5.81 \times 10^{-2}$ & $-9.29 \times 10^{-2}$ & $2.02 \times 10^{-1}$ & $-2.04 \times 10^{-1}$ \\
       $b_j$   & $2.11\times 10^3$ & $-2.04\times 10^4$ & $ 1.03\times 10^5$ & $-2.85\times 10^5$ & $3.27\times 10^5$ \\
    \hline
    \hline
    \end{tabular}
    \label{tab:tc_h0_surr}
\end{table*}

Note that in Fig.~\ref{fig:hybrid_hmem_vs_t}, the hybridized memory starts at $t=-10^4M$ at a value of around one tenth of the final memory offset, rather than zero for the surrogate memory without hybridization.
Using the PN expression for the memory in Eq.~\eqref{eq:3PN_memory} and the PN parameter as a function of time in Eq.~\eqref{eq:x-of-t}, we can estimate that to decrease the initial value of the memory from the surrogate by a factor of ten (to around one hundredth of the final offset), the surrogate would need to be evaluated for an amount of time $10^4$ times longer.
Although the computation of the memory from the surrogate for one waveform takes of order one second, increasing the length by this factor of $10^4$ would make the calculation take hours.
The fitting function in Eq.~\eqref{eq:h0_fit} can be evaluated in negligible time.
This indicates that there is a considerable advantage for using this fitting function when it is important to capture the initial offset in the memory.

Note also that the calculation of the 3PN memory signal in~\cite{Favata:2008yd} [our Eq.~\eqref{eq:3PN_memory}] uses additional multipole moments that are not those given in Table~\ref{tab:lm-modes-sur}.
We computed a PN memory signal using just the multipole modes in Table~\ref{tab:lm-modes-sur} evaluated at 3PN accuracy.
The relative difference between this PN memory signal and the full 3PN result in Eq.~\eqref{eq:3PN_memory} was of order $10^{-6}$.
Hybridizing the surrogate to this limited-multipole PN signal led to relative difference of the same order.
The smallness of this difference is why we opted to use the full 3PN result for the hybridization.

\subsection{Memory for extreme mass ratios}

We discuss several aspects of the results for the memory from nonspinning EMRI systems in this part.
We first demonstrate with the EMRISur1dq1e4 model~\cite{Rifat:2019ltp} that as the mass ratio becomes more extreme, a greater fraction of the memory signal accumulates during the inspiral rather than the merger and ringdown (consistent with the analytical argument given in Sec.~\ref{subsec:EMRIs}).
We then study the convergence of the EMRI memory signal that is calculated using the factorized oscillatory waveforms described in Sec.~\ref{subsec:EMRIs} as a function of the PN order and the multipole index $l$.

\subsubsection{EMRI surrogate results}\label{sec:EMRI_surrogate}

The results in this section using the EMRISur1dq1e4 surrogate model~\cite{Rifat:2019ltp} are intended to be illustrative of how the morphology of the memory signal changes between comparable and extreme mass ratios.
Thus, we will not use the results here in the construction of the memory fit in the following section, nor will we try to hybridize the EMRISur1dq1e4 surrogate memory signal with an appropriate PN model during the inspiral here either.\footnote{The EMRI surrogate is not hybridized and starts at a time $t/M = -10^4$ earlier than the merger time (the peak of the $l=2$, $m=\pm 2$ waveform). As the mass ratio becomes more extreme, the 3PN memory waveform in Eq.~\eqref{eq:3PN_memory} begins to lose accuracy at the starting time of the surrogate waveform. Instead, one could use the high-PN-order memory for EMRIs to hybridize, but that could also run into errors because it neglects higher-order terms in the symmetric mass ratio $\eta$. We are not aware of oscillatory waveform modes to second order in mass ratio and at high PN order, though once they are available, our work could be extended to include them.}

\begin{figure}
    \centering
    \includegraphics[width=0.48\textwidth]{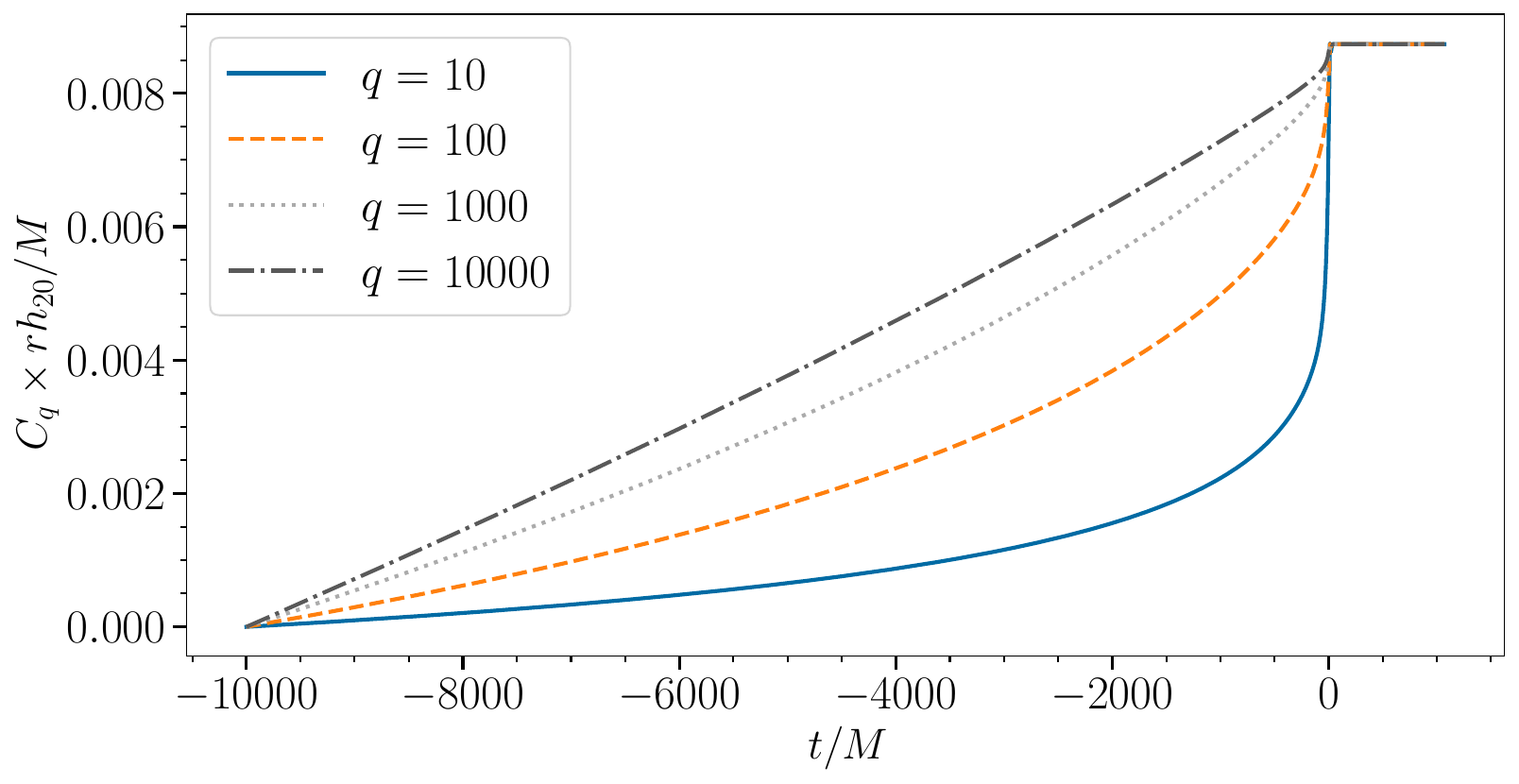}
    \caption{\textbf{Memory signal for comparable to extreme mass ratios versus time}:
    The time-domain memory signal computed using the EMRISur1dq1e4 surrogate model is shown for nonspinning BBH mergers with mass ratios $q=10$ (the solid blue curve), $q=10^2$ (the dashed orange curve), $q=10^3$ (the dotted light-gray curve) and $q=10^4$ (the dark-gray dashed-dotted curve). 
    As the mass ratio increases, a larger fraction of the final memory offset accumulates during the inspiral in a fixed range of time. 
    The memory signals have been scaled by the corresponding factors of $C_q$ defined in Eq.~\eqref{eq:Cq_scale_factor}, because the memory becomes smaller as the mass ratio becomes more extreme.
    The values were chosen to make the final memory strain equal for all mass ratios, thereby allowing all the curves to be plotted on the same scale and illustrating the accumulation of the memory more clearly over this fixed time interval.
    For reference, the values for the $C_q$ factors are $C_{10}=1$, $C_{10^2}=2.84\times 10^{1}$, $C_{10^3}=1.13\times 10^{3}$, and $C_{10^4}=7.08\times 10^{4}$. 
    The oscillatory waveform modes used to compute the memory signal are those listed in Table~\ref{tab:lm-modes-sur}.}
    \label{fig:EMRIMemory}
\end{figure}
The EMRISur1dq1e4 surrogate model contains the same waveform modes given in Table~\ref{tab:lm-modes-sur}, and it also includes modes for $l=5$ and $m=\pm 3$ and $\pm 4$.
In this section, we do not use these additional waveform modes in the EMRISur1dq1e4 surrogate to compute the $h_{20}$ memory signal via Eq.~\eqref{eq:Deltah20UV}, we use just the modes listed in Table~\ref{tab:lm-modes-sur}.
In Fig.~\ref{fig:EMRIMemory}, we compute the memory signal with the EMRISur1dq1e4 surrogate for nonspinning BBH systems with the mass ratios $q=10$, $10^2$, $10^3$ and $10^4$.
The merger (understood as the peak amplitude of the $l=2$, $m=\pm 2$ waveform modes) occurs at the time $t=0$ as with the comparable mass surrogate, and the surrogate model is not hybridized and does not extend earlier than the time $t/M=-10^4$, which is the earliest time at which the memory is plotted in Fig.~\ref{fig:EMRIMemory}.
The EMRI surrogate does not extend beyond a time $t/M = 130$, so we extended the length of time after $t=0$ by padding the memory waveform with the final value attained at $t/M = 130$.

The memory accumulated during a fixed time interval decreases with increasing the mass ratio.
Therefore, we scale the more extreme mass ratios shown in Fig.~\ref{fig:EMRIMemory} by a factor of $C_q$, which we defined as the ratio between the final memory strain $\Delta h_{20}$ for a BBH system with mass ratio $q=10$ and the final strain for the other with mass ratio $q$:
\begin{equation}\label{eq:Cq_scale_factor}
    C_q = \frac{\Delta h_{20}(q=10)}{\Delta h_{20}(q)} .
\end{equation}
Because the mass of the primary differs by at most ten percent in these different cases, the timescale of the merger is roughly the same for different mass ratios, and corresponds to the short range around $t=0$, where the memory accumulates most rapidly for the $q=10$ case.
This then allows one to see how much more memory accumulates during the inspiral phase as the mass ratio becomes more extreme.
This trend in Fig.~\ref{fig:EMRIMemory} becomes more pronounced for the common convention of EMRIs ($q \gtrsim 10^5$), where the memory offset during the merger and ringdown is suppressed from that during the inspiral by an additional factor of $\eta$, as discussed in Sec.~\ref{subsec:EMRIs}.
Figure~\ref{fig:EMRIMemory} provides a visual justification of why we will use a high-order PN approximation for the memory accumulation during the adiabatic inspiral phase to compute the memory signal from EMRIs and ignore the contribution from the transition to plunge, the plunge, and the merger and ringdown.

\subsubsection{Memory signal and memory offset} \label{subsubsec:EMRI-full}

Finally, we can compute the memory signal after using the prescription outlined in Sec.~\ref{subsubsec:eval-mem} with the required oscillatory multipole modes at the relevant PN orders discussed in Sec.~\ref{sec:hmem_EMRI_22PNl25}.
The result can be written schematically as
\begin{equation} \label{eq:h20series-schematic}
    h_{20} = \sum_{j=2}^{46}\sum_{k=0}^{k_\mathrm{max}(j)} h_{20,jk} v^j (\ln v)^k, 
\end{equation}
where $h_{20,jk}$ are numerical coefficients and $k_\mathrm{max}(j)$ is the maximum order of the powers of $\ln v$ that appear in the PN series at a given PN order $(j/2-1)$ relative to the leading $v^2$ term.
Note that only the powers of $v$, not $\ln v$, determine the PN order.
We do not list the coefficients $h_{20,jk}$ explicitly in this paper (there are 184 nonzero values at 22PN order), but we make this series data available in \textsc{Mathematica} format as an ancillary file associated with the arXiv version of this paper and on Zenodo~\cite{elhashash_2024_13132039}.
However, we do give the value of the final memory offset computed using all relevant $l\leq 25$ and up to 22PN order:
\begin{equation}\label{eq:Deltahmem_EMRI_value}
    \Delta h_{20}^{(25,22)} \approx  0.102414 \, \frac \mu r .
\end{equation}
The notation with the superscript $(25,22)$ indicates the multipole and PN accuracy of the expression, which we will use subsequently: i.e., $ h_{20}^{(\bar\ell,n)}$ denotes a memory signal evaluated with oscillatory $l$ modes up to $l=\bar \ell$ with these modes at the necessary accuracy to compute $h_{20}$ to 22PN order.\footnote{There is a potential ambiguity regarding the labeling of the multipole index $\bar \ell$ of the memory signal in the following sense.
As described in Sec.~\ref{sec:hmem_EMRI_22PNl25}, different elements of the calculation of $h_{20}$ at a fixed PN order require a different number of spherical-harmonic modes.
For example, the horizon term $(dE/dt)_H$ at 22PN order requires only harmonics up to $l=11$, the luminosity at infinity requires modes up to $l=24$, and the sum of modes that enter the memory expression in Eq.~\eqref{eq:Deltah20UV} requires modes up to $l=25$.
The label $\bar \ell = 25$, therefore, refers to the largest value of $l$ needed in \emph{some} aspect of the calculation to achieve the required PN order $n$; however, it does not imply that all the components of the calculation need to be evaluated at this multipole order to achieve the necessary PN accuracy.\label{fn:l-order}}
We discuss why we quote the result to six digits of accuracy in more detail in Secs.~\ref{subsubsec:multipole-orders} and~\ref{subsubsec:PN-orders}.

\begin{figure}
    \centering
    \includegraphics[width=0.48\textwidth]{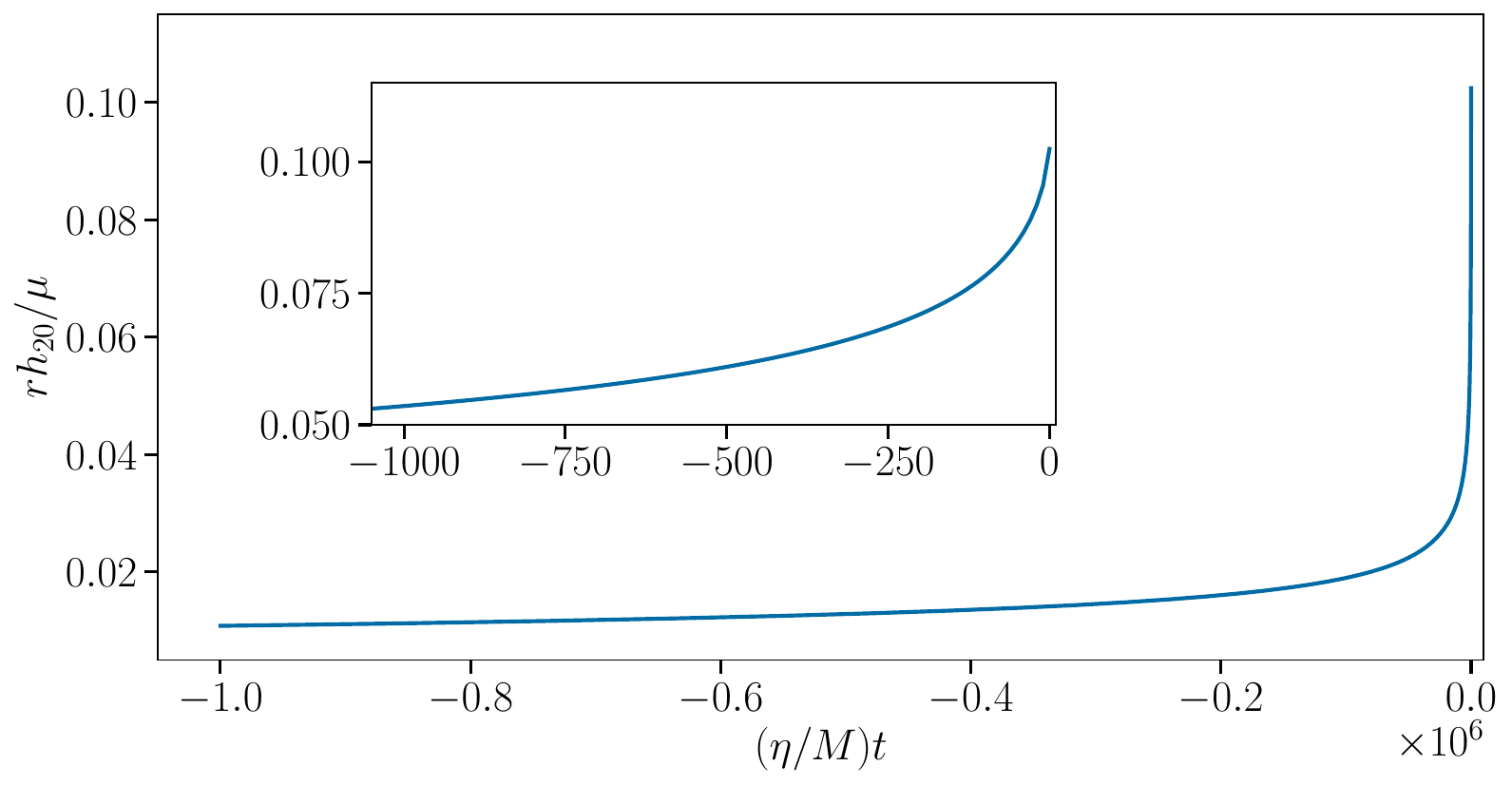}
    \caption{\textbf{Memory signal over different timescales}:
    The memory signal $h_{20}$ is shown as a function of the normalized time $(\eta/M)t$ for an EMRI with nonspinning components. 
    The memory is computed up to $22$PN order relative to the Newtonian memory using oscillatory modes with $l \leq 25$.
    The main panel shows a time span running from $-10^6 M/\eta$ to 0, whereas the inset focus on a shorter range from $-10^3 M/\eta$ to 0.
    The signal stops at ISCO (the time $t=0$), after which the memory accumulated is a small (order $\eta^2$) correction.
    Further discussion of the figure is given in the text of Sec.~\ref{subsubsec:EMRI-full}.}
    \label{fig:h_vs_t}
\end{figure}

We also show how the memory signal accumulates as a function of time in Fig.~\ref{fig:h_vs_t}.
We use the method for evaluating $v(t)$ described in Sec.~\ref{subsubsec:eval-mem}, which we then substitute into the final result in the form of Eq.~\eqref{eq:h20series-schematic}.
We show the memory signal starting from a time $t_i=-10^6 M/\eta$ until it reaches the ISCO, which is defined to be $t=0$.
For an EMRI with a primary BH mass of $10^6 M_\odot$ and a secondary BH with mass $10 M_\odot$, so that $\eta \approx 10^5$, then the range of Fig.~\ref{fig:h_vs_t} covers about $5\times 10^{11}$\,s or about $1.6\times 10^4$\,years.
While this is the time scale over which the memory accumulates about ninety-percent of its final value, it is also a long timescale from the perspective of GW detection.
We, therefore, also show the last $10^3 M/\eta$ of the signal in the inset, which corresponds to a roughly 16-year timescale for the same system (and is near the lower limit of the periods that pulsar timing arrays currently measure).
It also illustrates that roughly half of the final memory offset accumulates in this last decade and a half.
Note that for the same secondary BH of the same mass ($10 M_\odot$) but a primary of mass ten times smaller (respectively, larger), the relevant time spans would be 100 times shorter (respectively, longer), namely order a month (respectively, a millennium) for the $10^3 M/\eta$ duration.

Given the large number of terms that contribute to the mode $h_{20}$ at 22PN order [there are several thousand nonzero terms when evaluating the sums in Eq.~\eqref{eq:Deltah20UV}], and the fact that the PN series for the oscillatory waveform modes and the GW-luminosity have hundreds of terms when written in a form similar to the expression for the memory signal in Eq.~\eqref{eq:h20series-schematic}, it is useful to perform some consistency checks of our result.
The first was verifying that our result agrees with the linear in $\eta$ terms in the 3PN expression derived by Favata~\cite{Favata:2008yd} given in Eq.~\eqref{eq:3PN_memory}.
Next, we also perform some convergence analyses, in which we study the contributions to the memory signal as a function of increasing multipole index $l$ and PN order $n$.
The results of these two investigations are discussed in next in Secs.~\ref{subsubsec:multipole-orders} and~\ref{subsubsec:PN-orders} for the multipole and PN cases, respectively.

\subsubsection{Memory signal at different multipole orders} \label{subsubsec:multipole-orders}

To understand how much the different oscillatory multipoles contribute to the memory effect, we compare the complete memory signal at 22PN order with that computed by keeping $l$ values less than or equal to a given value $\bar \ell$. 
We define an absolute difference 
\begin{equation}\label{eq:deltalh20}
    \delta_{\bar{\ell}} h_{20} = |h_{20}^{(25,22)} - h_{20}^{(\bar{\ell},22)}|,
\end{equation}
which represents the contribution to the memory from the modes with $\bar \ell < l \leq 25$, at 22PN order.
For ease of notation, we will often drop the $(25,22)$ superscript on the memory signal with the highest PN order and number of multipoles, so that $h_{20}\equiv h_{20}^{(25,22)}$.
It will also be helpful to compute a fractional contribution of these higher $l$ modes, $\delta_{\bar{\ell}} h_{20} / h_{20}$, as well.
We computed $\delta_{\bar{\ell}} h_{20}$ for all $\bar \ell$ values from 2 to 24, but show only a subset of them in the figures below.
Because $h_{20}$ depends on the GW luminosities at infinity and at the horizon, we also truncate the sum over $l$ in Eqs.~\eqref{eq:dEdu_infinity} and~\eqref{eq:dEdu_horizon} at the corresponding value of $\bar \ell$.

As discussed in Sec.~\ref{sec:hmem_EMRI_22PNl25} and Footnote~\ref{fn:l-order} especially, different terms in the computation of the memory require different numbers of multipoles to be accurate to a given PN order.
For example, the power radiated down the future event horizon, $(dE/dt)_H$ only requires multipoles with $l\leq 11$.
Thus, when we consider values of $\bar \ell < 11$, we sum over multipoles in Eq.~\eqref{eq:dEdu_horizon} with $\bar \ell$ as the upper limit of the $l$ sum, as noted above, but when $\bar \ell$ satisfies $\bar \ell \geq 11$, we use the full sums in Eq.~\eqref{eq:dEdu_horizon} without any truncation nor with the addition of higher $l$ terms that would be of a higher PN order.

Our focus in this part will be on $h_{20}(v)$, for all values of $v$ from $0$ to $v_\ISCO = 1/\sqrt{6}$, on $h_{20}(t)$ for the last $10^3 M/\eta$ before the particle crosses the ISCO, and for the final memory offset $\Delta h_{20}$.
We also include some intermediate results on the different multipolar contributions to $\dot h_{20}$ and $dh_{20}/dv$ as a function of $v$ in Appendix~\ref{app:memory-integrand}.

\begin{figure*}
    \centering
    \includegraphics[width=0.48\textwidth]{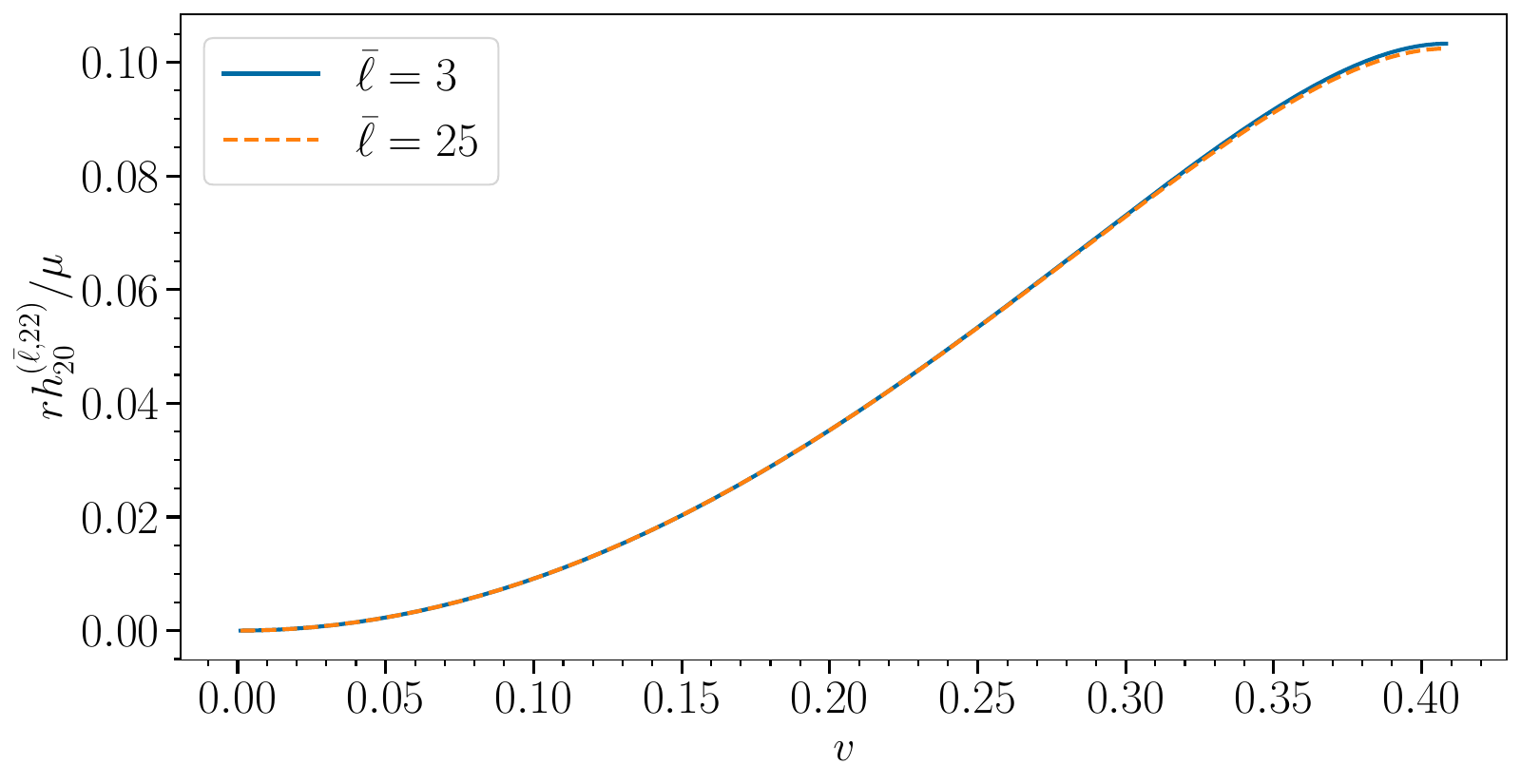}
    \includegraphics[width=0.48\textwidth]{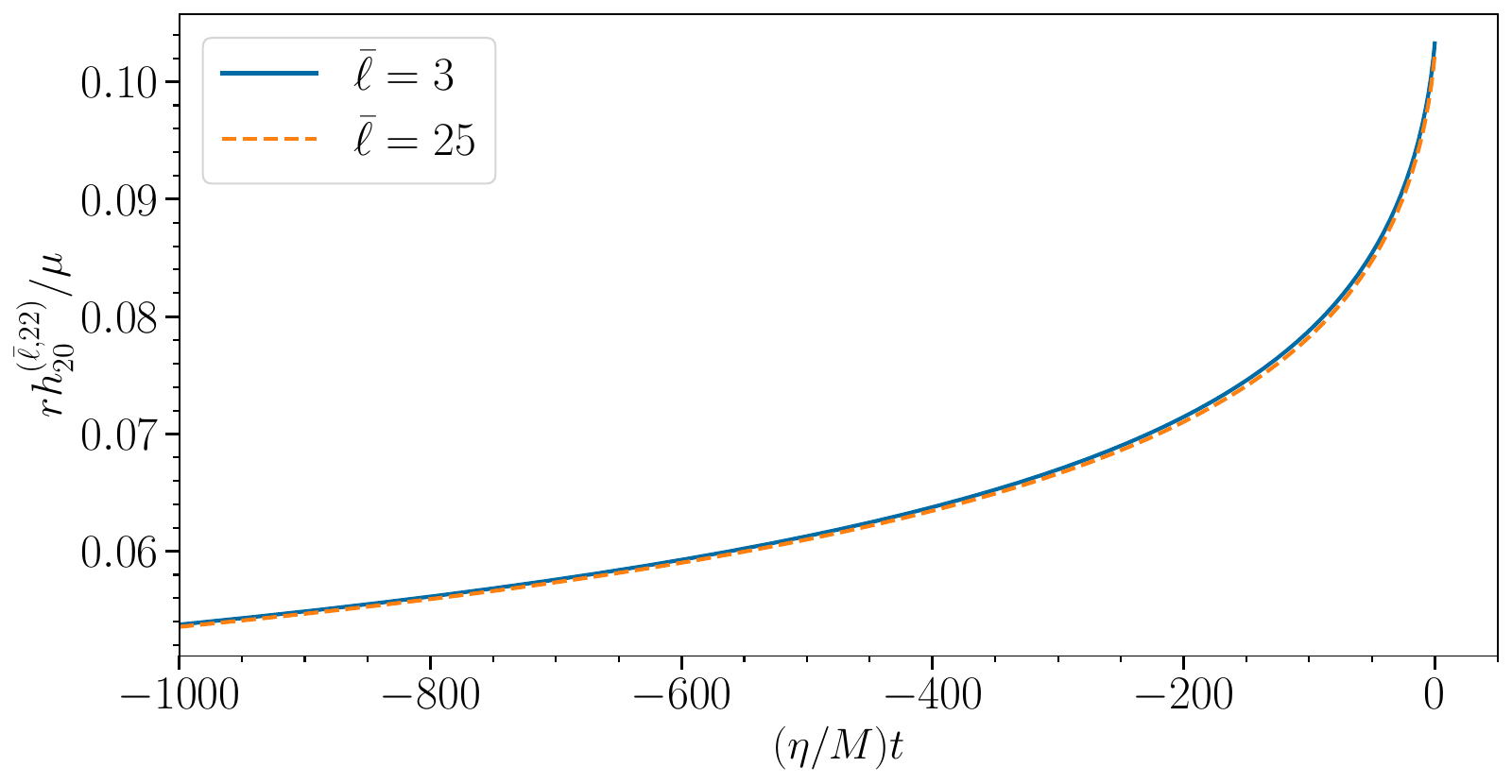}
    \includegraphics[width=0.48\textwidth]{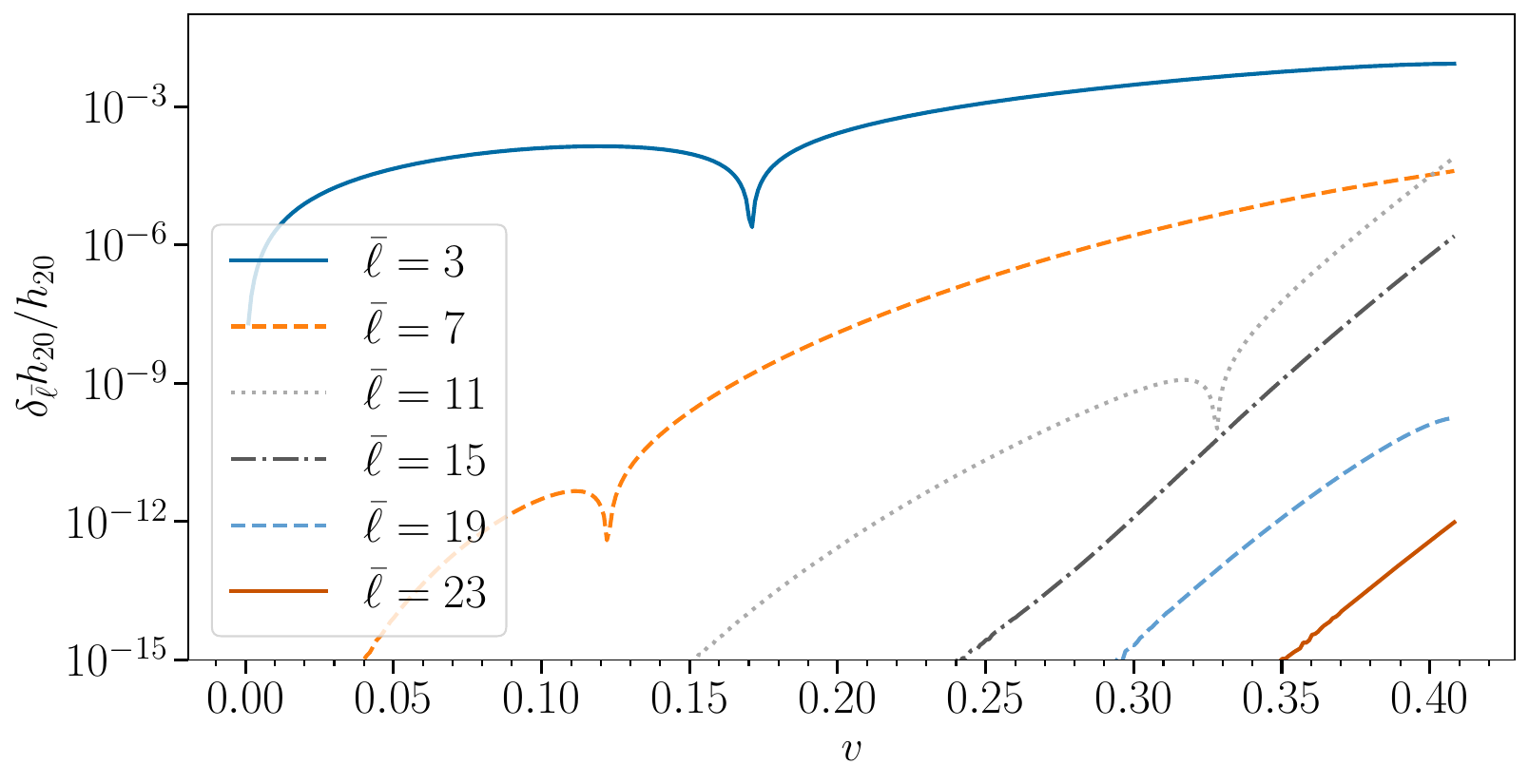}
    \includegraphics[width=0.48\textwidth]{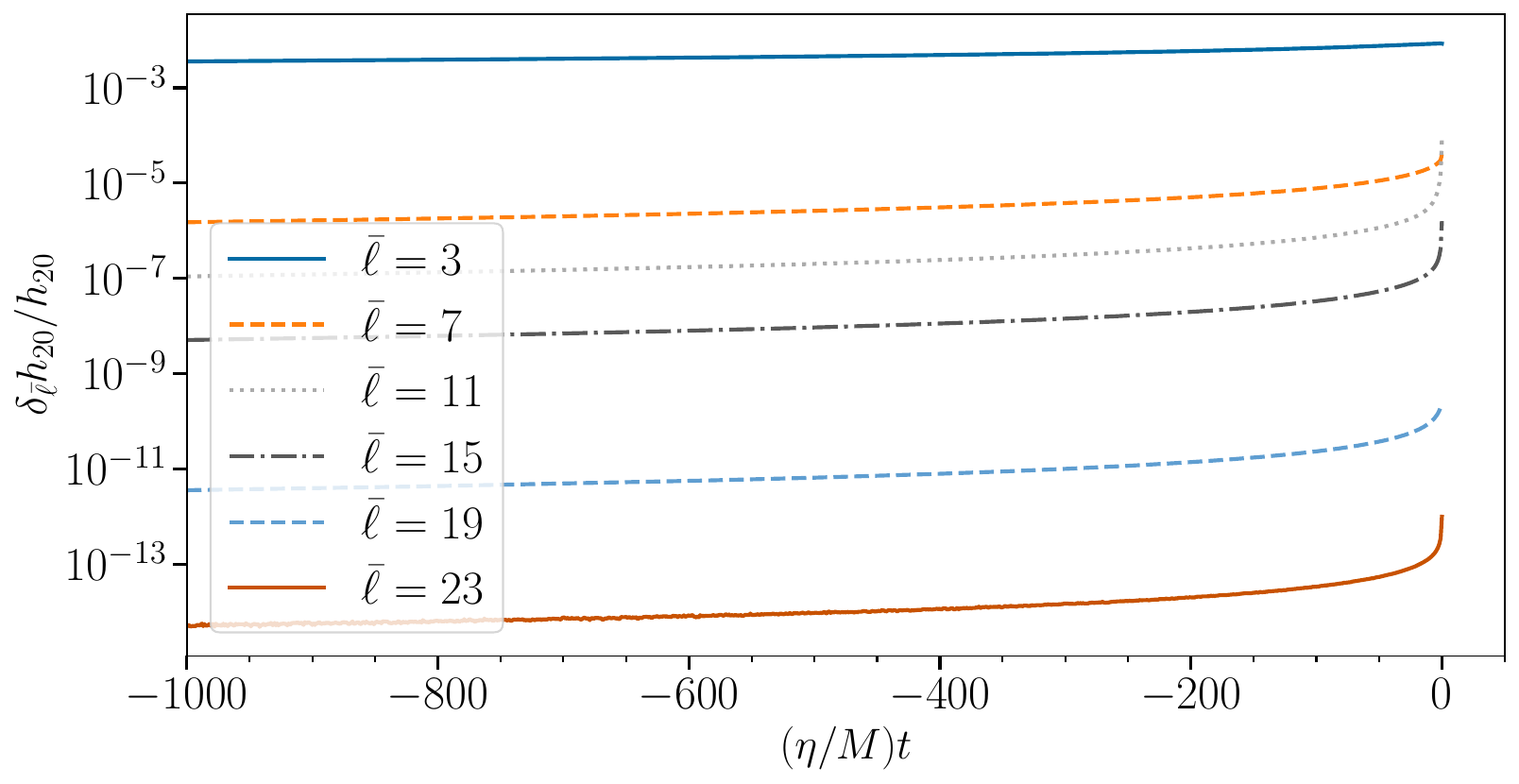}
    \caption{\textbf{Memory signals at different multipole orders versus velocity or time}:
    The top row shows the scaled memory signal $r h^{(\bar\ell,22)}_{20}/\mu$ for different total numbers of multipoles included.
    Including just up to octopole order ($\bar\ell = 3$) is shown as solid blue curves and all multiples up to the 25-pole ($\bar\ell = 25$) is shown as dashed orange.
    The bottom row shows the fractional contribution to the full 22PN expression for the memory from modes with $l > \bar\ell$ for six cases, $\bar{\ell}=3$, 7, 11, 15, 19, and 23, which are the solid dark-blue, dashed orange, dotted light-gray, dotted-dashed black, dashed light-blue, and solid brown curves, respectively.
    The left column shows these results versus velocity, with a lower velocity of $10^{-3}$ and a maximum velocity of $v_\ISCO =1/\sqrt{6}$.
    The right column shows the quantities versus time,  $(\eta/M)t$ over an interval spanning from $-10^3$ to 0.
    The initial time shown on the right corresponds to a velocity of $v\approx 0.25$.
    Further discussion of the implications of these figure panels are given in the text of Sec.~\ref{subsubsec:multipole-orders}.}
    \label{fig:h_vs_v_lmax}
\end{figure*}
In Fig.~\ref{fig:h_vs_v_lmax}, we show in the top row the scaled memory signal $r h_{20}^{(\bar \ell,22)}/\mu$, for different values of $\bar \ell$.
The solid blue curve, $\bar \ell = 3$, contains only up to octopole order, and the dashed orange, $\bar \ell = 25$, contains all the relevant multipoles.
The left panel shows this quantity versus velocity, from $10^{-3}$ to $v_\ISCO$, whereas the right panel shows it versus time over a range of $(\eta/M)t \in [-10^3,0]$, which corresponds to a velocity range of approximately $[0.25,v_\ISCO]$.
On the scale shown in the upper panels, the two different multipole orders are difficult to distinguish, which suggests that most of the memory signal comes from just the quadrupole and octopole oscillatory terms in the sum in Eq.~\eqref{eq:Deltah20UV}.

To determine the contributions of higher modes more quantitatively, we show in the bottom row the fractional contribution of the modes with $l > \bar \ell$ to the full memory signal ($rh_{20}/\mu$), which is the label on the vertical axis, $\delta_{\bar\ell} h_{20}/h_{20}$.
The left again shows the results as function of velocity, whereas the right shows them as a function of time; the ranges of both velocity and time are the same as in the respective panels above.
We show $\delta_{\bar\ell} h_{20}/h_{20}$ for six cases, $\bar\ell = 3$, 7, 11, 15, 19, and 23, which correspond to, respectively, the solid dark-blue, dashed orange, dotted light-gray, dotted-dashed black, dashed light-blue, and solid brown curves.
The solid blue curves correspond to the relative contributions from hexadecapole and higher moments, and they demonstrate that the contributions from all modes higher than quadrupole and octopole terms contribute at most a few percent to the total memory signal.
For certain applications, using just the quadrupole and octopole would be a justified approximation.
Even a more stringent requirement, such as requiring the fractional contributions from higher multipoles be $\lesssim 10^{-6}$ for all velocities or times, requires computing up to $\bar\ell = 15$, rather than $\bar\ell=25$.

As the PN approximation is most accurate in the limit of small $v$ and higher multipoles enter at higher PN orders, it is to be expected that the contributions of higher multipoles become larger as $v$ becomes larger (or $t$ approaches 0, for the time-domain curves).
The sharp v-shaped dips in the curves correspond to values of the velocity where the memory $rh_{20}^{(\bar \ell,22)}/\mu$ crosses the full 22PN ($rh_{20}/\mu$) expression (i.e., for different values of $v$ or $t$, the contribution of the higher multipoles with $l > \bar\ell$ can be either larger or smaller than the full 22PN-accurate expression with all the relevant multipoles included).
The time-domain results on the right illustrate the anticipated result that the majority of time during the EMRI's inspiral is spent at lower velocities (larger separations), so that the contributions from higher multipoles are most important for the shorter times near when the EMRI is close to the ISCO.

\begin{figure}
    \centering
    \includegraphics[width=0.48\textwidth]{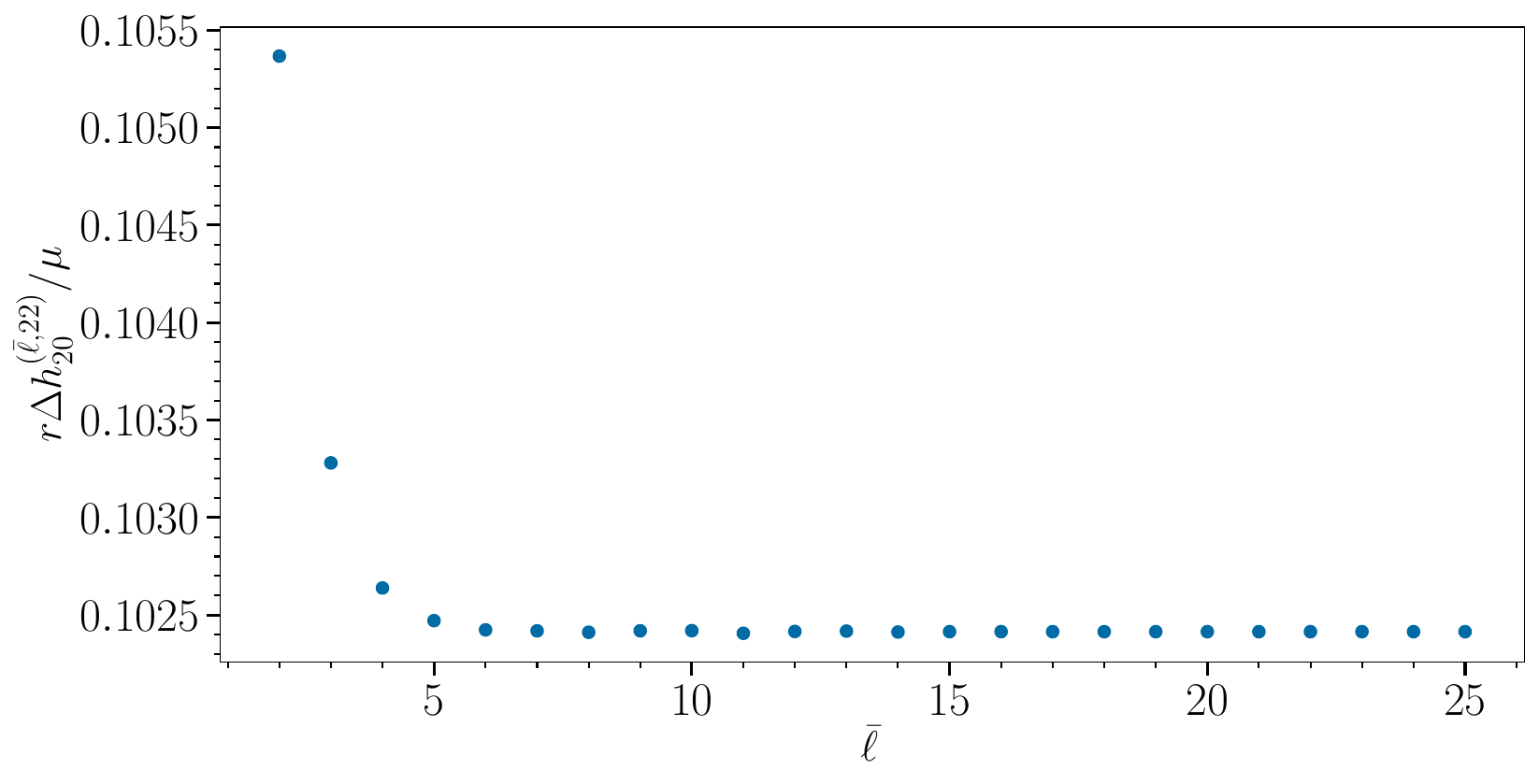}
    \includegraphics[width=0.48\textwidth]{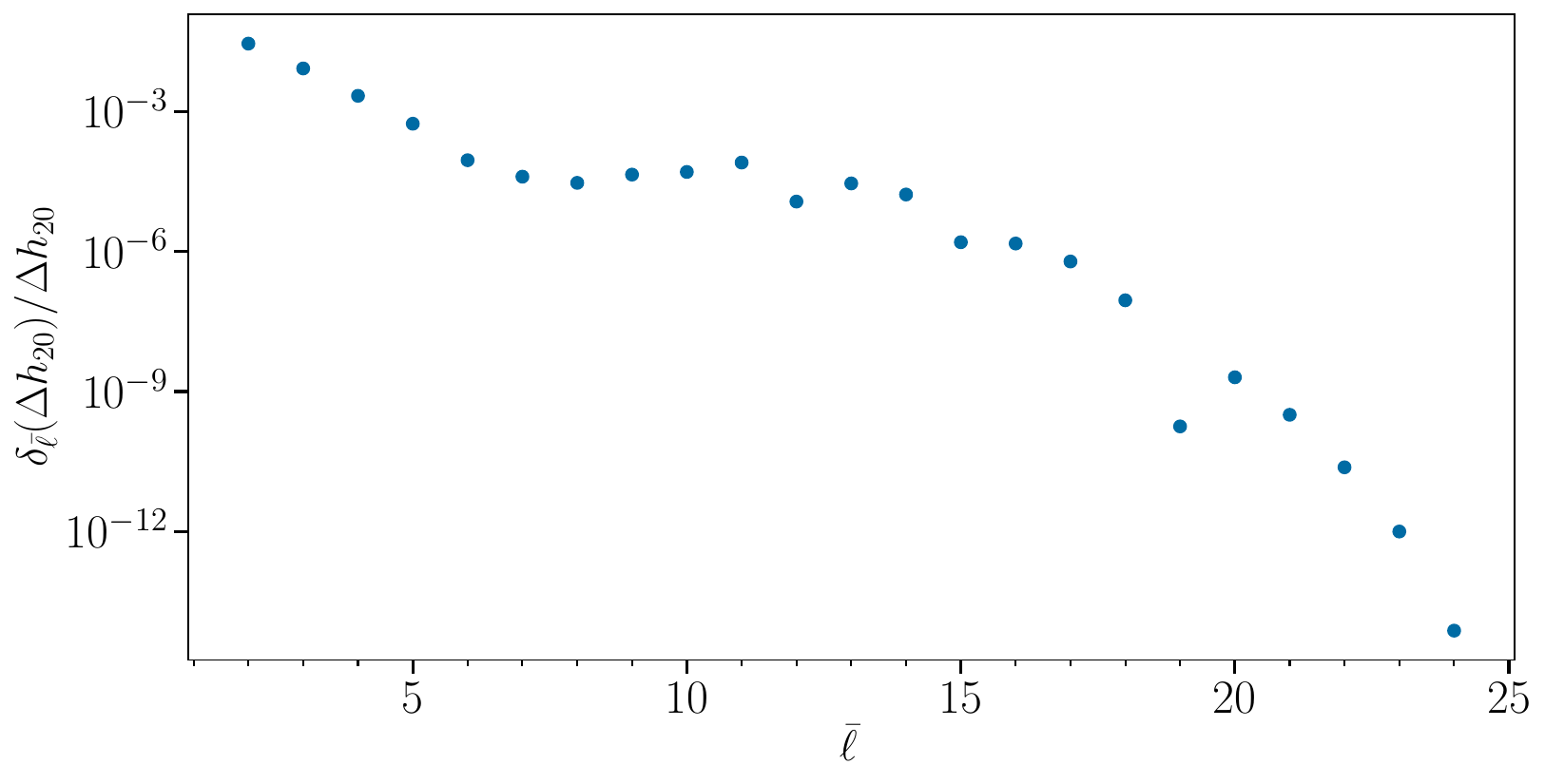}
    \caption{\textbf{Memory offset at ISCO versus highest multipole}:
    \emph{Top}: The total accumulated EMRI memory offset, $\Delta h_{20}^{(\bar\ell,22)}$, computed from different $l\leq \bar{\ell}$ modes is shown versus $\bar \ell$, for $\bar{\ell}=2$ to $\bar{\ell}=25$. 
    The values for $\bar\ell=3$ and $\bar\ell=25$ are the same as those shown in Fig.~\ref{fig:h_vs_v_lmax} at the final velocity $v_\mathrm{ISCO}=1/\sqrt{6}$. 
    \emph{Bottom}: The residual contribution to $\Delta h_{20}$ is shown for different values of $\bar\ell$ as a fraction of the total $\Delta h_{20}$.
    Further discussion of these panels is given in the text of Sec.~\ref{subsubsec:multipole-orders}.}
    \label{fig:DeltahmemEMRI_vs_lmax}
\end{figure}
We next turn to the memory offset at the ISCO in Fig.~\ref{fig:DeltahmemEMRI_vs_lmax}.
The value of $\Delta h_{20}^{(\bar\ell,22)}$ is shown in the top panel for $\bar{\ell}$ values from $\bar{\ell}=2$ up to $\bar{\ell}=25$.
Including higher multipoles tends to decrease the final value of the memory offset, though one should note that the overall vertical scale is small, and spans roughly a three percent relative difference in the values of the memory.
The change at higher multipoles is hard to resolve on the scale of the top panel; thus, we also plot the fractional change $\delta_{\bar\ell}(\Delta h_{20})/\Delta h_{20}$ in bottom panel of Fig.~\ref{fig:DeltahmemEMRI_vs_lmax}, where $\delta_{\bar\ell}(\Delta h_{20})$ is defined analogously to that for the time- or velocity-dependent memory signals in Eq.~\eqref{eq:deltalh20}.
The range of $\bar\ell$ in the lower panel runs from $\bar{\ell}=2$ to $\bar{\ell}=24$, because $\delta_{\bar\ell}(\Delta h_{20})$ involves a difference from the highest ($\bar \ell=25$) mode signal.
The quadrupole contribution to the memory offset is about three percent larger than the final value with all multipoles, which could be sufficient for low-accuracy applications.
The trace of the points has a nontrivial structure.
There is a rapid decrease in the fractional contribution for the next few $\bar\ell$, up to about $\bar\ell=7$.
From $\bar \ell = 7$ until $\bar \ell=14$, there a plateau-like structure in the trace of the points.
This feature arises because successive multipole contributions have comparable amplitudes; however, above $\bar\ell=14$ the amplitudes of successive multipoles decrease raplidly leading to the faster fall-off with $\bar\ell$. 
To obtain a more stringent accuracy requirement, then including a larger number of $\bar \ell$ is required.

Both Figs.~\ref{fig:h_vs_v_lmax} and~\ref{fig:DeltahmemEMRI_vs_lmax} indicate that the highest $l$ contributions to the memory at 22PN order contribute a relatively small amount to the memory signal and final offset.
As there are $2l+1$ modes for each $l$, the number of terms in Eq.~\eqref{eq:Deltah20UV} grows as a function of $l$, but the increase in terms is negated by a larger decrease in the amplitudes of these terms.

\subsubsection{Memory signal at different post-Newtonian orders} \label{subsubsec:PN-orders}

Because the PN series for each of the oscillatory modes becomes increasingly complex as the PN order increases, it is also useful to determine the contributions of the higher PN terms to the total memory signal and the final memory offset, both to better understand the convergence of the PN series and to establish how many PN orders are required to compute the memory to a given precision.

Computing the memory to a given PN order also limits the multipolar order at which the memory is computed.
Specifically, because an $(l,m)$ mode of the strain has a time derivative that scales as $v^{l+\epsilon_p+3}$ for nonspinning binaries, then from Eq.~\eqref{eq:Deltah20UV}, the lowest PN-order contribution from a given $l$ to the memory signal $h_{20}$ will enter at a PN order $n$ of $\lfloor n \rfloor = l-3$ (for $l>3$) and $\lfloor n \rfloor=l-2$ (for $l \leq 3$).
The notation $\lfloor x \rfloor$ is the floor function of $x$, which is necessary to account for half-integer PN orders.
Thus, multipoles with $l > \lfloor n \rfloor + 3$ (for $n >1$) and $l > 2 \lfloor n + 1 \rfloor$ (for $n \leq 1$) will not contribute when working at $n$-PN order, which implies that limiting the PN order simultaneously limits the multipole order.
To denote the required corresponding truncation in $l$, we define
\begin{equation} \label{eq:bar-ell-n}
    \bar \ell_n \equiv \begin{cases}
        2\lfloor n + 1\rfloor & n \leq 1 \\
        \lfloor n \rfloor + 3 & n > 1
    \end{cases} .
\end{equation}
We then introduce a notation for the residual contribution to $h_{20}$ from PN orders greater than $n$:
\begin{equation}\label{eq:deltanh20}
    \delta_{n} h_{20} = |h_{20} - h_{20}^{(\bar\ell_n,n)}|,
\end{equation}
where, as before, $h_{20}$ is the 22PN memory computed from $\bar{\ell}=25$ and $h_{20}^{(\bar\ell_n,n)}$ is the memory computed up to $\bar{\ell}=\bar \ell_n$ and accurate to a relative $n$-PN order.\footnote{As discussed in Footnote~\ref{fn:l-order}, the $\bar\ell_n$ should be interpreted as the largest $l$ value needed in Eq.~\eqref{eq:Deltah20UV}, not that needed in the GW luminosities at the horizon or at infinity involved in the calculation of the memory, which require a lower order in $l$ to achieve the same PN accuracy.}

\begin{figure*}
    \centering
    \includegraphics[width=0.48\textwidth]{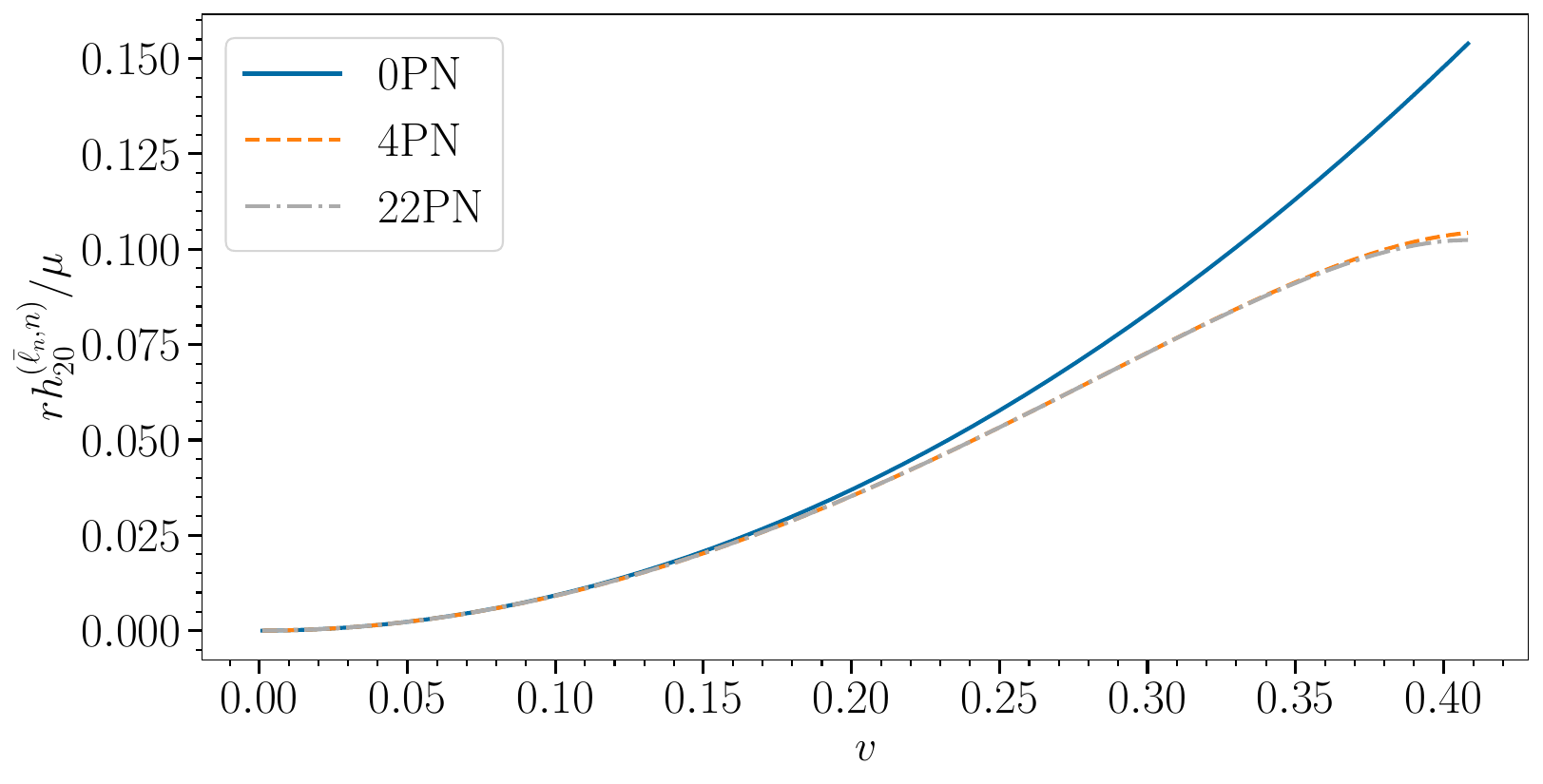}
    \includegraphics[width=0.48\textwidth]{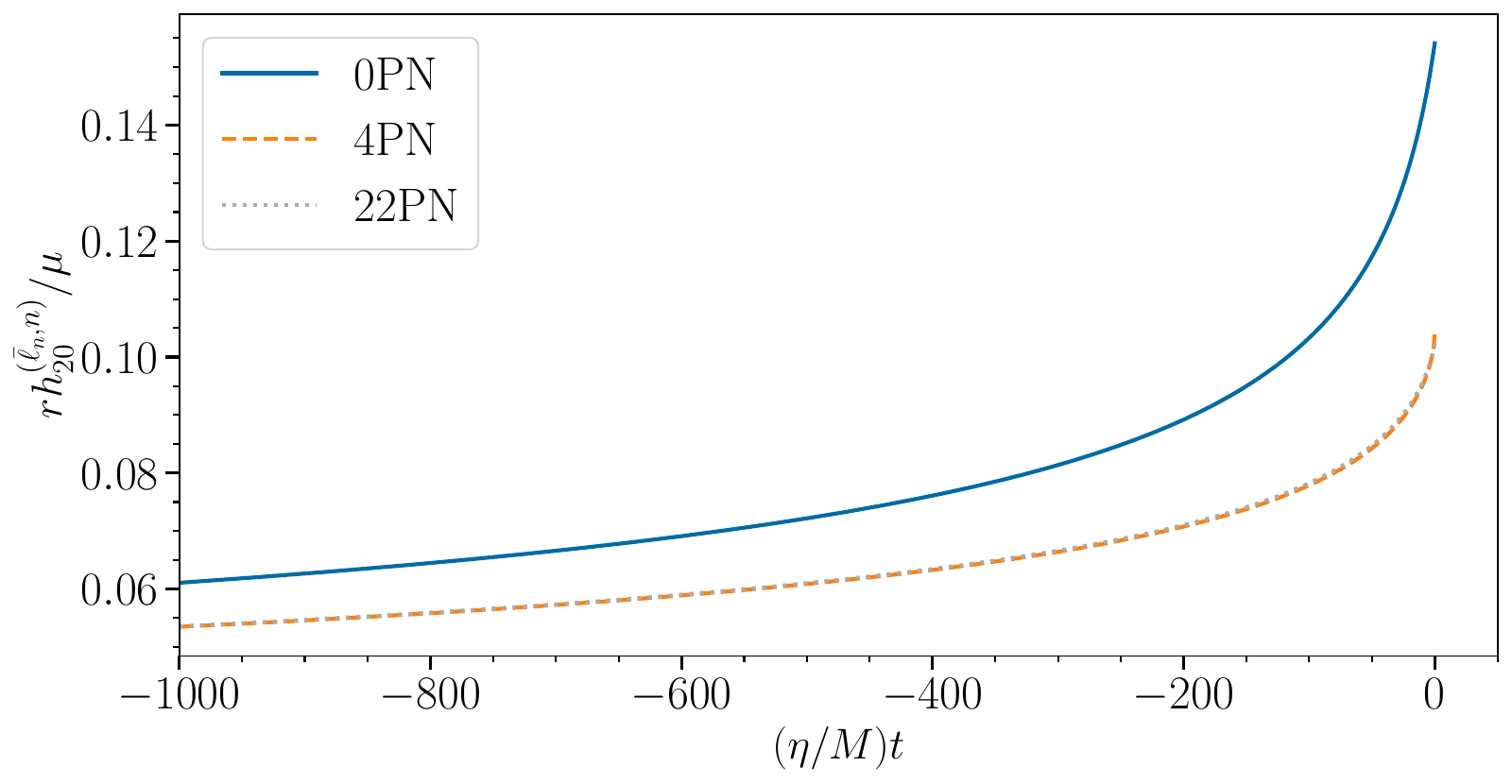}
    \includegraphics[width=0.48\textwidth]{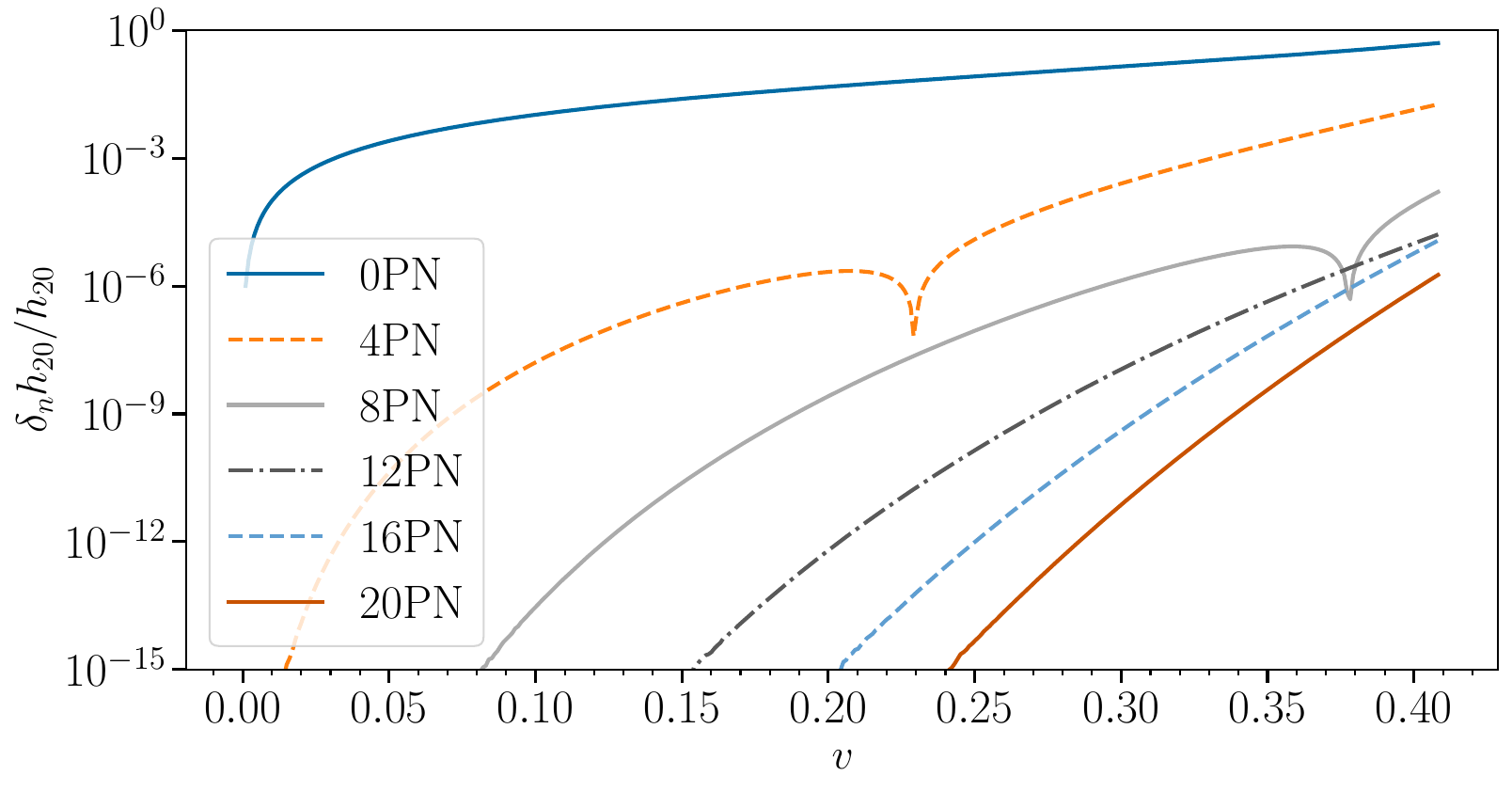}
    \includegraphics[width=0.48\textwidth]{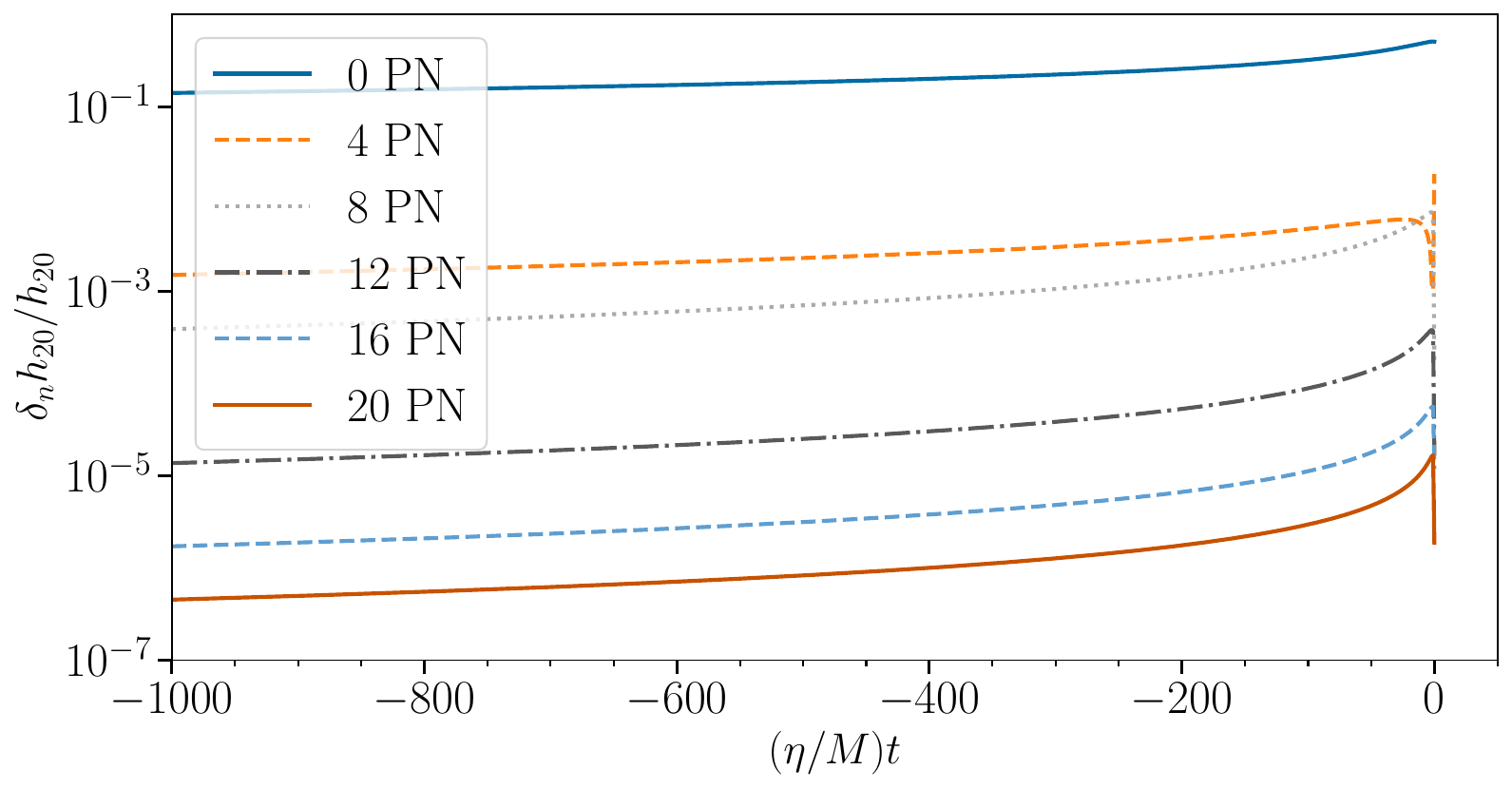}
    \caption{\textbf{Memory signals at different PN orders versus velocity or time}:
    This figure is analogous to Fig.~\ref{fig:h_vs_v_lmax}, but the different curves now represent the memory signal computed up to a fixed PN order, which also requires truncating at a multipolar order given in Eq.~\eqref{eq:bar-ell-n}.
    The top panels now show $h_{20}^{(\bar\ell_n,n)}$ at three different PN orders, $n=0$, 4, and 22, which are the solid blue, dashed orange, and dotted-dashed light gray curves, respectively.
    The time or velocity ranges are identical to those in Fig.~\ref{fig:h_vs_v_lmax}.
    The bottom panels now show the fractional contribution to the full memory signal $h_{20}$ that come from PN orders greater than $n=0$, 4, 8, 12, 16, and 20, which are depicted as the solid dark-blue, dashed orange, dotted light-gray, dotted-dashed black, dashed light-blue, and solid brown curves, respectively.
    Further discussion of the panels in this figure is given in the text of Sec.~\ref{subsubsec:PN-orders}.
    }
    \label{fig:h_vs_v_pn}
\end{figure*}
In Fig.~\ref{fig:h_vs_v_pn}, we show the memory-signal quantities analogous to those in Fig.~\ref{fig:h_vs_v_lmax}, but the top panels now show the memory signal $h_{20}^{(\bar\ell_n,n)}$ and the bottom panels show the fractional contribution $\delta_{n} h_{20}/h_{20}$ for different PN orders $n$ (rather than the fixed PN order of 22 and different multipole orders $\bar\ell$ in Fig.~\ref{fig:h_vs_v_lmax}).
The top panels show $h_{20}^{(\bar\ell_n,n)}$ versus velocity and time for the PN orders $n=0$, 4, and 22 as the solid blue, dashed orange, and dotted-dashed light-gray curves, respectively.
The ranges of velocity on the left and times on the right are identical to those in Fig.~\ref{fig:h_vs_v_lmax}.

The Newtonian (0PN) curve is much larger than the 22PN memory signal, especially at velocities close to $v_\ISCO$.
The reason for this can be understood from $dh_{20}/dv$, which is a linear function of $v$ in the Newtonian limit, but which has a peak near $v\approx 0.3$ for higher PN orders (see Appendix~\ref{app:memory-integrand}).
The 4PN-accurate memory signal, however, differs from the 22PN signal by at most a few percent, which could make it a useful approximation for lower-accuracy applications.
To understand the role of higher PN orders more quantitatively, it is useful to consider the bottom panels.

The bottom panels of Fig.~\ref{fig:h_vs_v_pn} display the fractional contribution to the full memory signal $h_{20}$ that come from a PN orders greater than $n=0$, 4, 8, 12, 16, and 20, which are depicted as the solid dark-blue, dashed orange, dotted light-gray, dotted-dashed black, dashed light-blue, and solid brown curves, respectively.
Similarly to the equivalent panels in Fig.~\ref{fig:h_vs_v_lmax}, the errors grow as velocity and time increase; however, the contributions at larger $v$ or $(\eta/M)t$ at higher PN orders are more significant than the high multipole terms.
Even for $n=20$ (PN orders greater than 20, namely), the contribution to the 22PN memory signal near $v_\ISCO$ is an order $10^{-6}$ correction (to be compared with the $10^{-12}$-level contribution for $\bar \ell =23$ in Fig.~\ref{fig:h_vs_v_lmax}).
Thus, neglecting higher PN orders is likely a larger source of error than is neglecting higher multipoles at a fixed PN order.

\begin{figure}
    \centering
    \includegraphics[width=0.48\textwidth]{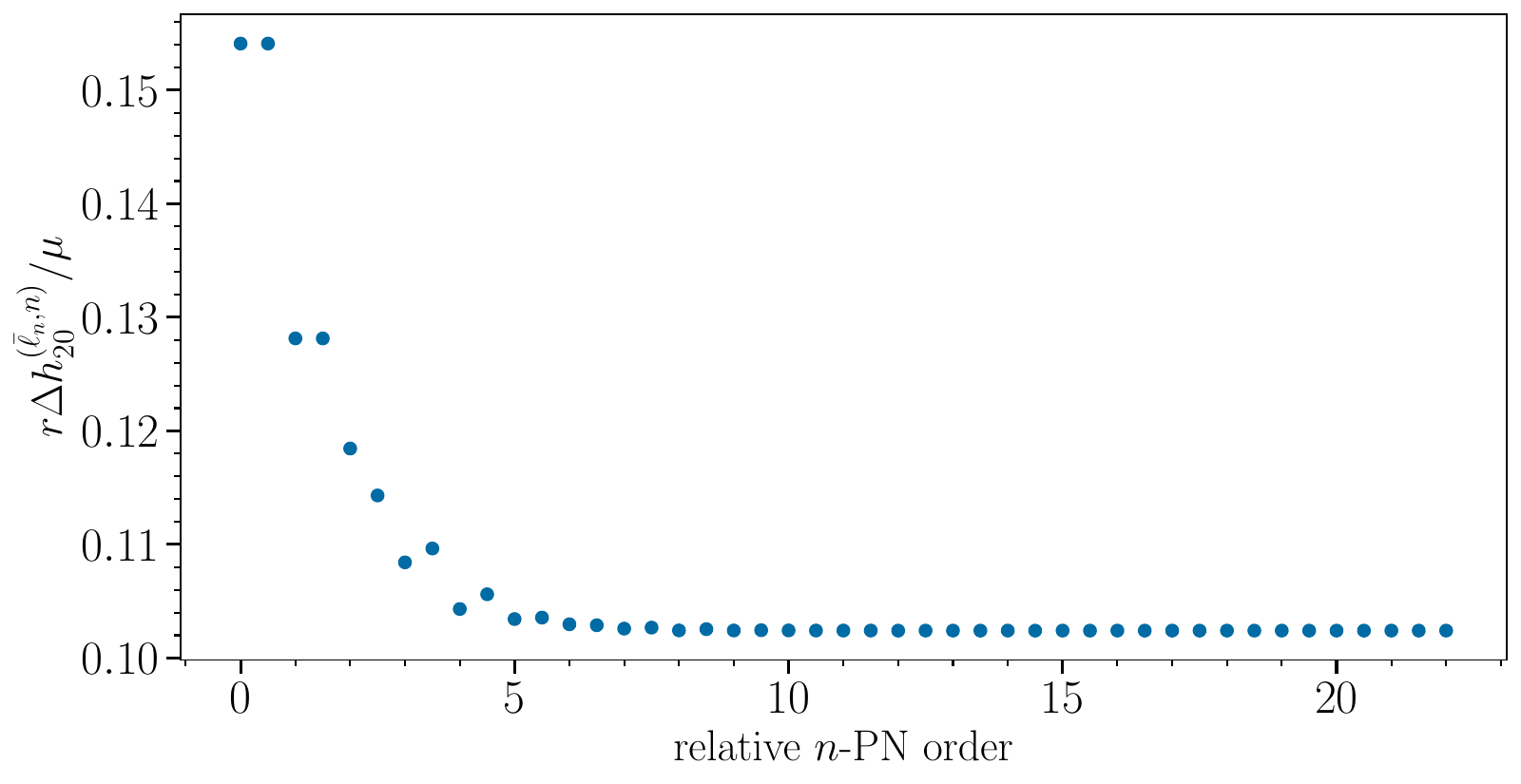}
    \includegraphics[width=0.48\textwidth]{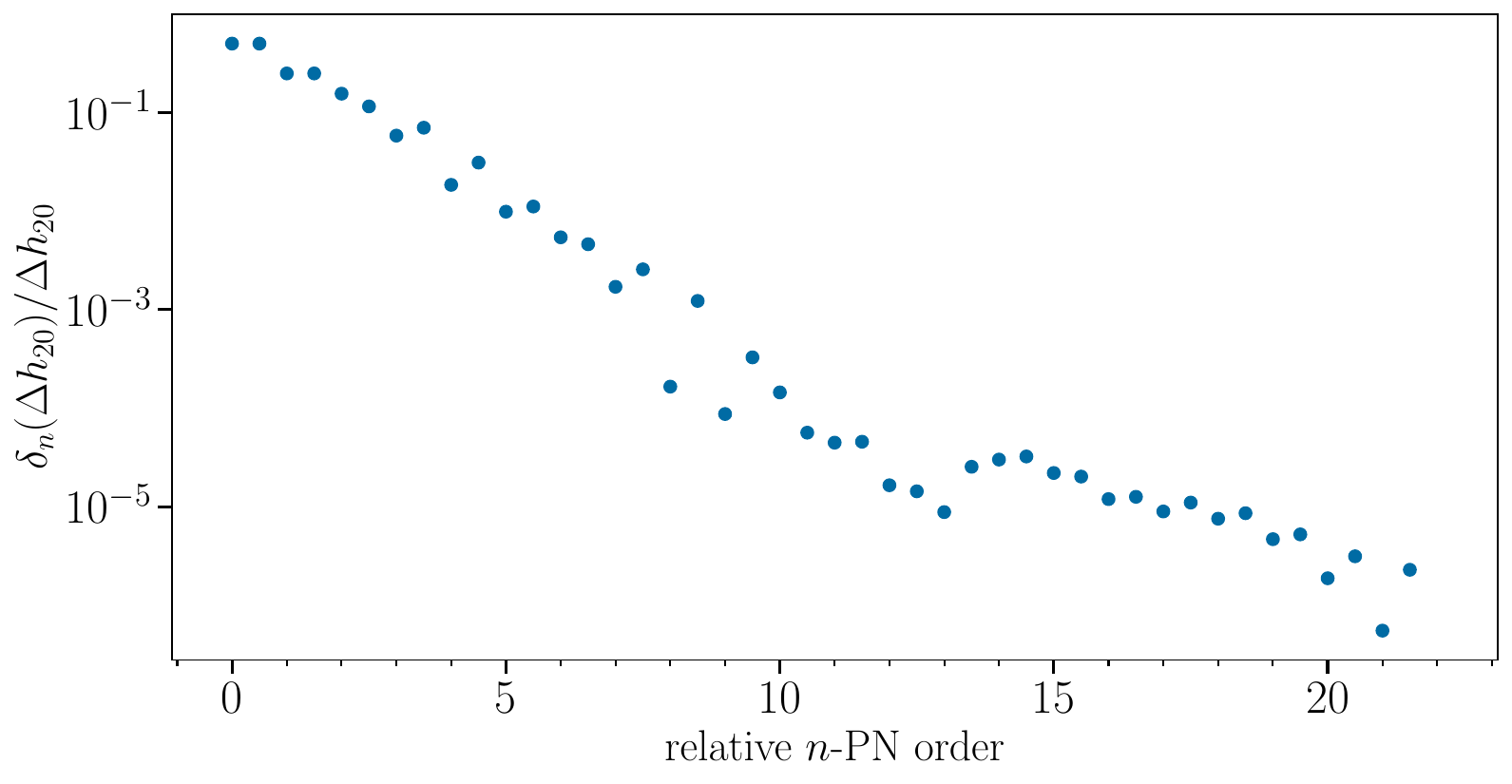}
    \caption{\textbf{Memory offset at ISCO versus PN order}:
    This figure is analogous to Fig.~\ref{fig:DeltahmemEMRI_vs_lmax}, but it focuses on the PN order rather than the multipole index.
    memory offset for different PN orders.
    \emph{Top}: The total accumulated EMRI memory offset at ISCO, $r\Delta h_{20}^{(\bar\ell_n,n)}/\mu$ is shown for all $n$-PN orders from Newtonian ($n=0$) to $n=22$ in increments of $1/2$ PN orders (single powers of $v$). 
    The $n=0$, 4 and 22 cases can be obtained from the final value of the memory as a function of velocity or time in Fig.~\ref{fig:h_vs_v_pn}.
    \emph{Bottom}: The fractional residual contribution to the total accumulated EMRI memory from PN orders greater than $n$, $\delta_n (\Delta h_{20})/h_{20}$, which is computed from Eq.~\eqref{eq:deltanh20} evaluated at ISCO. 
    Further discussion of this figure is given in the text of Fig.~\ref{subsubsec:PN-orders}.}
    \label{fig:Deltah_vs_PN}
\end{figure}
In Fig.~\ref{fig:Deltah_vs_PN}, we show the analogue for different PN orders of the results in Fig.~\ref{fig:DeltahmemEMRI_vs_lmax} for different multipole orders.
Specifically, in the top panel, we show $r\Delta h_{20}^{(\bar\ell_n,n)}/\mu$, the final memory offset at the ISCO, as a function of different PN orders $n$, relative to the leading $O(v^2)$ term.
We include all half PN orders from the Newtonian memory (0PN) and up to 22PN.
Note that the 0.5PN has the same value as the 1PN, because there are no 0.5PN terms in the leading $l=2$, $m=\pm2$ oscillatory waveform, and the 1.5PN term is the same as the 1PN term, because of a cancellation between terms in the GW luminosity at infinity and the tail terms in the leading $l=2$, $m=\pm2$ oscillatory waveform when evaluating the memory integral (this is the reason there is no 1.5PN term in the 3PN memory signal in~\cite{Favata:2008yd}).
Above roughly 7PN order, the difference between the next successive PN orders is difficult to resolve on the scale of the figure.

To better understand the importance of the higher PN-order contributions to the memory, we show $\delta_n (\Delta h_{20})/h_{20}$, the relative residual contribution to the full 22PN-order memory from PN orders greater than $n$ in the bottom panel of Fig.~\ref{fig:Deltah_vs_PN}.
Comparing this with the analogous bottom panel in Fig.~\ref{fig:DeltahmemEMRI_vs_lmax}, one again can see that higher PN contributions tend to be more significant than higher multipole-order terms.
For example, considering the furthest-right point at 21.5PN order, this point can be interpreted as showing that all 22PN order terms contribute a relative contribution to the memory of order $10^{-6}$.
Following the general trend in the bottom panel of Fig.~\ref{fig:Deltah_vs_PN}, we can estimate that the contributions of the 22.5PN and 23PN terms would be of order $10^{-7}$ at the smallest.
This is the reason why we only quote six digits of accuracy for the coefficient of the final memory offset at ISCO in Eq.~\eqref{eq:Deltahmem_EMRI_value}.
The bottom panel also shows that if one requires a relative accuracy of the memory of order only $10^{-4}$, one could instead work at 10PN order.

\section{Fit for the final memory strain} \label{sec:memory-fit}

\begin{table*}[htb]
    \centering
    \caption{Coefficients for the $\Delta h_{20}(\eta)$ polynomial fits in Eq.~\eqref{eq:Deltahmem_fit}.
    Similar to the fits described in Table~\ref{tab:tc_h0_surr}, these fits were constructed through a least-squares procedure using $50$ BBH systems with mass ratios equally spaced in $\eta$ for a range of mass ratios with the range of validity of the NRHybSur3dq8: $1\leq q\leq 8$. 
    The data in the first row, the comparable-mass-ratio fit labeled ``comp,'' allows the term linear in $\eta$ to be a free coefficient determined by least-squares fitting.
    The data in the second row, the fit labeled ``EMRI,'' fixes the linear term to be the value computed in the EMRI limit, and given in Eq.~\eqref{eq:Deltahmem_EMRI_value}.
    It is a five-parameter rather than a six-parameter fit.}
    \begin{tabular}{lcccccc}
    \hline
    \hline
       Fit type & $c_1$ & $c_2$ & $c_3$ & $c_4$ & $c_5$ & $c_6$ \\
       \hline
       comp & 0.113875 & 0.421532 & 2.44125 & -5.90547 & 33.6768 & -23.0496 \\
       EMRI   & 0.102414 & 0.770384 & -1.70081 & 18.1139 & -34.4687 & 52.7548 \\
    \hline
    \hline
    \end{tabular}
    \label{tab:Delta-hmem-fit}
\end{table*}
In this section, we will make least-squared polynomial fits of the final memory strain for nonspinning BBH mergers that can be applied to different ranges of mass ratios.

The first fit is computed using just comparable mass-ratio ($1\leq q\leq 8$) data from the hybridized 3PN memory signal with the NRHybSur3dq8 surrogate model (which follows the procedure described in Sec.~\ref{sec:memory_comp_hybridization}).
We compute the hybridized memory for 100 nonspinning BBH systems with mass ratios between $q=1$ and $q=8$, uniformly spaced in the symmetric mass ratio parameter $\eta$.
We then fit fifty of the hybridized final memory values to a sixth-order polynomial in the symmetric mass ratio $\eta$, using a linear least-squares fit, and we use the other fifty to test how well the fit compares with points not used to construct the fit.
We denote the coefficients in this fit by $c_j$, so that the final memory offset can be written as 
\begin{align}\label{eq:Deltahmem_fit}
    \Delta h_{20}(\eta) = \frac Mr \sum_{j=1}^6 c_j \eta^j .
\end{align}
We do not include a constant ($\eta$-independent) term in the fit, because the memory offset should go to zero in the test-particle limit of $\eta\rightarrow 0$.
The values of the coefficients are given in the first row of Table~\ref{tab:Delta-hmem-fit}, and plots of the fit and its residual are given in Appendix~\ref{app:memory-fit}.

Note, however, that the comparable-mass fit has a term linear in $\eta$ that does not agree with the numerical coefficient of the EMRI final memory offset in Eq.~\eqref{eq:Deltahmem_EMRI_value}.
This means that the fit, which we denote $\Delta h_{20}^\comp(\eta)$, should not be extrapolated to the EMRI limit, because it will be larger than the output of the EMRI calculation by roughly ten percent.
We can construct another fit, which we denote by $\Delta h_{20}^\EMRI$, that has the correct EMRI limit if we instead fix the linear in $\eta$ coefficient in the fit to be the value of the coefficient in Eq.~\eqref{eq:Deltahmem_EMRI_value}, $c_1 = 0.102414$, (similarly to how we fixed the coefficient $c_0 = 0$); then, we can still fit for the remaining five coefficients.
The results of the fit are shown in the second row of Table~\ref{tab:Delta-hmem-fit}.
This fit has the correct EMRI limit and agrees with the comparable-mass-ratio results (as we show in Fig.~\ref{fig:Deltah_vs_eta_EMRI} below); thus, it could be used to interpolate between the EMRI and comparable-mass-ratio limits to give an estimate of the final memory strain from IMRI systems.
We discuss this further in Appendix~\ref{app:memory-fit}.

\begin{figure}
    \centering
    \includegraphics[width=0.48\textwidth]{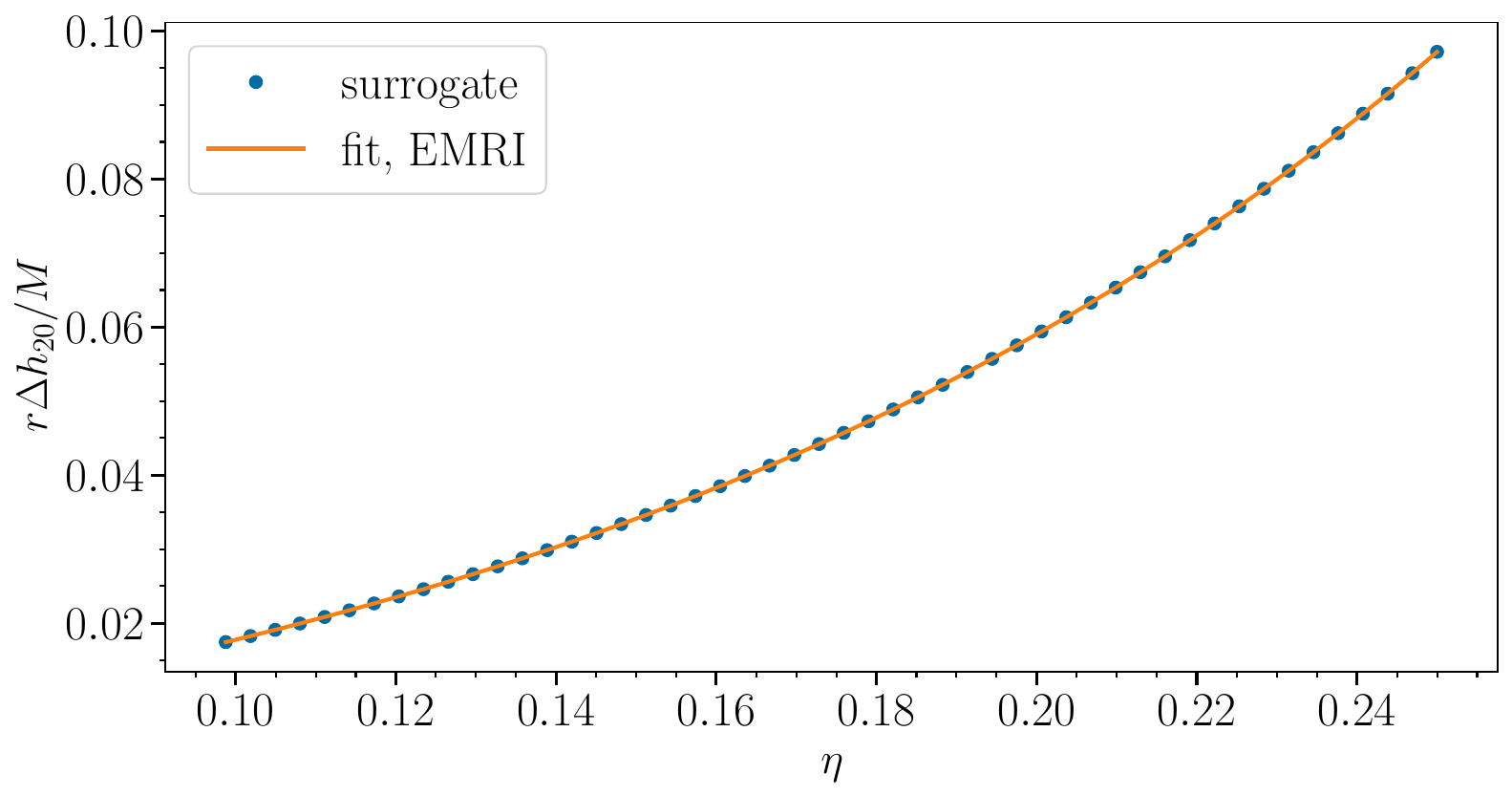}
    \includegraphics[width=0.48\textwidth]{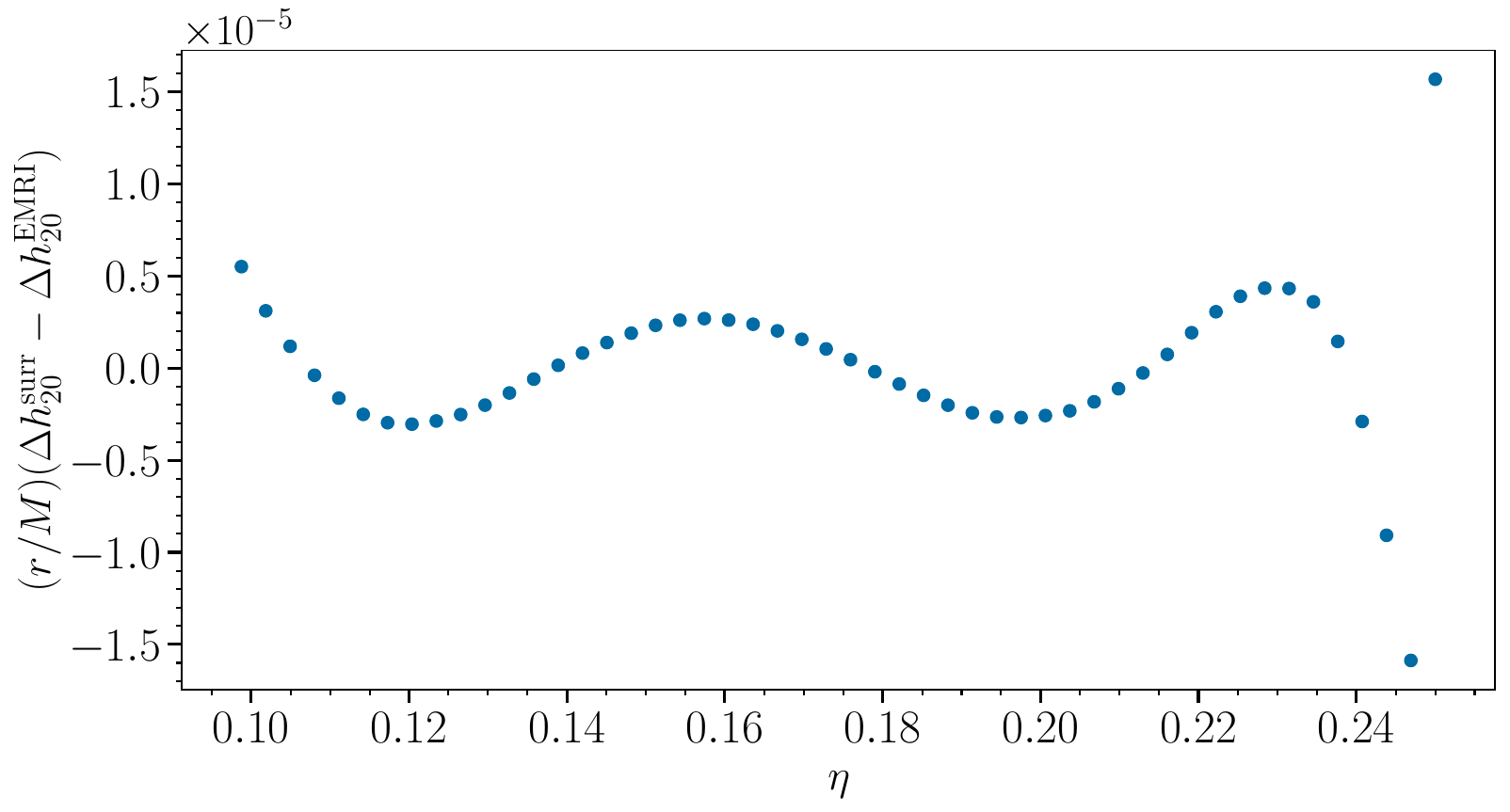}
    \caption{\textbf{Final memory offset fit and residual}:
    The blue points in both the top and bottom panels are for fifty values of the mass ratio between $q=1$ and $q=8$ that are distinct from the 50 values used to construct the fit. 
    \emph{Top}: The final memory computed from the hybridized surrogate in Sec.~\ref{sec:memory_comp_hybridization}, $\Delta h_{20}^\mathrm{surr}$, is depicted with blue dots, whereas the fit for the final memory strain, $\Delta h_{20}^\EMRI$, is the solid orange curve. 
    This fit fixes the term that is linear in $\eta$, so that it agrees with the EMRI calculation in that limit. 
    \emph{Bottom}: The blue points here depict the residual between the fit and the fifty values of the mass ratio that are used to test the fit.
    The maximum absolute error is of order $10^{-5}$ and occurs near the equal-mass-ratio case ($\eta=0.25$); however given the smaller value of the memory at smaller $\eta$, the largest relative error occurs near the smallest values of $\eta$ shown ($q=8$).}
    \label{fig:Deltah_vs_eta_EMRI}
\end{figure}
To verify that the fit for the final memory offset that includes EMRI data works well in the comparable mass ratio regime, we plot the memory offset computed from the hybridized surrogate described in Sec.~\ref{sec:memory_comp_hybridization} with fifty comparable mass-ratio values between $q=1$ and $q=8$ that were not used to construct the fit.
These are the blue points in the top panel of Fig.~\ref{fig:Deltah_vs_eta_EMRI}, and the solid orange curve is the polynomial fit that includes the EMRI data, $\Delta h_{20}^\EMRI$.
The residuals for the same fifty values of $\eta$ are shown in the bottom panel of Fig.~\ref{fig:Deltah_vs_eta_EMRI}.
Adding the EMRI data slightly made very little difference in the accuracy of the fit (see Appendix~\ref{app:memory-fit} for the analogous figure for the fit $\Delta h_{20}^\comp$), but it now recovers the correct EMRI limit.

\begin{figure}
    \centering
    \includegraphics[width=0.48\textwidth]{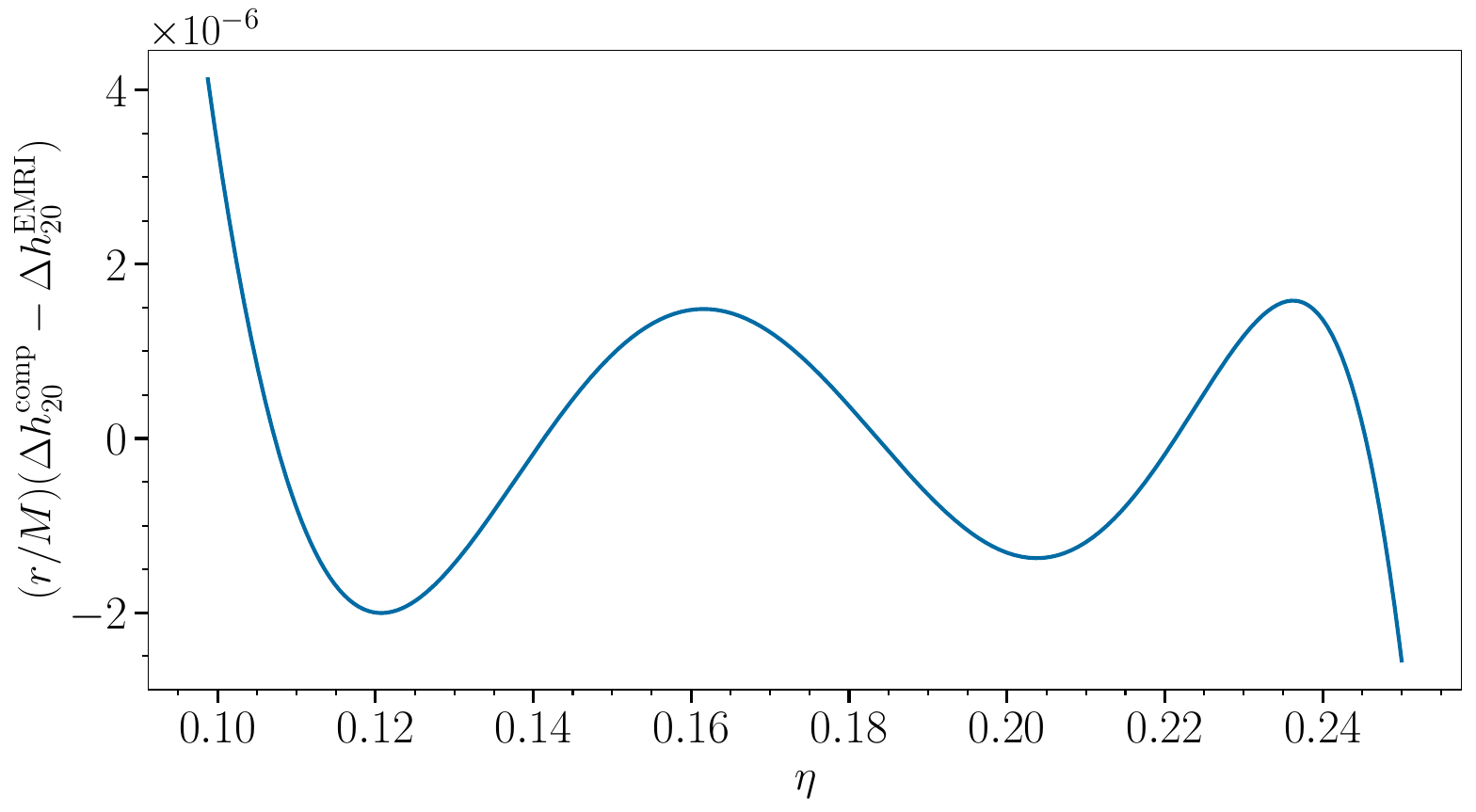}
    \includegraphics[width=0.48\textwidth]{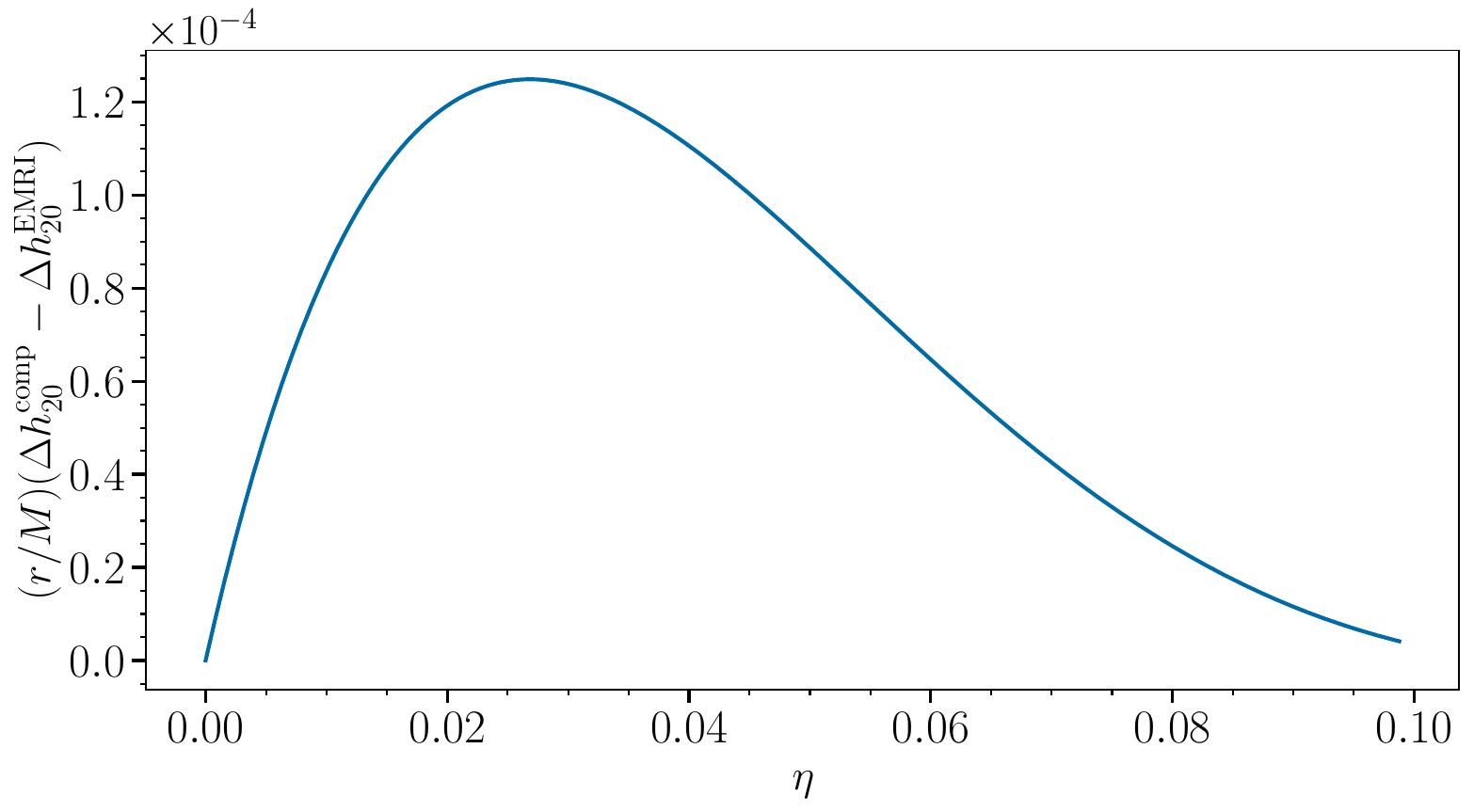}
    \caption{\textbf{Difference between the two polynomial memory-offset fits}:
    \emph{Top}: The difference between the final memory strain fit $\Delta h_{20}^\comp$ that does not use the EMRI calculation, and the fit $\Delta h_{20}^\EMRI$ that does, is plotted versus $\eta$ as a continuous function of $\eta$ for values of $q$ between 1 and 8.
    The differences between the two fits are smaller than the differences between the fit and the surrogate itself.
    \emph{Bottom}: The same difference between the final memory strain fits $\Delta h_{20}^\comp$ and $\Delta h_{20}^\EMRI$ as in the top panel, but now for values of $\eta$ in the IMRI regime. 
    The comparable-mass fit consistently estimates a larger value of the final memory offset.}
    \label{fig:Deltah_vs_eta_compEMRI}
\end{figure}
The difference between the two fits is shown in Fig.~\ref{fig:Deltah_vs_eta_compEMRI} for two different ranges of $\eta$.
The top panel spans the same range of $\eta$ as shown in Fig.~\ref{fig:Deltah_vs_eta_EMRI}; comparing with the bottom panel there, shows that the difference between the two fits is actually smaller than the difference between the hybridized-surrogate final memory offset and the fit.
For comparable mass ratios, either fit would work equally well.
The bottom panel of Fig.~\ref{fig:Deltah_vs_eta_compEMRI} shows the same difference as the top panel, but for a different range of $\eta$ corresponding to mass ratios greater than $q=8$.
The fit with EMRI data consistently gives smaller values for the memory offset than the comparable-mass-only fit.
Given the small number of NR simulations or second-order self-force calculations, there are not many robust waveforms that could be used to determine which fit performs better in this regime.
However, we do give in Appendix~\ref{app:memory-fit} an estimate for the memory in this regime by hybridizing the 3PN inspiral memory signal with the EMRISur1dq1e4.

\section{Conclusions} \label{sec:conclusions}

This paper focused on modeling two aspects of the GW memory signal from nonspinning BBH mergers on quasicircular orbits: 
\begin{enumerate}
    \item[(i)] a 22PN-order calculation of the memory signal as a function of time or PN parameter for EMRIs
    \item[(ii)] and two polynomial fits for the final memory offset after ringdown that used numerical-relativity data in one case, exclusively, and in the other case, NR data in combination with the result of the EMRI calculation.
\end{enumerate}
We focused on the nonlinear memory effect, and we used the continuity equation for the supermomentum to compute the memory signal from oscillatory ($m\neq 0$) waveform modes without the memory effect.
To obtain these results, we had to make some improvements in calculating the memory offset for comparable mass ratios systems, and we needed to perform a new calculation of the memory effect from EMRI systems.

For comparable mass ratios, most of the memory accumulates during the late inspiral, merger and ringdown; however, there is a nontrivial contribution from the earlier inspiral.
While the numerical-relativity hybrid surrogate model NRHybSur3dq8 has oscillatory waveform multipoles that were hybridized with Effective-One-Body waveforms to produce arbitrary-length signals, evaluating the surrogate for long enough to resolve the offset accumulated during the early inspiral is time consuming.
Instead, we hybridized the surrogate memory signal with a 3PN memory waveform.
The 3PN waveform accounts for the memory accumulated at all times before the starting time of the surrogate signal, and it is faster to evaluate over long times.
We computed a polynomial fit for this initial offset that must be added to the surrogate memory at the starting time (and also a polynomial fit for the PN time of coalescence, which is a free parameter in the 3PN waveform used in the hybridization).
This fit is valid for mass ratios from one to eight.
The data contained within this fit feeds into the other polynomial fits that we construct for the final memory offset.

For EMRIs, the memory offset, to leading order in the symmetric mass ratio, accumulates just during the inspiral.
We computed the EMRI memory up to 22PN order relative to the Newtonian memory, using resummed, factorized waveforms for EMRIs that were computed by Fujita in \cite{Fujita:2012cm} for a test particle orbiting a Schwarzschild BH on a quasicircular orbit.
To have an expression for the $l=2$, $m=0$ harmonic of the memory at 22PN order, that requires using oscillatory modes up to $l=25$.
To compute the time dependence of the memory signal, we assumed that the test particle evolves adiabatically between geodesics with different energies such that the change in the geodesic energy is equal to the energy radiated to infinity and into the horizon (which made use of the GW luminosities computed in~\cite{Fujita:2012cm}).
Our 22PN order result, restricted to 3PN order, agrees with the 3PN expression used in the hybridization procedure discussed above, when the 3PN expression is restricted to linear order in the symmetric mass ratio $\eta$.

We also investigated how the EMRI memory signal behaves as a function of the number of oscillatory modes $l$ modes and of the post-Newtonian order.
The largest contribution to the memory comes from the quadrupole ($l=2$) and octopole ($l=3$) oscillatory modes, with the higher oscillatory modes contributing at most a few percent to the total memory signal; for certain applications, the quadrupole and octopole modes would be sufficient.
Different multipole orders did not contribute uniformly: the final memory offset, for example, adding modes between $l=4$ and $l=7$ caused the memory to converge rapidly towards the full $l=25$ expression; including modes between $l=8$ and $l=14$ produced little change in the relative accuracy; the convergence sped up above this value of $l$ again.
We performed a similar analysis of the contributions of different PN orders to the full 22PN expression (though note that truncating the series at lower PN orders also truncates the highest multipole order needed in the computation).
The memory signal and the final memory offset do converge to the 22PN expression as higher PN orders are included, but the rate at which it converges as a function of PN order is slower than the analogous convergence with the highest multipole $l$ included in the calculation.
In this sense, ignoring higher PN orders has a larger effect than ignoring higher $l$ oscillatory modes.

We then used the EMRI and comparable mass calculations to construct two polynomial fits in $\eta$ for the final memory strain for nonspinning BBH mergers.
Both fits used comparable mass-ratio data with mass ratios within the range $1\leq q\leq 8$, which were computed from our hybridization of numerical-relativity surrogate.
The first fit used only this data, and it performed well in its range of validity, but it overestimated the memory in the EMRI limit when it was extrapolated to small $\eta$.
To incorporate the result of the EMRI calculation into a new fit, we used the same comparable-mass data, but we fixed the coefficient linear in $\eta$ to the computed value of the EMRI memory offset.
This polynomial fit has the correct EMRI limit and it performed similarly to the first fit in the comparable-mass regime.
Incorporating information about EMRIs into the fit that largely used comparable mass-ratio data from numerical-relativity simulations allowed the memory to be interpolated, rather than extrapolated into the IMRI regime, where there are fewer reliable waveforms (though see, e.g.,~\cite{Wardell:2021fyy,Albertini:2022rfe,Albertini:2022dmc,vandeMeent:2023ols} for notable exceptions).

We can foresee a few applications of the results of this paper.
When it is necessary to know the amount of memory accumulated during the inspiral (for example, in setting initial data in Cauchy-characteristic-extraction simulations) the polynomial fit of the memory accumulated during the early inspiral could be useful.
While EMRIs are a target GW source for the LISA detector, the long timescale over which the memory accumulates for most EMRI systems would make the signal fall out of LISA's frequency range; however for lighter IMRI systems the frequency ranges are more consistent with what LISA could measure.
The study~\cite{Islam:2021old} focused on very light IMRIs (which were challenging for LIGO and Virgo to detect, but could be detected by Einstein Telescope and Cosmic Explorer); it would be interesting to revisit this study in the context of LISA.
To use the analytical waveforms here, however, it would be useful to generalize the calculations to spinning primaries, and also eccentric systems, to have a wider coverage of the parameter space of binaries.\footnote{While this paper was under review, a pre-print~\cite{Cunningham:2024dog} appeared which computed the memory for spinning binaries and included post-adiabatic corrections, too.
Reference~\cite{Cunningham:2024dog} also provides further discussion about computing the memory signal from EMRIs on eccentric orbits.}
The fit for the memory offset could also be useful for interpreting the results of a potential pulsar-timing-array detection of a burst with GW memory, as discussed in Sec.~\ref{sec:intro}.
Finally, this paper is a first step towards developing a stand-alone waveform model for the nonlinear gravitational wave memory effect for BBH mergers.
The fit will feed into the construction of both time- and frequency-domain waveform models, which could be used for searches for the memory effect.

\begin{acknowledgments}
A.E.\ and D.A.N.\ acknowledge support from NSF Grants No.\ PHY-2011784 and No.\ PHY-2309021.
They thank Niels Warburton for discussions about the EMRI calculation during the early stages of this project.
They also thank an anonymous referee for helpful feedback on this paper.
\end{acknowledgments}

\appendix

\section{Properties of the memory integrand} \label{app:memory-integrand}

In this appendix, we give a few supplementary results on the memory integrands $dh_{20}/dt$ and $dh_{20}/dv$, which highlight other features of the contributions of higher multipole and higher PN terms to the memory signal.

\subsection{Memory integrand at different multipole orders}

\begin{figure*}
    \centering
    \includegraphics[width=0.48\textwidth]{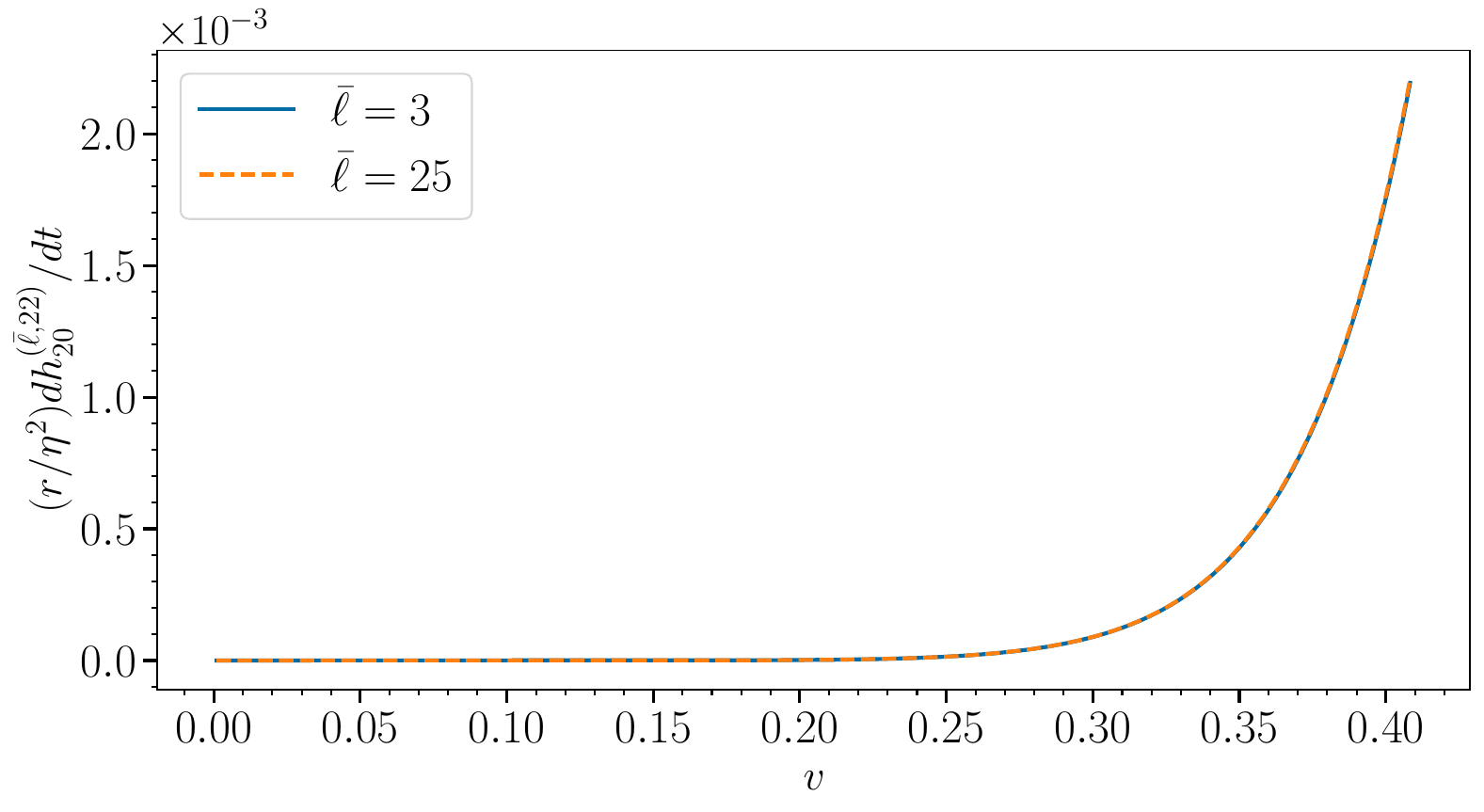}
    \includegraphics[width=0.48\textwidth]{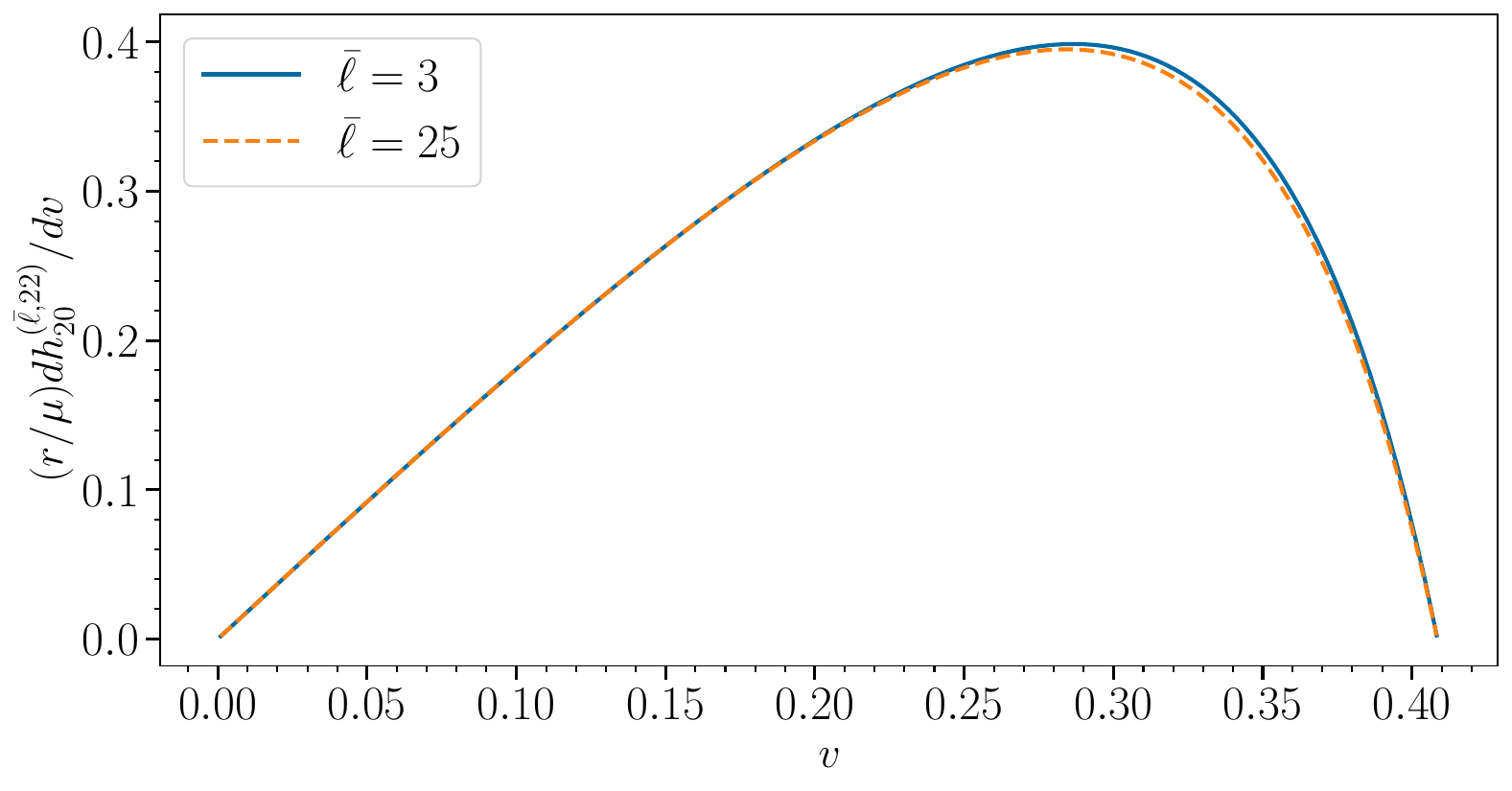}
    \includegraphics[width=0.48\textwidth]{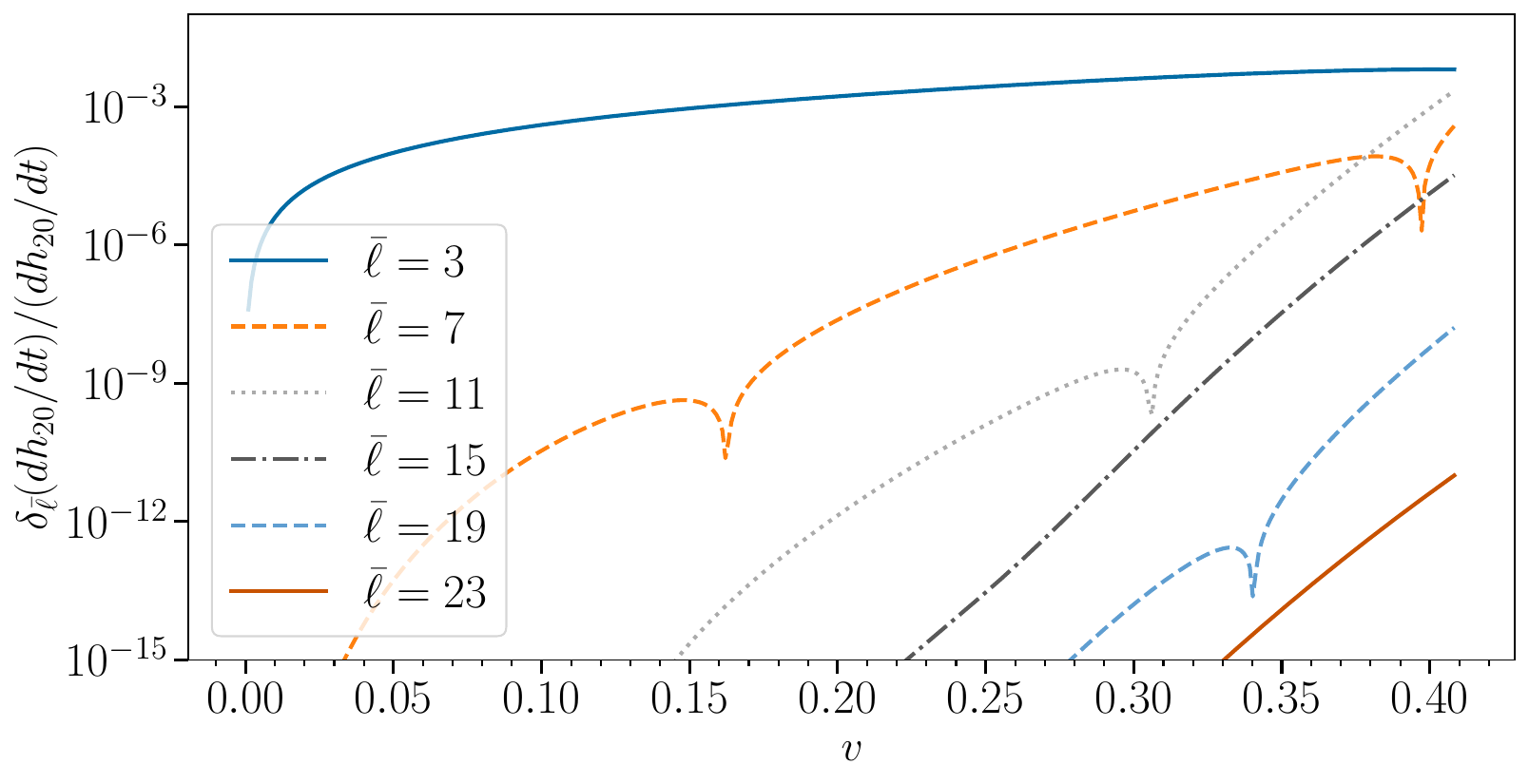}
    \includegraphics[width=0.48\textwidth]{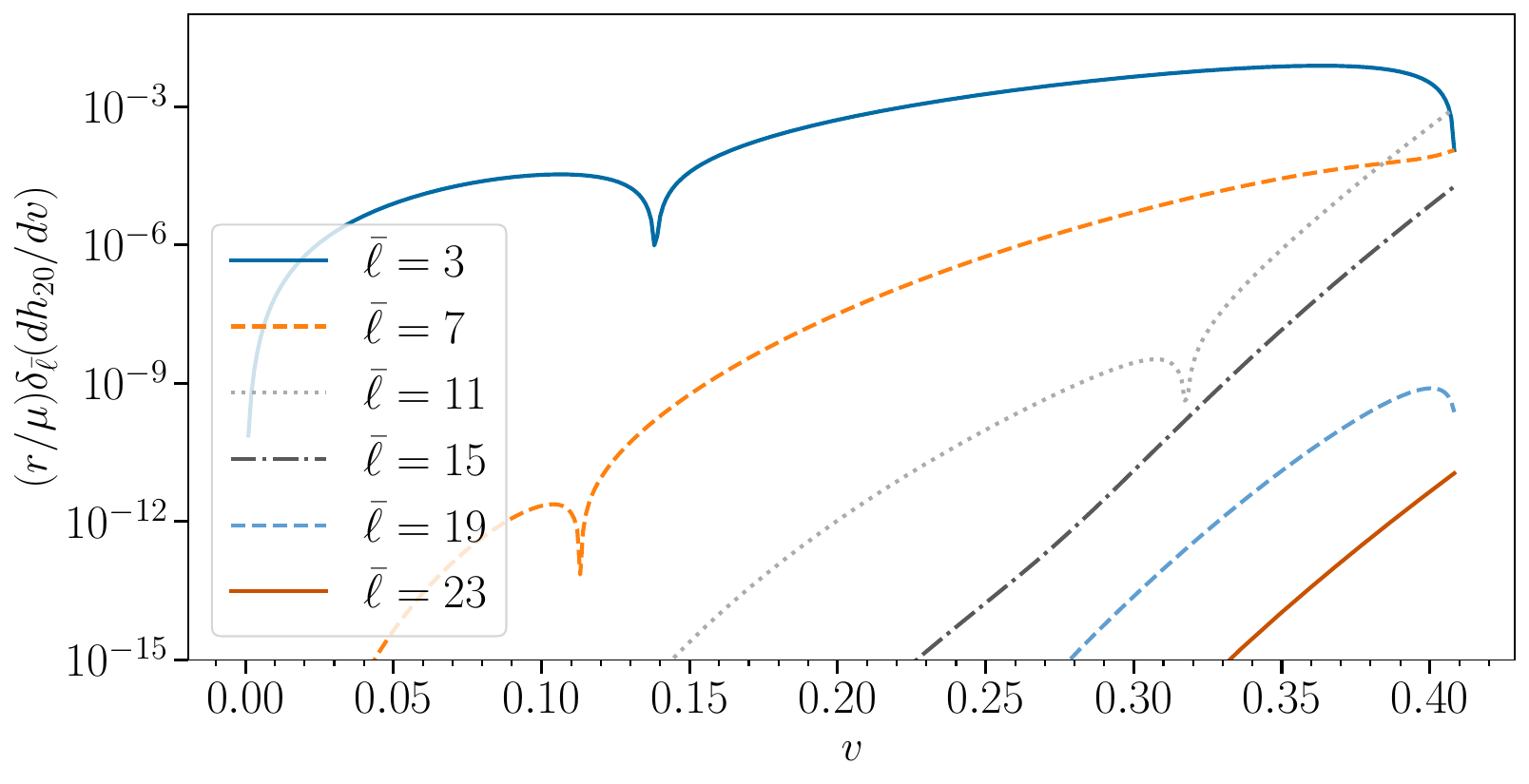}
    \caption{\textbf{Memory integrand versus velocity for different multipole orders}:
    This figure is the analogue of the left column of Fig.~\ref{fig:h_vs_v_lmax}, but here the left column shows $(r/\eta^2)dh_{20}^{(\bar\ell,22)}/dt$ in the top panel and the relative contribution from higher modes, $\delta_{\bar\ell}(dh_{20}/dt)/(dh_{20}/dt)$, in the bottom panel, where the $\delta_{\bar\ell}$ notation is defined analogously to the expression for the memory signal in Eq.~\eqref{eq:deltalh20}.
    The right column shows the integrand of the memory $(r/\mu) dh_{20}^{(\bar\ell,22)}/dv$ in the top panel when the integral is performed over velocity rather than time (as in the left column).
    The fractional relative contribution from higher multipoles is in the bottom panel of the right column.
    The coloring and line styles of the curves correspond to the same values of $\bar \ell$ as in Fig.~\ref{fig:h_vs_v_lmax}, and the values of the velocity depicted on the horizontal axis is also the same.}
    \label{fig:dhdt_vs_v_lmax}
\end{figure*}
In Fig.~\ref{fig:dhdt_vs_v_lmax}, the integrand for the memory (for an integral over time), $(r/\eta^2)dh_{20}^{(\bar\ell,22)}/dt$, is shown on the top left, and the integrand for an integral over velocity, $(r/\mu) dh_{20}^{(\bar\ell,22)}/dv$ is shown in the top-right panel, for different values of $\bar\ell$.
The velocity values and the line styles for the curves are identical to those in the left panels of Fig.~\ref{fig:h_vs_v_lmax}. 
The bottom panels show the relative contributions of higher multipoles to the full memory computed with all multipoles up to $\bar\ell=25$ at 22PN order.
The notation used is analogous to that defined in Eq.~\eqref{eq:deltalh20}, with the memory integrand replacing the memory signal.

The main feature to highlight in Fig.~\ref{fig:dhdt_vs_v_lmax} is that while $dh_{20}/dt$ is a monotonically increasing function with $v$, the velocity integrand $dh_{20}/dv$ has a peak near $v\approx 0.3$ for all multipole orders.
This arises from the terms $(dE/dv)/(dE_\GW/dt)$ used to convert the integral with respect to time to one with respect to velocity.
Otherwise, there are not any substantive differences between the relative importance of the higher spherical-harmonic modes from the memory integrand versus the memory signal.

\subsection{Memory integrand at different post-Newtonian orders} \label{subapp:PN}

\begin{figure*}
    \centering
    \includegraphics[width=0.48\textwidth]{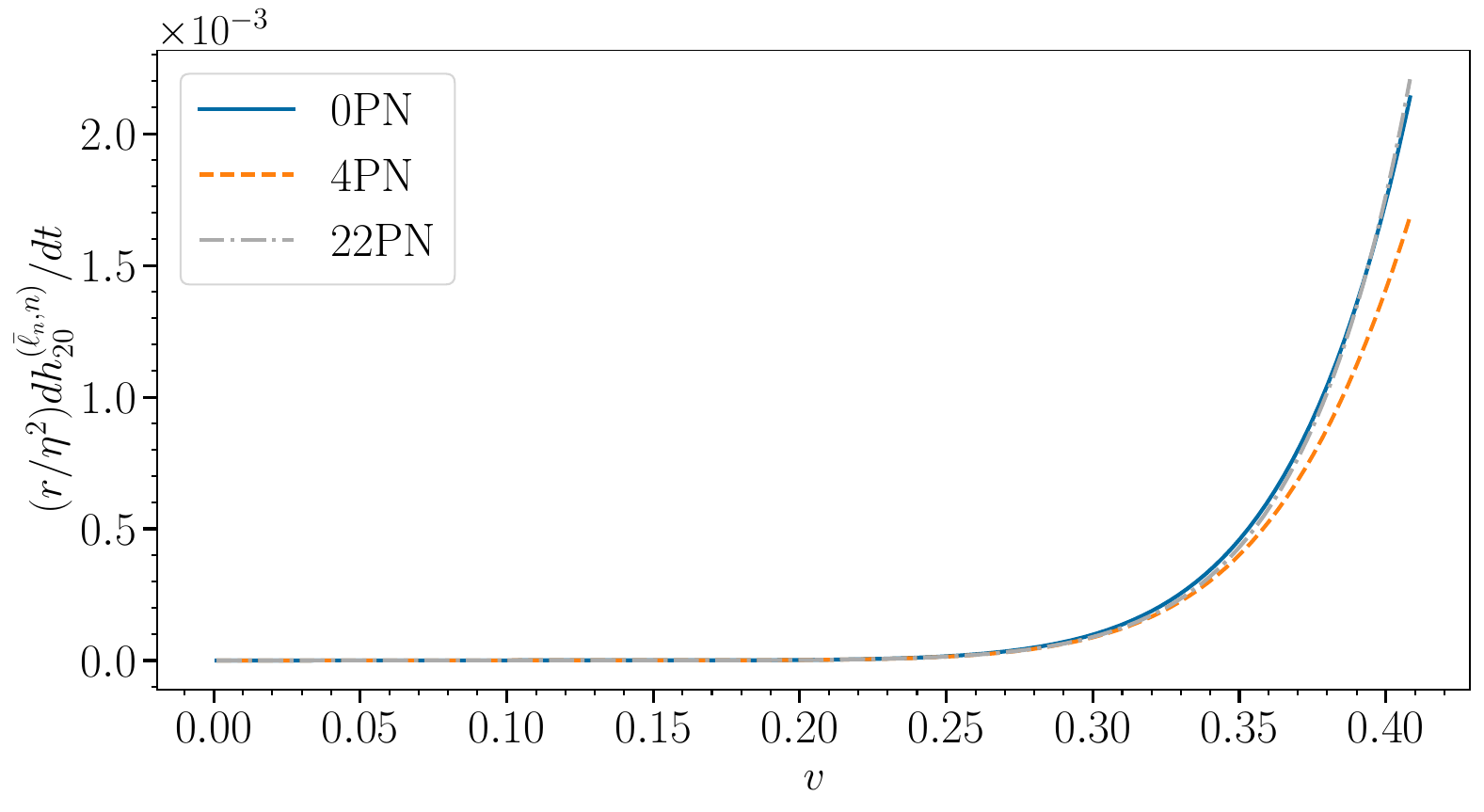}
    \includegraphics[width=0.48\textwidth]{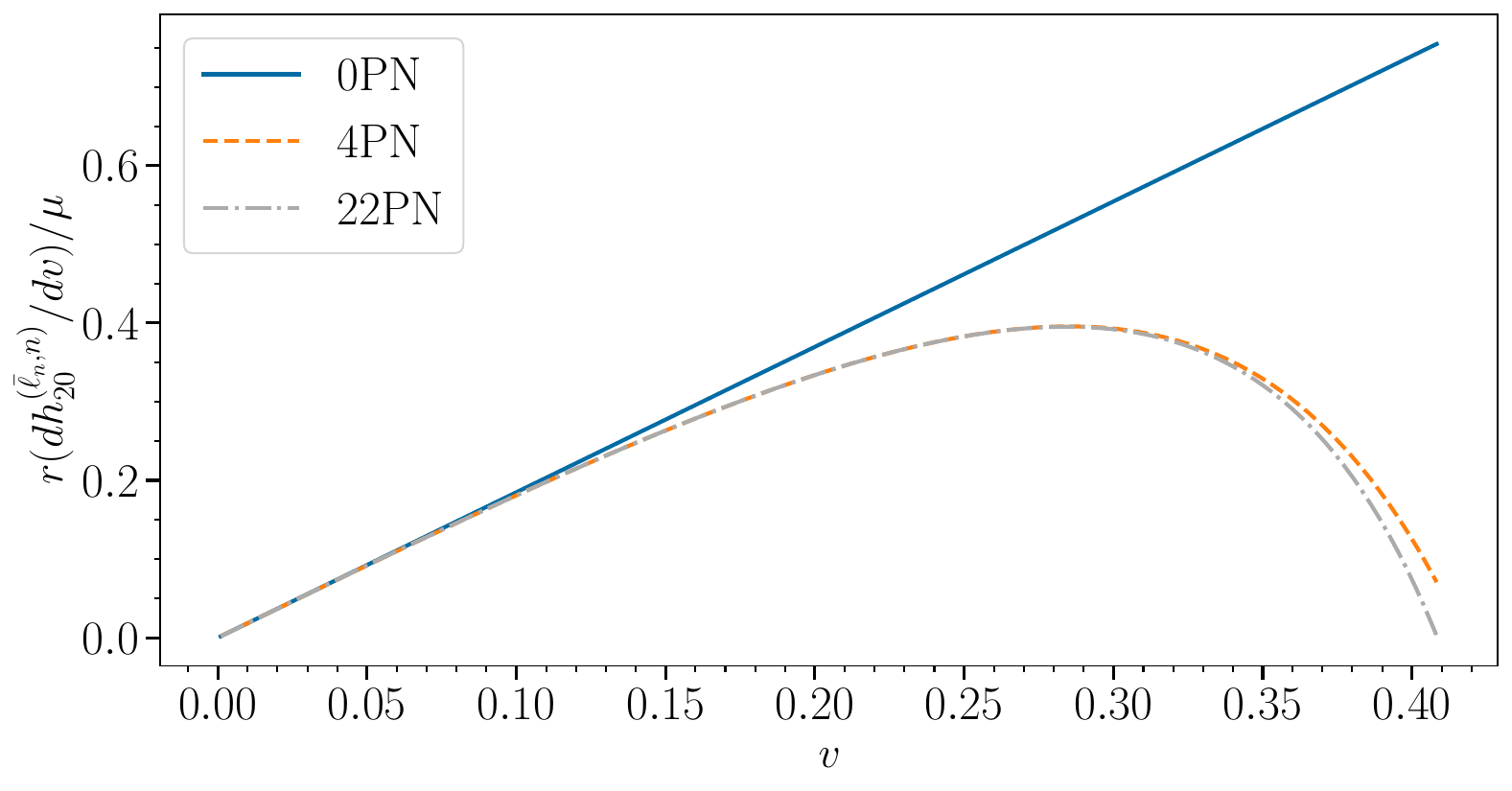}
    \includegraphics[width=0.48\textwidth]{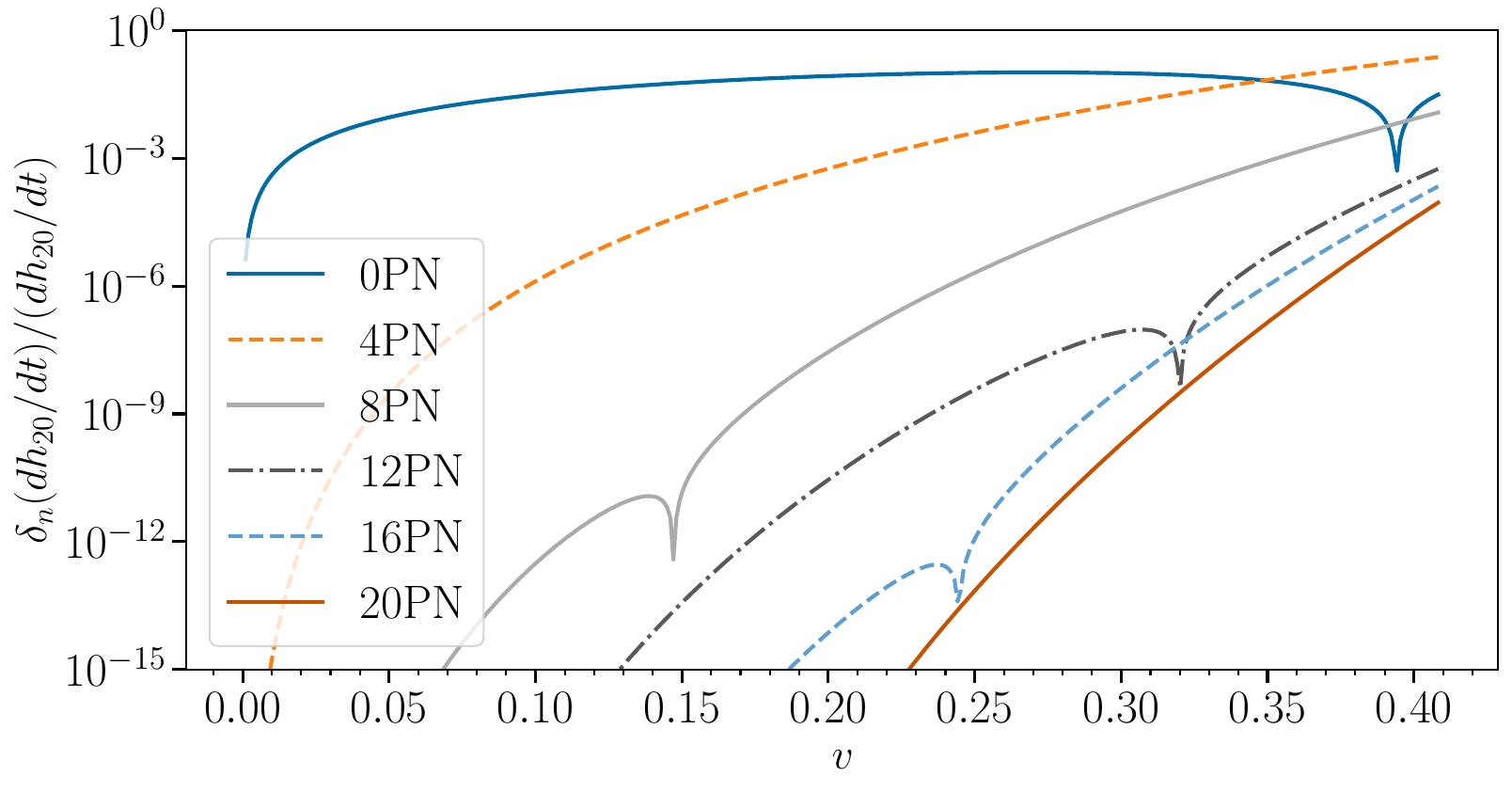}
    \includegraphics[width=0.48\textwidth]{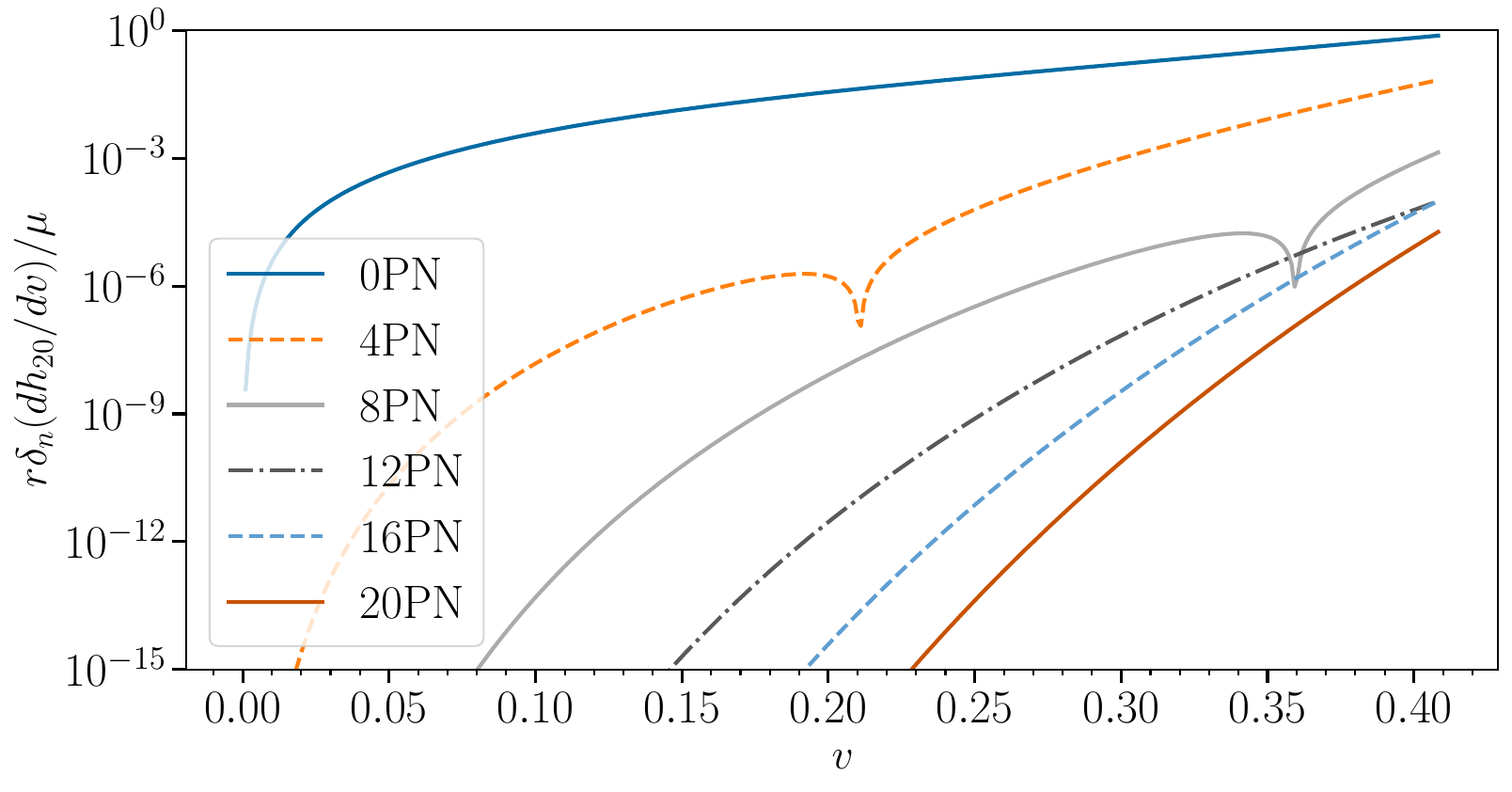}
    \caption{\textbf{Memory integrand versus velocity for different PN orders}:
    This figure is the analogue of Fig.~\ref{fig:dhdt_vs_v_lmax} for PN order rather than multipole order.
    It uses the same interval of velocities and has the same line styles and colors for the different curves as is used in Fig.~\ref{fig:h_vs_v_pn}.
    Additional discussion of this figure is given in the text of Appendix~\ref{subapp:PN}.}
    \label{fig:dhdt_vs_v_pn}
\end{figure*}
Figure~\ref{fig:dhdt_vs_v_pn} has several similarities with both Fig.~\ref{fig:dhdt_vs_v_lmax} and the left column of Fig.~\ref{fig:h_vs_v_pn}.
Like Fig.~\ref{fig:dhdt_vs_v_lmax}, it focuses on the memory integrand with respect to time in the left column and velocity in the right column.
Like Fig.~\ref{fig:h_vs_v_pn}, it focuses on the contributions and results from different PN orders, and it uses the same interval of velocities and the same line styles and colors there.
However, the memory integrand $(r/\eta^2)dh_{20}^{(\bar\ell_n,n)}/dt$ does have some more unusual properties.
In the top-left panel, the Newtonian (0PN) curve agrees better with the 22PN result than the 4PN result does at higher velocities above $v\approx 0.35$, though not below.
However, comparing this with $(r/\mu)dh_{20}^{(\bar\ell_n,n)}/dv$, the 4PN result is closer to the 22PN result for all values of $v$ shown.
Otherwise, there is a relatively clear trend of higher PN orders contributing less to the total value of the memory.
Thus, we suspect that the improved performance of the Newtonian signal at larger velocities is coincidental.

Another feature worth noting about the top-right panel is that at Newtonian order $dh_{20}/dv$ is proportional to $v$, which is why the 0PN curve is a straight line in the upper right panel.
Higher PN orders do not have this property, and instead have a peak around $v\approx 0.3$.
This was noted in Sec.~\ref{subsubsec:PN-orders} as the reason for why the Newtonian memory is notably larger than all the other higher PN cases shown.

\section{Further analysis of the final memory fit} \label{app:memory-fit}

\begin{figure}
    \centering
    \includegraphics[width=0.48\textwidth]{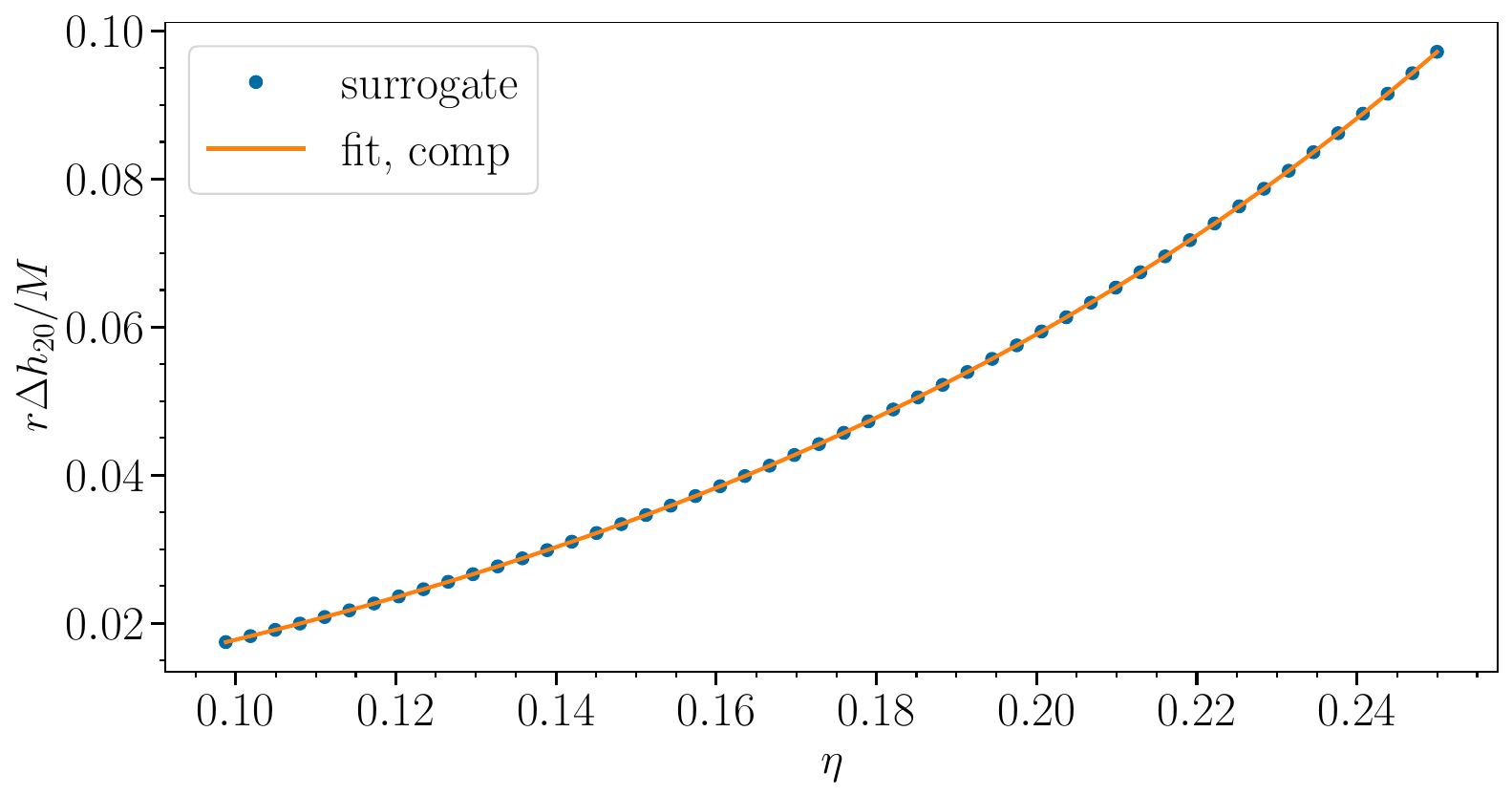}
    \includegraphics[width=0.48\textwidth]{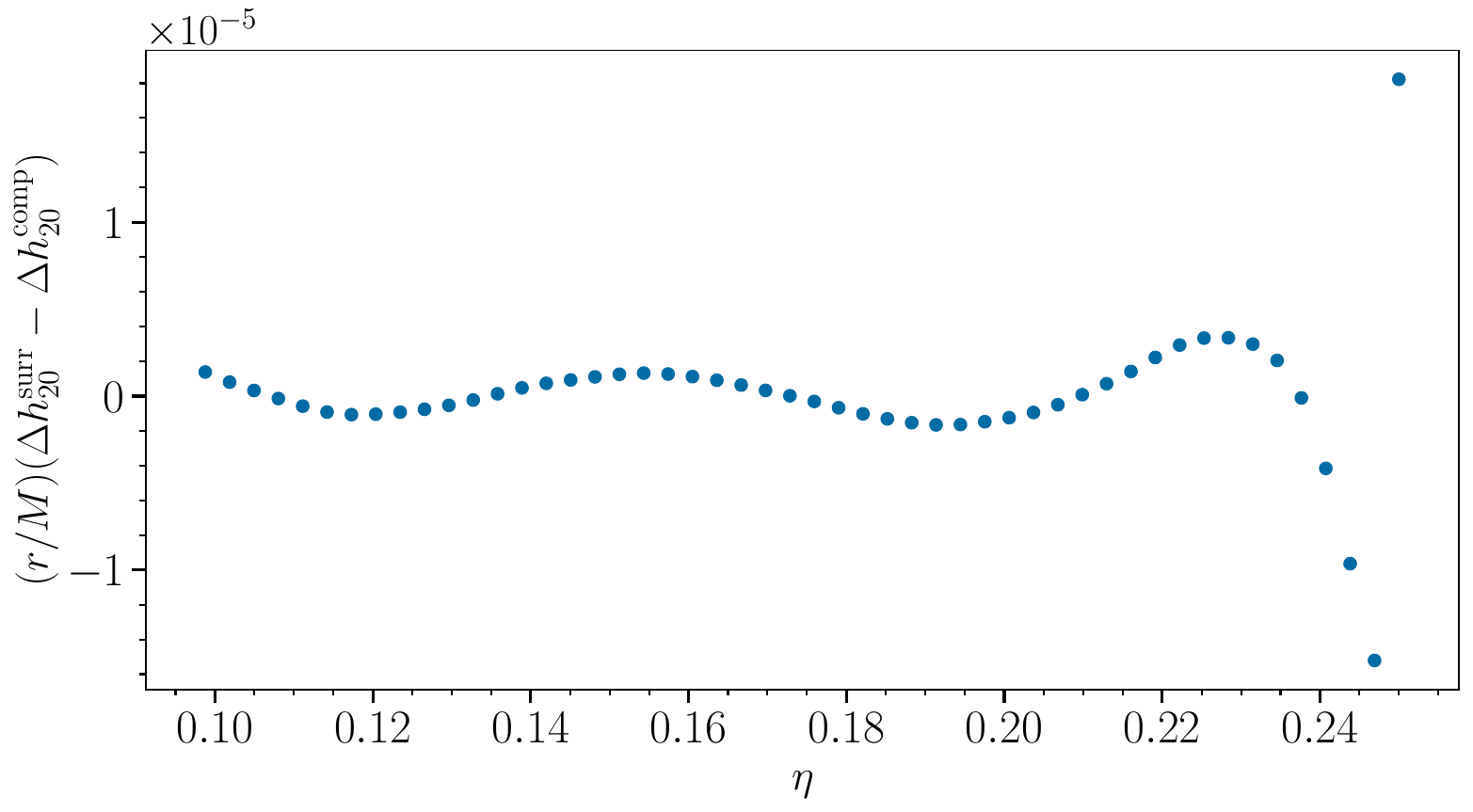}
    \caption{\textbf{Final memory offset fit and residuals}:
    This figure is identical to Fig.~\ref{fig:Deltah_vs_eta_EMRI}, except we plot the final memory computed from the comparable-mass-ratio fit, $\Delta h_{20}^\comp$ as the solid orange line in the top panel, and the residual  between $\Delta h_{20}^\comp$ and $\Delta h_{20}^\surr$ is shown in the bottom panel.}
    \label{fig:Deltah_vs_eta_comp}
\end{figure}
Figure~\ref{fig:Deltah_vs_eta_comp} is similar to Fig.~\ref{fig:Deltah_vs_eta_EMRI} in the main text, but the fit used is $\Delta h_{20}^\comp$, which does not use any information from the EMRI calculation, whereas the fit $\Delta h_{20}^\EMRI$ in Fig.~\ref{fig:Deltah_vs_eta_EMRI} does.
We include this figure primarily for completeness.

\begin{figure}[thb]
    \centering
    \includegraphics[width=0.48\textwidth]{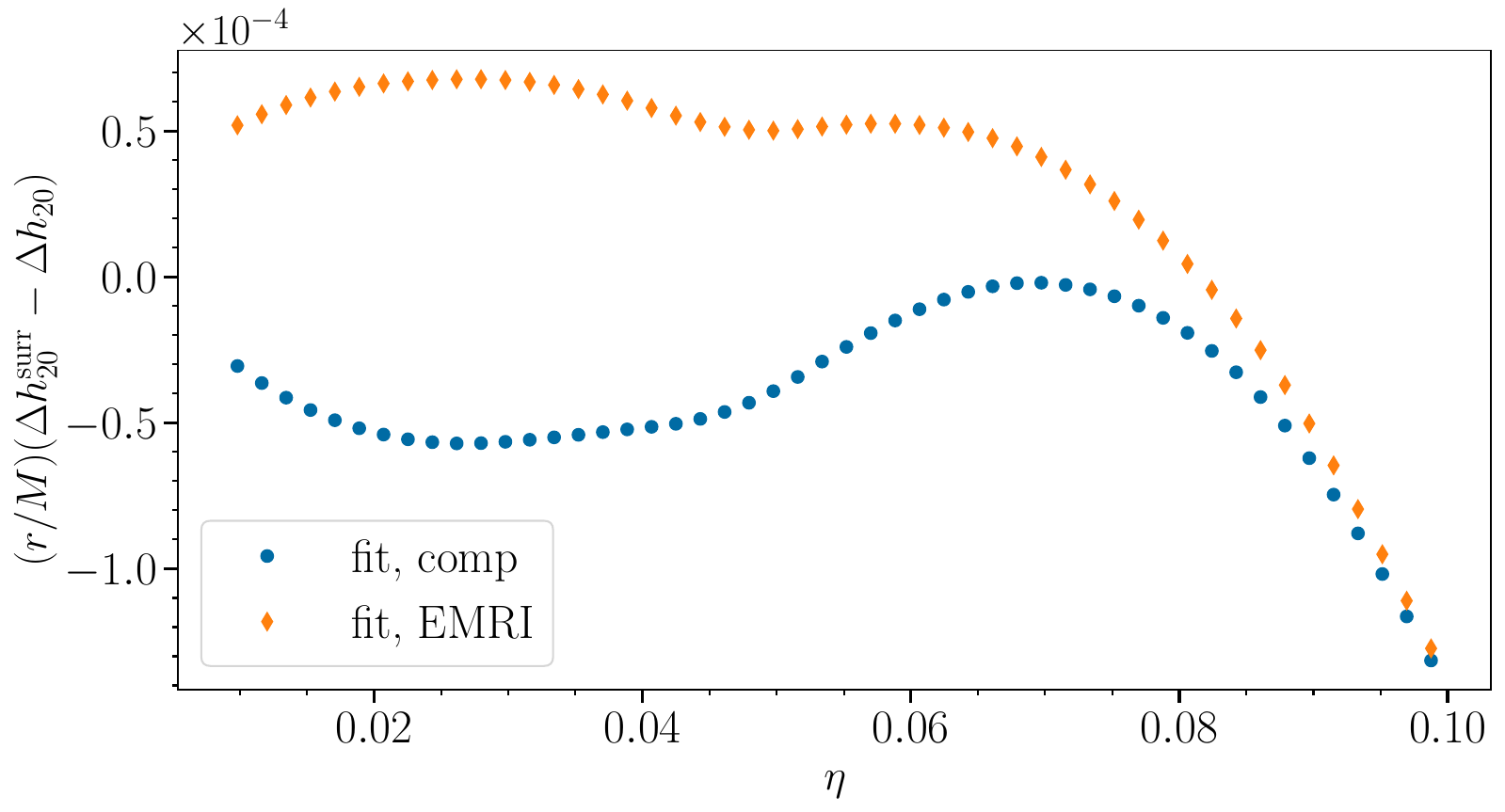}
    \caption{\textbf{Residuals between two fitting functions and an estimate for the memory from IMRIs}:
    We discuss how we compute the memory from hybridizing the 3PN inspiral memory with EMRISur1dq1e4 surrogate model to obtain the $\Delta h_{20}^\surr$ in the text of Appendix~\ref{app:memory-fit}.
    We use this value to compute the absolute error between it and the comparable mass fit $\Delta h_{20}^\comp$ (the blue filled circles) or the EMRI fit  $\Delta h_{20}^\EMRI$ (the orange filled diamonds).
    At larger $\eta$ values, both fits tend to predict a larger final memory offset than our IMRI estimate, but at smaller $\eta$ values, the EMRI surrogate is smaller, whereas the comparable fit is larger.}
    \label{fig:Deltah_vs_emrieta_compEMRI}
\end{figure}
A useful check of the EMRI fit would be to compare it with an accurate calculation of the memory effect for values of $\eta$ between 0 and 0.1 (i.e., mass ratios greater than $q=8$) to determine how the two fits perform.
Specifically, we would like to determine if performing extrapolation by using $\Delta h_{20}^\comp$ is more or less accurate in this range of $\eta$ than performing interpolation from the EMRI data point ($\eta\rightarrow 0$) to the next largest value of $\eta \approx 0.1$ ($q=8$) from the hybridized surrogate.
Using the EMRISur1dq1e4 surrogate model cannot address this, because it is not hybridized and spans a time interval of about $10^4 M$, whereas the EMRI calculation showed that long time spans during the inspiral are required to obtain an accurate calculation of the final memory offset accumulated from a widely separated EMRI.
We cannot simply hybridize with the PN-expanded EMRI memory signal, because it neglects order $\eta^2$ corrections, which become important as $\eta$ is in the 0.01 to 0.1 range.

Instead, we opt to use the 3PN memory, which has nonlinear corrections in $\eta$, to hybridize with EMRISur1dq1e4 surrogate model, so as to obtain an estimate of the final memory offset for IMRI mass ratios.
Specifically, we compute the memory for 50 BBH systems with mass ratios in the interval $8\leq q \leq 100$, uniformly spaced in $\eta$, using the EMRISur1dq1e4 surrogate model.
We then hybridize the surrogate memory with a 3PN waveform, over a length of time of $10^3M$, between $-10^4M\leq t\leq -9\times 10^3 M$, as the EMRI surrogate does not extend before $-10^4M$.
This allows us to compute data to compare against both of our fitting functions $\Delta h_{20}^\comp$ and $\Delta h_{20}^\EMRI$.
These residuals are shown in Fig.~\ref{fig:Deltah_vs_emrieta_compEMRI}.

At the largest values of $\eta$ shown ($q=8$) both fits give larger values of the memory than our estimate with the hybridized EMRISur1dq1e4 data, which fit the hybridized NRHybSur3dq8 data to a higher precision.
The EMRISur1dq1e4 surrogate was trained on EMRI simulations and rescaled to match the numerical-relativity results at less extreme mass ratios; however, based on this memory calculation, it does not seem to be accurate to more that one percent.
At the smallest $\eta$ value of roughly $0.01$, both fits have a comparable magnitude of the error, but the EMRI is consistently lower than the EMRI surrogate estimate, whereas the comparable mass fit is larger.
Given the limitations of the data and waveform models for computing the memory in this parameter space of $\eta$, such a consistency check of the fits will likely have to wait for more numerical relativity or second-order self-force calculations to make a more definitive assessment of the accuracy of these two fitting functions in this regime.

\end{document}